\documentclass[12pt]{article}
\usepackage{a4wide}
\usepackage{amssymb}
\usepackage{amsfonts}
\usepackage{graphicx}
\usepackage{caption}
\usepackage{amsmath}
\usepackage{color}
\begin{document}
{\renewcommand{\thefootnote}{\fnsymbol{footnote}}
\medskip
\begin{center}

{\LARGE  Minisuperspace results\\ for causal dynamical triangulations }\\
\vspace{1.5em}

Bekir Bayta\c{s},${^{a}}$\footnote{e-mail address: {\tt bub188@psu.edu}}
Martin Bojowald,${^{a}}$\footnote{e-mail address: {\tt bojowald@gravity.psu.edu}}
Sean Crowe,${^{a}}$\footnote{e-mail address: {\tt stc151@psu.edu}} and
Jakub Mielczarek${^{b,c}}$ \footnote{e-mail address: {\tt jakub.mielczarek@uj.edu.pl}} \\
\vspace{0.5em}
${^{a}}$Institute for Gravitation and the Cosmos,\\
The Pennsylvania State
University,\\
104 Davey Lab, University Park, PA 16802, USA\\
\vspace{0.5em}
${^{b}}$ CPT, Aix-Marseille Universit\'e, Universit\'e de Toulon, CNRS, \\
F-13288 Marseille, France\\
\vspace{0.5em}
${^{c}}$ Institute of Physics, Jagiellonian University, \\ 
{\L}ojasiewicza 11, 30-348 Cracow, Poland
\vspace{1.5em}
\end{center}
}
\setcounter{footnote}{0}

\begin{abstract}
  Detailed applications of minisuperspace methods are presented and compared
  with results obtained in recent years by means of causal dynamical
  triangulations (CDTs), mainly in the form of effective actions. The analysis
  sheds light on conceptual questions such as the treatment of time or the
  role and scaling behavior of statistical and quantum
  fluctuations. In the case of fluctuations, several analytical and numerical
  results show agreement between the two approaches and offer possible
  explanations of effects that have been seen in causal dynamical
  triangulations but whose origin remained unclear. The new
  approach followed here suggests ``CDT experiments'' in the form of new
  simulations or evaluations motivated by theoretical predictions, testing
  CDTs as well as the minisuperspace approximation.
\end{abstract}

\section{Introduction}

Causal Dynamical Triangulations (CDTs) \cite{Ambjorn:2012jv,Loll} are an
attempt to compute gravitational transition amplitudes by utilizing a
discretized version of path integrals.  The novelty of the approach with
respect to Euclidean Dynamical Triangulations (EDTs) \cite{Ambjorn:2013eha} is
a built-in notion of causality, which guarantees that branching of the path
integrals does not occur --- the topology of a spatial slice is preserved in
time. In EDTs, by contrast, so-called baby universes can form, leading to
divergences and incorrect semiclassical behavior. However, while CDTs enforce
causality in every discrete time step, the formulation is still Euclidean in
the sense of the Wick rotation: they utilize a path integral of $e^{-S/\hbar}$
rather than $e^{iS/\hbar}$. As a consequence, the solutions obtained within
CDT represent \emph{instanton} configurations.

The analytic continuation to Euclidean signature is performed for technical
reasons, allowing one to transform the complex propagator into a real
partition function. This step opens up the possibility of using numerical
methods developed to study statistical systems, such as Monte Carlo
simulations. In particular, performing a sequence of Markov moves the system
under consideration can be equilibrated. The maximal-entropy state corresponds
to the instanton trajectory in the path integral formulation and its thermal
fluctuations can be associated with quantum fluctuations around the classical
path \cite{Ambjorn:2008wc}.

Over the last two and a half decades the CDT approach has been investigated in
1+1, 2+1 and 3+1 dimensions. The especially interesting case of 3+1
dimensional CDT has been studied by performing advanced computer simulations
with hundreds of thousands of simplices. Most of the simulations were
performed assuming the spatial topology to be that of the three-sphere, $S^3$
\cite{Ambjorn:2004qm}. However, studies have recently been extended to the
case of toroidal topology, $S^1 \times S^1 \times S^1$
\cite{CDTTorusKin,CDTTorusKin2,CDTTopology, CDTTorus}. In each case, a
contribution from the cosmological constant, $\Lambda$, is included in the
gravitational action.

The first significant conclusion resulting from the simulations is the
nontrivial phase structure of the theory. It has been shown that, depending on
the type of order parameter used, three or more phases of the gravitational
field can be identified. The different phases are separated by transition
lines some of which are of the first and others of higher order
\cite{Ambjorn:2011cg,Ambjorn:2012ij}. In this article, we will focus our
attention on the so-called phase $C$, which shares properties with classical
space-time. In particular, it has been shown that the evolution of the
averaged 3-volume is, in this phase, consistent with the results for the
4-dimensional de Sitter instanton \cite{Ambjorn:2004qm}.  Furthermore, an
analysis of both fractal dimension and the so-called spectral dimension
revealed the correct large scale dimension of the geometry
\cite{Ambjorn:2005db}. It has also been noticed that the spectral dimension
undergoes \emph{dimensional reduction} at short scales, reflecting quantum
properties of the semiclassical space-time. It is worth stressing at this
point that an analogous semiclassical space-time is not present in the EDT
formulation, unless some non-trivial measure is introduced
\cite{Ambjorn:2013eha}.

CDTs have therefore been successful in producing results for a full 3+1
dimensional approach to quantum gravity. Minisuperspace models, by comparison,
are much more restricted and questions remain as to how reliably they capture
properties of full quantum gravity. Nevertheless, there is promise in
combining these approaches, both to interpret results from CDTs and
potentially to use CDTs to test the validity of minisuperspace truncations and
approximations. Here, we initiate a detailed study of the first aspect, using
derivations of different properties of quantum fluctuations in minisuperspace
models to interpret effective actions extracted from CDTs. We begin with a
discussion of one major difference between CDTs and the traditional
minisuperspace treatment in the way they deal with the problem of time.

\section{Implications of gauge-fixed time}
\label{s:Fix}

One of the characteristic features of CDTs, as they have mainly been used to
find effective actions, is the fixing of a time gauge. (A
foliation-independent formulation of CDTs also exists
\cite{CDTFol,CDTFolDeSitter}, on which we will briefly comment at the end of
this section.) For technical reasons, one chooses a preferred time coordinate
and the implied constant-time foliation in order to be able to define
observables, such as the spatial 3-volume, and to extract their time evolution
from numerical simulations.  The evolution of homogeneous geometries derived
from CDTs then also refers to a preferred choice of time, which should be
taken into account in a comparison with minisuperspace models.

As usual, fixing the gauge comes at a price. Breaking time reparameterization
invariance reduces the number of constraints in the Hamiltonian formulation of
general relativity. In homogeneous models, for instance, one loses the
Friedmann equation. More generally, if the scalar (Hamiltonian) constraint is
no longer imposed, the theory includes an additional physical (scalar) degree
of freedom of the gravitational field. Nevertheless, homogeneous geometries
extracted from CDTs seem to be in agreement with minisuperspace considerations
in general relativity, see for instance Ref.~\cite{Ambjorn:2004qm}.  The
purpose of this section is to discuss this issue for the example of a
compact (Euclidean) universe with positive cosmological constant $\Lambda$,
which has been studied in detail with CDTs.

\subsection{Minisuperspace solutions with time reparameterization invariance}

The minisuperspace counterpart of the model with spatial topology $S^3$ and
positive cosmological constant, $\Lambda>0$, in CDTs is expected to be the de
Sitter space-time. In general relativity, the minisuperspace Lagrangian
(without gauge-fixing) is
\begin{equation}
L = \frac{3 V_0}{8\pi G} \left( - \frac{a \dot{a}^2}{N}+Na - \frac{\Lambda}{3}
  a^3 N \right),   
\label{LagranDS}
\end{equation}
where the coordinate volume is finite and equal to $V_0=2\pi^2$, $N$ is the
lapse function, $a$ is the scale factor and $G$ is the Newton's constant. The
associated canonical momenta are
\begin{eqnarray}
p_{N} &:=& \frac{\partial L}{\partial \dot{N}} = 0, \\
p_{a} &:=&  \frac{\partial L}{\partial \dot{a}} = - \frac{3V_0}{4\pi G} \cdot
\frac{a \dot{a}}{N}\,.
\end{eqnarray} 
The condition $p_{N}=0$ is a primary constraint, which tells us that $N$ is a
non-dynamical variable, or a Lagrange multiplier. Performing the Legendre
transformation, the Hamiltonian of the model is found to be
\begin{equation}
H = \dot{N} p_N + \dot{a} p_a - L =N \left[ -\frac{2\pi G}{3V_0}
  \frac{p_a^2}{a}-\frac{3V_0}{8\pi G}\left(a- \frac{\Lambda}{3} a^3
  \right)\right]\,. 
\end{equation}  
The Poisson bracket
\begin{equation}
\{f , g\} := \frac{\partial f}{\partial N}\frac{\partial g}{\partial p_N} -
\frac{\partial f}{\partial p_N}\frac{\partial g}{\partial N} + \frac{\partial
  f}{\partial a}\frac{\partial g}{\partial p_a}- \frac{\partial f}{\partial
  p_a}\frac{\partial g}{\partial a}
\end{equation}
allows us to write Hamilton's equations $\dot{f}=\{f,H\}$ for arbitrary
phase-space functions $f$. Considering the basic variables $a$ and $p_a$, we
obtain:
\begin{align}
\dot{a} &=\frac{\partial H}{\partial p_a} =- N \frac{4\pi G}{3V_0}
\frac{p_a}{a}, \label{adot} \\  
\dot{p}_a &=-\frac{\partial H}{\partial a} = -N \left[  \frac{2\pi G}{3V_0}
  \frac{p_a^2}{a^2} 
-\frac{3V_0}{8\pi G}\left(1-\Lambda a^2\right)\right]\,. \label{padot}
\end{align}

Furthermore, we have the secondary constraint $0=\dot{p}_N=-\partial
  H/\partial N$, or
\begin{equation}
  -\frac{2\pi
  G}{3V_0} \frac{p_a^2}{a} 
-\frac{3V_0}{8\pi G}\left(a- \frac{\Lambda}{3} a^3
\right)=0 \label{secondaryconstraint} 
\end{equation}
which, together with Eq.~(\ref{adot}), leads to the Friedmann equation 
\begin{equation}
\left( \frac{\dot{a}}{Na} \right)^2 =
\frac{\Lambda}{3}-\frac{1}{a^2}\,. \label{Friedmann} 
\end{equation}   
If we choose the lapse function $N=1$ (proper time), the general solution to
this equation has the well-known hyperbolic form
\begin{equation}
a(t)= a_0 \cosh (t/a_0), \label{dSSolution}
\end{equation}
where $a_0 := \sqrt{3/\Lambda}$. 

We obtain the Euclidean version of this solution by performing a Wick rotation
$t \mapsto i \tau$. The Lorentzian Friedmann--Robertson--Walker
  line element
\[
 {\rm d}s^2= -{\rm d}t^2+a(t)^2 \left(\frac{{\rm d}r^2}{1-r^2}+ r^2({\rm
     d}\vartheta^2+\sin^2\vartheta{\rm d}\varphi^2)\right)
\]
of a closed model with $a(t)$ given in (\ref{dSSolution}) is then equivalent
to the line element
\begin{equation} \label{Sphere}
  {\rm d}s^2= a_0^2\left({\rm d}\sigma^2+\cos^2\sigma \left({\rm d}\psi^2+
      \sin^2\psi ({\rm d}\vartheta^2+\sin^2\vartheta{\rm
        d}\varphi^2)\right)\right) 
\end{equation}
of a 4-sphere with radius $a_0$, where the angle $\sigma :=\tau/a_0$ takes 
values in the range $-\pi/2\leq\sigma\leq \pi/2$.  Written for the case of the 
volume variable, the solution is
\begin{equation}
V(\tau)=V_0 a^3(\tau) = 2\pi^2 a_0^3 \cos^3
\left(\tau/a_0\right)\,, \label{Vsolution} 
\end{equation}
where $\tau \in \left[- \frac{\pi}{2}a_0,\frac{\pi}{2}a_0\right]$. This cosine
cubed solution has been well confirmed as an effective behavior, averaging
over a large number of simulations performed within CDTs
\cite{CDTDeSitter,CDTEffAc}.  Although this result is encouraging, it is also
puzzling because (\ref{Vsolution}) is a consequence of the constraint equation
(\ref{secondaryconstraint}), which is not present in a gauge-fixed theory such
as CDTs.  We will try to explain this outcome in the next subsection.

\subsection{Minisuperspace solutions with gauge-fixed time}  
\label{s:GaugeFixedSol}

Let us now revisit the derivation of classical minisuperspace equations, but
now fixing the time gauge from the very beginning such that $N=1$.  Using this
value in the Lagrangian (\ref{LagranDS}), the only remaining variable is
the scale factor $a$, with canonical momentum
\begin{equation}
p_{a} = - \frac{3V_0}{4\pi G} \cdot a
\dot{a}\,.
\end{equation}
The resulting gauge-fixed Hamiltonian is
\begin{equation}
H_{\rm GF} =-\frac{2\pi G}{3V_0} \frac{p_a^2}{a}-\frac{3V_0}{8\pi G}\left(a-
  \frac{\Lambda}{3} a^3 \right)\,. \label{HamiltonianGF}
\end{equation} 
Consequently, Hamilton's equations for the phase space variables are
\begin{align}
\dot{a} &=\frac{\partial H_{\rm GF}}{\partial p_a} =-\frac{4\pi G}{3V_0}
\frac{p_a}{a} \label{adotGF} \\  
\dot{p}_a &=-\frac{\partial H_{\rm GF}}{\partial a} = - \frac{2\pi G}{3V_0}
\frac{p_a^2}{a^2} 
+\frac{3V_0}{8\pi G}\left(1-\Lambda a^2\right)\,. \label{padotGH}
\end{align}  

There is no Friedmann equation now.  However, combining Eqs.~(\ref{adotGF})
and (\ref{padotGH}), we obtain the second-order Raychaudhuri equation
\begin{equation}
\frac{\ddot{a}}{a}=-\frac{1}{2}\left( \frac{\dot{a}}{a}
\right)^2-\frac{1}{2a^2}+\frac{\Lambda}{2}\,.  
\label{Raychaudhuri}
\end{equation}
A similar equation can, of course, be derived also in the case without gauge
fixing. However, by imposing the Friedmann equation, one of the two constants
of integration in solutions of the Raychaudhuri equation is fixed.  If we have
only the second-order Raychaudhuri equation (\ref{Raychaudhuri}), the general
solution has two unknown constants of integration.

Let us try to find solutions to Eq.~(\ref{Raychaudhuri}). For this purpose, it
will be useful to introduce the Hubble parameter
\begin{equation}
\mathbb{H}: =\frac{\dot{a}}{a}\,,
\end{equation}
so that Eq.~(\ref{Raychaudhuri}) can be written as
\begin{equation}
  \dot{\mathbb{H}} =a \mathbb{H} \frac{{\rm d}\mathbb{H}}{{\rm d}a}= -
    \frac{3}{2}\mathbb{H}^2-\frac{1}{2a^2}+\frac{\Lambda}{2}\,.  \label{dotH} 
\end{equation}
In order to find the relation $\mathbb{H}(a)$, we note that any solution of
the Friedmann equation must also be a solution to the Raychaudhuri
equation. Therefore, we may write the expression for $\mathbb{H}^2$ as a
combination of the part that is expected from the Friedmann equation and some
unknown remainder which we parametrize by a function $f(a)$:
\begin{equation}
\mathbb{H}^2 =  \frac{\Lambda}{3}-\frac{1}{a^2} + f(a)\,. \label{Hsquare}
\end{equation}
We emphasize that this parametrization does not restrict the generality of our
considerations, solving (\ref{Raychaudhuri}).  Inserting (\ref{Hsquare}) in
(\ref{dotH}), we find that $f(a)$ has to obey
\begin{equation}
\frac{{\rm d}f}{{\rm d}a} =-3 \frac{f}{a}\,,  
\end{equation} 
which can directly be integrated to $f(a) = c/a^3$ with a constant of
integration $c$. Therefore, the first integration of the Raychaudhuri equation
(\ref{Raychaudhuri}) leads to the Friedmann-like equation
\begin{equation}
\left( \frac{\dot{a}}{a} \right)^2=\frac{\Lambda}{3}-\frac{1}{a^2} +
\frac{c}{a^3} \label{EffectiveFriedmann} 
\end{equation}
with an extra term that has the same form as the matter contribution from
dust.  The presence of the scalar constraint in general relativity fixes the
free constant of integration to be equal to zero, $c=0$. However, in the
gauge-fixed case relevant for CDTs, the value of $c$ remains undetermined. Is
there a way to select a particular value of $c$ within CDTs?

In order to answer this question, we should take into account the fact that
computer simulations require finiteness.  Instanton configurations simulated
in CDTs therefore have a finite spatial extension,
in addition to a periodic (or finite-range) time variable. These conditions
guarantee that the total 4-volume of the instanton is finite and can be
discretized with a finite number of 4-simplices. We may then select solutions
to the Euclidean version of Eq.~(\ref{EffectiveFriedmann}) such that the
constraint
\begin{equation}
V_4 = 2\pi^2 \int {\rm d}\tau  a^3(\tau) < \infty \label{ConditionV4}
\end{equation}
is satisfied. Taking into account the periodicity of the time variable in
CDTs, the condition (\ref{ConditionV4}) is satisfied when the value of the
3-volume, or equivalently the scale factor $a$, is bounded from above.  When
the maximal value of $a(\tau)$ is reached the Euclidean version of the Hubble
parameter, $a^{-1}{\rm d}a/{\rm d}\tau$, as well as the right-hand side
of 
\begin{equation}
\left(  \frac{{\rm d}a}{{\rm d}\tau}\right)^2=
1-\frac{\Lambda}{3}a^2-\frac{c}{a} \label{EuclidFriedmann} 
\end{equation}
(the Wick rotated version of (\ref{EffectiveFriedmann})) are equal to zero,
leading to the condition
\begin{equation}
\frac{\Lambda}{3}a^3-a+c=0\,. 
\end{equation}

\begin{figure}[h]
\begin{center}
\includegraphics[width=10cm]{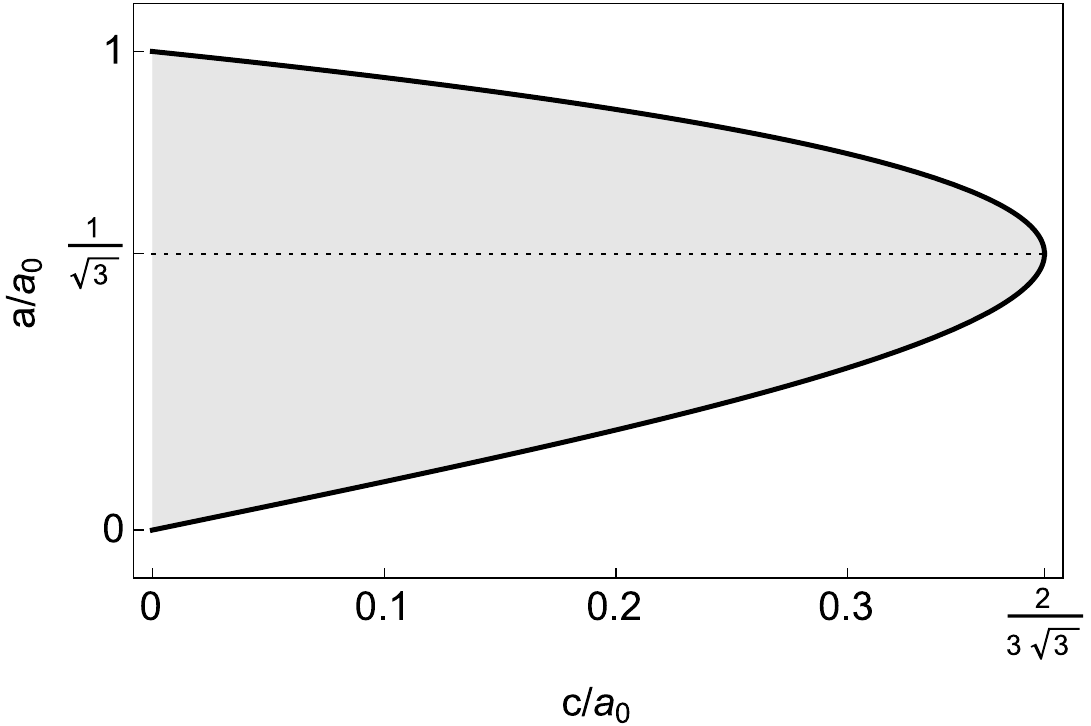}
\caption{The shadowed region represents allowed values of the scale factor as
  a function of the integration constant $c$. For $c=0$, which overlaps with
  the case without gauge-fixing, $a\in[0,a_0]$.  The range
  $[a_{\text{min}},a_{\text{max}}]$ is decreasing with increasing $c$, reaching
  a single point with $a_{\text{min}}=a_{\text{max}}=a_0/\sqrt{3}$ at
  $c_{\text{max}}:=2a_0/3\sqrt{3}$.}
\label{SolRegion}
\end{center}
\end{figure}

This equation indicates that the Euclidean version of the Friedmann equation
has real solutions for $a\geq 0$ if $c \leq 2/3\sqrt{\Lambda}$. Furthermore,
regularity of ${\rm d}a/{\rm d}\tau$ at the boundaries in time is guaranteed
if $c \geq 0$. (The term $-c/a$ in Eq.~(\ref{EuclidFriedmann}) then tends to
$+\infty$ while $a\rightarrow0^{+}$, a limit which cannot be approached on
solutions of Eq.~(\ref{EuclidFriedmann}) thanks to positivity of the left-hand
side).  In summary, well-behaved instanton solutions are obtained for
\begin{equation}
0 \leq   c  \leq  \frac{2}{3} \frac{1}{\sqrt{\Lambda}}\,. 
\end{equation} 
In this case, the values of the scale factor are confined to the interval
$[a_{\text{min}}(c),a_{\text{max}}(c)]$, as shown in Fig.~\ref{SolRegion}.

\begin{figure}[h]
\begin{center}
\includegraphics[width=10cm]{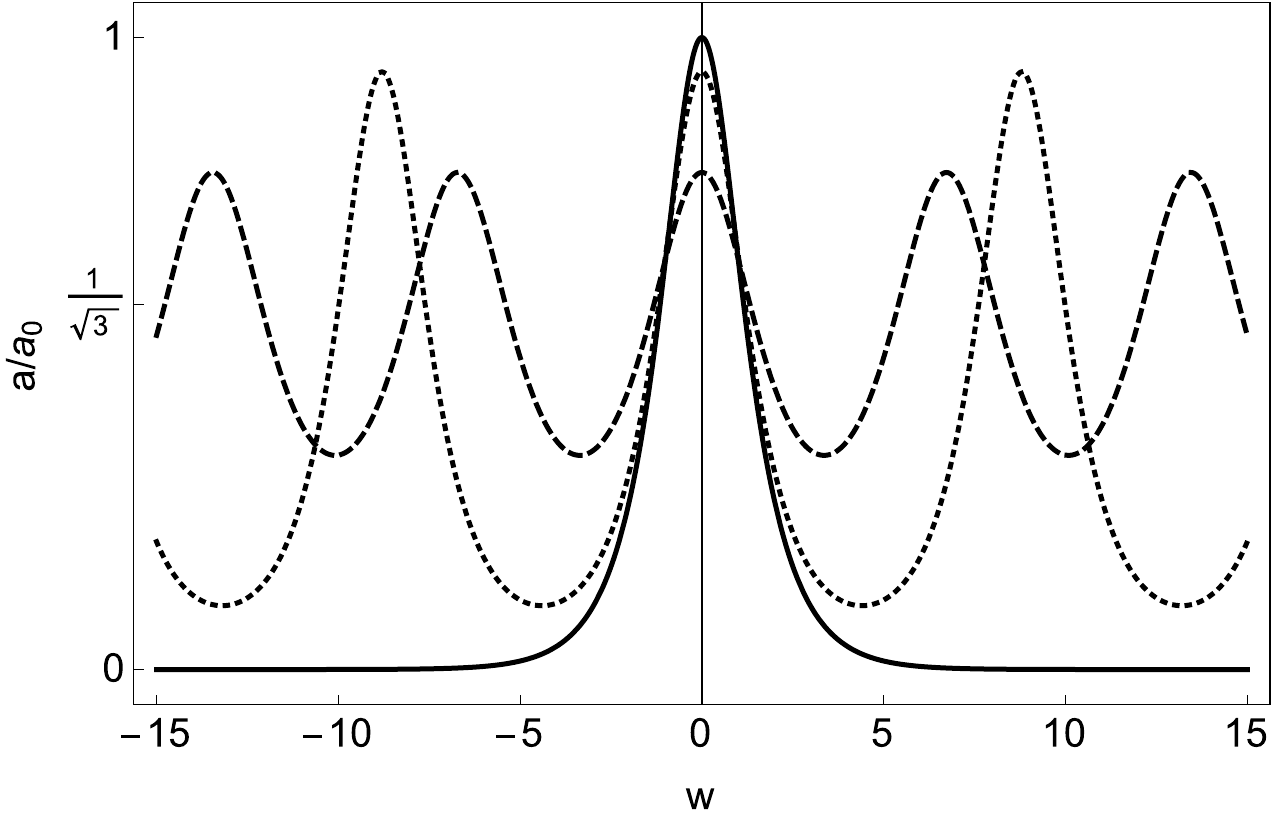}
\caption{Sample solutions to Eq.~(\ref{EuclidFriedmann}) for $c =0$ (solid
line), $c =0.1 a_0$ (dotted line) and $c =0.2 a_0$ (dashed line) as a
function of the time parameter $w$ given by Eq.~(\ref{Timexw}). 
Here, $w_0=0$.}
\label{SolSF}
\end{center}
\end{figure}

A general solution to Eq.~(\ref{EuclidFriedmann}) for $a\geq 0$ is derived in
the Appendix in terms of Weierstrass's elliptic function, given by
Eq.~(\ref{WeierstrassSol3}) as a function of a new time variable $w$ related
to the time variable $\tau$ via
\begin{equation}
 \tau = \int_0^w a(w') {\rm d}w'. \label{Timexw}
\end{equation}
Sample solutions for different values of $c$ are shown in Fig.~\ref{SolSF}.
Importantly, while for $c=0$ the instanton is a single peak, the solutions
for $c>0$ have oscillatory form. However, for a single (central) period there
is no qualitative difference between the different solutions for $c \in
\left[0, 2a_0/3\sqrt{3}\right]$. When the value of $c$ is increasing, the
amplitude of oscillations is decreasing. For the limiting case
$c_{\text{max}}=2a_0/3\sqrt{3}$ there is only a constant solution $a=
a_0/\sqrt{3}$. The minimal value of the scale factor
$a_{\text{min}}$ is a monotonic function of $c$ (see the bottom boundary of
the shadowed region in Fig. \ref{SolRegion}) and tends to zero for
$c\rightarrow 0$, which corresponds to the solution satisfied by the scalar
constraint.

CDT solutions do not exhibit oscillatory behavior of the 3-volume but rather
correspond to the case with $c=0$.  One may speculate that this happens for
the reason that this solution is selected either by the dynamics of CDTs or by
boundary conditions. While passing to Euclidean path integrals and the
statistical-physics formulation, only paths that satisfy the condition
$\Sigma(\tau_i) =\Sigma(\tau_f)$ are considered, where $\Sigma$ is a spatial
configuration and $\tau_i$ and $\tau_f$ are initial and final times
respectively, so that the time interval is:
\begin{equation}
\tau_f - \tau_i = \frac{\hslash}{k_{\text{B}} T}={\rm constant}.  
\end{equation} 
In CDT simulations, instead of considering the finite time interval
$[\tau_i,\tau_f]$ together with the symmetric boundary condition
$\Sigma(\tau_i) =\Sigma(\tau_f)$, periodicity of the imaginary time 
variable $\tau$ is introduced. (The time domain becomes $S^1$.) 
At the level of the scale factor $a$ (expressed it terms of time $w$) 
this implies that regularity conditions for $a$ and its derivatives 
should be satisfied, which we evaluate here for the first 
two orders
\begin{align}
a|_{w_i} &= a|_{w_f}, \label{cond1} \\
\left. \frac{{\rm d}a}{{\rm d}w} \right|_{w_i}  &= \left. \frac{{\rm d}a}{{\rm
      d}w} \right|_{w_f}, \label{cond2}  
\end{align}
where $w_i = w(\tau_i)$ and $w_f = w(\tau_f)$. The first condition can be
imposed for all solutions by fixing the value of an integration constant
$w_0$; see the Appendix. In Fig.~\ref{SolSF}, $w_0$ is fixed to zero, so that
all solutions are even with respect to the point $w=0$. The condition
(\ref{cond2}) is less trivial to satisfy and requires fixing the second
constant of integration, $c$. Without loss of generality, let us assume now
that $w_f=-w_i>0$. The first observation is that the condition (\ref{cond2})
cannot be satisfied for small values $w_f \sim 1$, except for the case of a
constant solution $a=a_0/\sqrt{3} $ with $c=c_{\text{max}}
=2a_0/3\sqrt{3}$. For $w_f \gg 1$ there is a finite set of values of $c$ for
which the boundary conditions (\ref{cond1}) and (\ref{cond2}) are satisfied
 (values of $c$ such that $a(w)$ has local extrema at $\pm
  w_f$). Furthermore, in general, if one tries different values of $w_f$, one
  would have to change the values of $c$ for which the periodic boundary
  conditions are satisfied. There are two exceptions, the mentioned constant
  solution for $c=c_{\text{max}}$ and the case with $c=0$. (In the latter
  case, (\ref{cond2}) is not fulfilled exactly but, for large enough $w_f$,
  can easily hold within the numerical accuracy of a CDT.)  One can summarize
  this result as follows:

For $w_f \gg 1$ there is only one non-constant solution which satisfies
  the periodic boundary conditions (\ref{cond1}) and (\ref{cond2}) with fixed
  values of the constants of integration $w_0$ and $c$. This solution
  corresponds to the case with $c=0$, which is the one satisfying the scalar
  constraint (Friedmann equation).

This observation may not provide the sole explanation why CDT simulations
select the solutions satisfying the scalar constraint. However, the results
presented provide an interesting hint as to this enigmatic feature of CDT. (It
is interesting to note that CDT simulations in a framework that aims to relax
the imposition of a fixed time foliation \cite{CDTFol,CDTFolDeSitter} indicate
a behavior with local minima near $\tau_i$ and $\tau_f$, closer to one of our
background solutions with $c\not=0$. However, as the authors of
\cite{CDTFol,CDTFolDeSitter} point out, differences appear in a region of
small volume in which discretization effects may be non-negligible.)  The
issue deserves further deepened investigations. Moreover, a comparison with
the toroidal model is of interest, in which case we will develop a rather
different picture in Sec.~\ref{s:Toroidal}.  
 
\section{Fluctuations around the instanton solution}

In addition to background solutions, CDTs are able to derive statistical
fluctuations around them. These results allow a comparison with quantum
minisuperspace models, in which the background volume $V$ can be identified
with the expectation value of a volume operator in an evolving state, and
statistical fluctuations are replaced by quantum fluctuations.
 
\subsection{Spherical model}

We will focus on the specific instanton solution associated with the choice
$c=0$ in Eq.~(\ref{EffectiveFriedmann}). An analytic continuation to complex
time, suitable for Euclidean solutions, can be performed in several possible
ways. In particular, the two choices $t \rightarrow \pm i \tau$ may be
considered. The corresponding instanton solution is independent of the choice
of the sign: In both cases,
\begin{equation}
\bar{a}(t)= a_0 \cos (\sigma)\,,
\label{sol2}
\end{equation}
where for later convenience we have introduced the parameter 
\begin{eqnarray} \label{stau}
\sigma := \tau/a_0
\end{eqnarray}
such that $\sigma \in \left[- \pi/2,\pi/2\right]$. 

However, the action $S=\int {\rm d}t L$ matters not only for background
solutions but also for the derivation of fluctuations. Performing the Wick
rotation $t \rightarrow \pm i \tau$, the action associated with the
gauge-fixed model is
\begin{equation} \label{ClassAc}
S(t = \pm i \tau ) =\pm i \frac{3\pi}{4G} \int {\rm d}\tau  a \left[
  \left(\frac{{\rm d}a}{{\rm d}\tau}\right)^2+1 - \frac{a^2}{a_0^2}  \right]\,. 
\end{equation} 
The Euclidean action $S_{\rm E}$ is then defined such that the Feynman
amplitude $e^{iS/\hslash}$ reduces to the Boltzmann weight
$e^{-S_{\rm E}/\hslash}$, or $S_{\rm E} \equiv -i S(t = \pm i \tau
)$. The Euclidean action associated with the instanton trajectory (\ref{sol2})
for $\sigma \in \left[- \pi/2,\pi/2\right]$ is then
\begin{eqnarray} 
S_{\rm E}^{\rm C} = \pm \frac{3\pi}{4G} a_0 \int_{- \pi/2}^{\pi/2}
{\rm d}\sigma \ a \left[ \left(\frac{1}{a_0}\frac{{\rm d}a}{{\rm
        d}\sigma}\right)^2+1  
-\left(\frac{a}{a_0} \right)^2  \right]  
 = \pm \frac{\pi}{G} a_0^2 =\pm \frac{3 \pi}{\Lambda G}\,. 
\label{ClassicalAction} 
\end{eqnarray}
Because the action on half the considered trajectory is associated with the
probability of tunneling through the potential barrier from $a=0$ to $a=a_0$,
we can compare (\ref{ClassicalAction}) with the tunneling probability $P
\propto e^{-3 \pi/(2 \hslash \Lambda G)}$ obtained in
Refs.~\cite{Vilenkin:1982de,Linde:1983mx}: $e^{-\frac{1}{2} S_{\rm E}^C/\hslash}
= e^{\mp 3 \pi/(2 \hslash \Lambda G)}$.  Consistency of the results requires that 
the Wick rotation $t \rightarrow + i \tau$ be chosen.

\subsubsection{Fluctuations}

We are now ready to consider\footnote{Some of the results presented in this
  subsection have been discussed by Jakub Wn\c{e}k in his Bachelor thesis
  ``Quantum aspects of the de Sitter space'' (Jagiellonian University, 2017)
  who based on calculations and a notebook originally made by his supervisor,
  one of the authors of this article (JM).}  fluctuations around the instanton
trajectory
\begin{equation} \label{bara}
\bar{a}(\sigma)=a_0\cos \sigma\,.
\end{equation}
 We introduce $y(\sigma)$ such that
\begin{equation}
a(\sigma) =\bar{a}(\sigma)+y(\sigma),
\end{equation}
together with the boundary conditions $y(-\pi/2)=0=y(\pi/2)$. Here we are
using zero boundary conditions for fluctuations, so that values of the scale
factor are fixed at the boundaries. The Dirichlet boundary conditions of this
kind are consistent with the scalar constraint (Friedmann equation) and a
procedure of derivation of a propagator. However, such choice can be
generalized to the case with non-vanishing quantum fluctuations of the scale
factor at the boundaries. Furthermore, periodic boundaries with conditions
of the type (\ref{cond1}) and (\ref{cond2}) can also be
considered. Presumably, such choice would be closer to the case of
CDTs. Nevertheless, here the standard boundary conditions for fluctuations
(which satisfy the scalar constraint) will be considered. 

In this case, the
Euclidean action can be expanded in $y$:
\begin{eqnarray}
S_{\rm E} &=& \frac{3\pi}{4G} a_0^2 \int_{- \pi/2}^{\pi/2}
{\rm d}\sigma
\left(\frac{a}{a_0}\right)\left[ \left(\frac{1}{a_0}\frac{{\rm d}a}{{\rm
        d}\sigma}\right)^2+1  
-\left(\frac{a}{a_0} \right)^2  \right]  \nonumber \\
&=& S_{\rm E}^{\rm C}-\frac{3\pi}{4G} \int_{- \pi/2}^{\pi/2} {\rm
  d}\sigma \cos^2(\sigma)\ y \left[ \frac{{\rm d}}{{\rm d}\sigma}\left(
    \frac{1}{\cos (\sigma)} 
    \frac{{\rm d}}{{\rm d}\sigma}  \right) 
+ \frac{3}{\cos (\sigma)}  \right]y \nonumber \\
&+&\frac{3\pi}{4G} \frac{1}{a_0} \int_{- \pi/2}^{\pi/2} {\rm d}\sigma \ y
\left[  \left(\frac{{\rm d}y}{{\rm d}\sigma}\right)^2 - y^2\right]\,, 
\end{eqnarray}
where $S_{\rm E}^{\rm C}$ is the classical action (\ref{ClassicalAction}) with
the plus sign.  By introducing a new time variable $z=\sin \sigma$, we obtain
\begin{equation}
S_{\rm E} = S_{\rm E}^{\rm C}+ \frac{3\pi}{4G} \int_{-1}^{1} {\rm d}z\ y(z)
\hat{L} y(z) +\mathcal{O}(y^3) 
\end{equation}
with the differential operator 
\begin{equation}
\hat{L}  = (z^2-1)\frac{{\rm d}^2}{{\rm d}z^2}-3\,. 
\end{equation}

We now solve the eigenproblem for the operator $\hat{L}$ defined on
$L^2([-1,1], {\rm d}\mu)$,
\begin{equation}
\hat{L} \phi_n = \lambda_n \phi_n 
\end{equation}  
together with the boundary conditions $\phi_n(-1)=0=\phi_n(1)$. The operator
$\hat{L}$ belongs to the class of Sturm--Liouville operators, which guarantees
its self-adjointness. The eigenvectors are orthonormal with respect to the
measure ${\rm d}\mu = w(z) {\rm d}z$, where
\begin{equation}
w(z)= \frac{1}{1-z^2}\,.
\end{equation}

The equation $\hat{L} u = \lambda u $ can be written as
\begin{equation}
(1-z^2)\frac{{\rm d}^2\phi}{{\rm d}z^2}+d \phi = 0,
\label{EqLU}
\end{equation}
where we defined $d=3+\lambda$.  By introducing the
variable $\xi =\frac{1}{2}(1+z)$, this equation reduces to a special case of
the Heun equation,
\begin{equation}
\xi(1-\xi)\frac{{\rm d}^2\phi}{{\rm d}\xi^2}+d \phi = 0\,.
\end{equation}
The general solution to (\ref{EqLU}) can therefore be expressed in terms of
hypergeometric functions:
\begin{eqnarray}
\phi(z)&=&   {_2F_1} \left(-\frac{1}{4}-\frac{1}{4}\sqrt{1+4d},
  -\frac{1}{4}+\frac{1}{4}\sqrt{1+4d},\frac{1}{2}; z^2   \right) c_{1}
\nonumber  \\
      &&+\:  {_2F_1} \left(\frac{1}{4}-\frac{1}{4}\sqrt{1+4d},
        \frac{1}{4}+\frac{1}{4}\sqrt{1+4d},\frac{3}{2}; z^2   \right) c_{2}  i
      z
\end{eqnarray}
with constants of integration $c_1$ and $c_2$. The boundary conditions
$\phi(\pm 1) =0$ lead to
\begin{equation}
\underbrace{ \left( \begin{array}{cc}  2/A  & -i/B  \\
      2/A & i/B \end{array} \right)}_{M}  
\left( \begin{array}{c} c_1 \\ c_2 \end{array} \right) =
\left( \begin{array}{c} 0 \\ 0 \end{array} \right) 
\label{sys1}
\end{equation}
for
\begin{eqnarray}
A &=& \Gamma\left[\frac{3}{4}-\frac{1}{4} \sqrt{1+4 d}\right]
\Gamma\left[\frac{3}{4}+\frac{1}{4} \sqrt{1+4 d} \right]  \\
B &=& \Gamma\left[\frac{5}{4} -\frac{1}{4}\sqrt{1+4 d}\right]
\Gamma\left[\frac{5}{4}+\frac{1}{4}\sqrt{1+4 d}\right]\,. 
\end{eqnarray} 
The system of equations (\ref{sys1}) has a nontrivial solution only if
\begin{equation}
0=\det M= \frac{4i}{\Gamma\left[\frac{3}{4}-\frac{1}{4} \sqrt{1+4d}\right]
  \Gamma\left[\frac{3}{4}+\frac{1}{4} \sqrt{1+4d} \right] 
\Gamma\left[\frac{5}{4} -\frac{1}{4}\sqrt{1+4d}\right]
\Gamma\left[\frac{5}{4}+\frac{1}{4}\sqrt{1+4d}\right]}\,. 
\end{equation}
This equation, in turn, is fulfilled if
\begin{equation}
\sqrt{1+4d} = 2n+1 \ \ \text{for} \ \ n=1,2,3,4,... .
\end{equation}
The eigenvalues of $\hat{L}$ are therefore
\begin{equation}
\lambda_n=-3+n(n+1) \ \ \text{for} \ \ n=1,2,3,4,... .  
\end{equation}

Interestingly, the first eigenvalue is negative, $\lambda_1=-1$, while the
others are positive.  The corresponding eigenfunctions are
\begin{eqnarray}
\phi_n(z)=   {_2F_1} \left(-\frac{1}{2}(1+n),\frac{n}{2},\frac{1}{2}; z^2
\right) c_{1}  
      + {_2F_1} \left(-\frac{n}{2},\frac{1}{2}(1+n),\frac{3}{2}; z^2   \right)
      c_{2}  i z\,. 
\end{eqnarray}
From the solutions of (\ref{sys1}) we obtain $c^{(n)}_{1}=0$ for
even $n$ and $c^{(n)}_{2}=0$ for odd $n$. This result enables us to
classify eigenfunctions according to their parity:
\begin{eqnarray}
\phi^{\rm e}_{n_{\rm e}}(z) &=&  {_2F_1}
\left(-\frac{1}{2}(1+n_{\rm e}),\frac{n_{\rm e}}{2},\frac{1}{2}; z^2   \right)
c_{1}  \ \ 
\text{for} \ \ n_{\rm e}=1,3,5,7, ... , \\   
\phi^{\rm o}_{n_{\rm o}}(z) &=& {_2F_1}
\left(-\frac{n_{\rm o}}{2},\frac{1}{2}(1+n_{\rm o}),\frac{3}{2}; z^2   \right) c_{2}
i z  
\ \ \text{for} \ \ n_{\rm o}=2,4,6,8, ... \,.  
\end{eqnarray}
Expressing $n_{\rm e} = 2m+1$ and $n_{\rm o}=2m$ with $m\in
\mathbb{N}$, we have
\begin{eqnarray}
\phi^{\rm e}_{m}(z) &=&  {_2F_1} \left(-m-1,m+\frac{1}{2};\frac{1}{2}; z^2
\right) c_{1}  \ \ \text{for} \ \ m=0,1,2,3,4,... ,  \label{phie1}\\   
\phi^{\rm o}_{m}(z) &=& {_2F_1} \left(-m,m+\frac{1}{2};\frac{3}{2}; z^2
\right) c_{2}  i z  \ \ \text{for} \ \ m=1,2,3,4,... \label{phio1}
\end{eqnarray}
with the corresponding eigenvalues 
\begin{eqnarray}
\lambda_m^{\rm e} &=& -3+(2m+1)(2m+2)     \ \ \text{for} \ \ m=0,1,2,3,4,... 
\\ 
\lambda_m^{\rm o}  &=& -3+2m(2m+1)  \ \ \text{for} \ \ m=1,2,3,4,... \,.
\end{eqnarray}

It is possible to express the eigenfunctions in terms of Jacobi polynomials
$P_n^{(\alpha,\beta)}(z)$, employing the relation
\begin{equation}
{_2F_1} \left(-n,\alpha+1+\beta+n ;\alpha+1; \rho\right) = \frac{n!}{
  (\alpha+1)_n}P_n^{(\alpha,\beta)}(1-2\rho)
\end{equation}  
where $(\alpha)_n$ is the Pochhammer symbol.  
Normalization is performed using
\begin{equation}
\int_{-1}^{1} P_m^{(\alpha,\beta)}(\rho)
P_n^{(\alpha,\beta)}(\rho)(1-\rho)^{\alpha}(1+\rho)^{\beta}{\rm d}\rho =  
\frac{2^{\alpha+\beta+1}}{2n+\alpha+\beta+1} 
\cdot \frac{\Gamma(n+\alpha+1)\Gamma(n+\beta+1)}{n!\: \Gamma(n+\alpha+\beta+1)}
\delta_{mn}\,. \nonumber
\end{equation}
The normalized odd eigenfunctions are
\begin{equation}
\phi_m^{\rm o}(z) = \sqrt{\frac{m(4m+1)}{(2m+1)}}\: z\:
P_m^{\left(\frac{1}{2},-1\right)}(1-2z^2) 
\end{equation}
while the normalized even eigenfunctions are
\begin{equation}
\phi_m^{\rm e}(z) = \sqrt{\frac{(4m+3)(m+1)}{(2m+1)}}\:
P_{m+1}^{\left(-\frac{1}{2},-1\right)}(1-2z^2)\,,
\end{equation}
and they obey the orthonormality relation
\begin{equation}
\int_{-1}^{1}   \frac{\phi_n(z)\phi_m(z)}{1-z^2}{\rm d}z = \delta_{nm}\,.
\end{equation}
Using this condition, the normalized version of the eigenfunctions
(\ref{phie1}) and (\ref{phio1}) can be found:
\begin{align}
\phi_m^e(z) &= \sqrt{\frac{(4m+3)(m+1)}{(2m+1)}} \:
P_{m+1}^{\left(-\frac{1}{2},-1\right)}(1-2z^2)  \ \ \text{for} \ \
m=0,1,2,3,4,... ,  \label{phie2} \\ 
\phi_m^o(z) &= \sqrt{\frac{m(4m+1)}{(2m+1)}} \:z\:
P_m^{\left(\frac{1}{2},-1\right)}(1-2z^2) \ \ \text{for} \ \
m=1,2,3,4,... .   \label{phio2} 
\end{align}

There are some important features emerging from the performed analysis.  The
first is the presence of the negative eigenvalue $\lambda^e_0=-1$ associated
with the eigenfunction:
\begin{equation}
\phi_0^e(z) = \frac{\sqrt{3}}{2}(1-z^2)=\frac{\sqrt{3}}{2} \cos^2s. 
\end{equation}    
The eigenfunction leads to a negative contribution 
\begin{equation}
\frac{3\pi}{4G} \int_{-1}^{1} {\rm d}z\ \phi_0^e(z) \hat{L} \phi_0^e(z) =
-\frac{3\pi}{4G} \int_{-1}^{1} {\rm d}z\ \phi_0^e(z)^2=-\frac{3\pi}{5G} < 0
\end{equation}
to the action. Therefore, the second-order variation may lower the value of
the action below the value corresponding to the classical
trajectory. Furthermore, one can verify that the $\mathcal{O}(y^3)$ term also
contributes negatively. This peculiar behavior is a consequence of the
negative kinetic term in the gravitational action and may be associated with
the presence of the conformal mode \cite{Ambjorn:2002gr}.  In what follows,
the contribution from this eigenvalue will not be taken into
account. Secondly, there is no translational mode in the spectrum of the
$\hat{L}$ operator, which would be associated with the zero eigenvalue. This
is due to the fact that the instanton solution is defined on a finite domain
of the imaginary time $\tau$.
 
\subsubsection{Green function} 
\label{s:Green}

The Euclidean path integral for the model under consideration can be written
as
\begin{equation}
Z= \int Dy e^{-S_{\rm E}/\hslash} =e^{-S^{\rm C}_{\rm E}/\hslash} \int
Dy\exp\left(-{\textstyle\frac{1}{2}} \smallint_{-1}^{1} {\rm d}z\ y(z) \hat{M}
  y(z) +\mathcal{O}(y^3)\right)\, , 
\end{equation}
where we defined 
\begin{equation}
\hat{M} = \frac{3\pi}{2 G \hslash} \hat{L}  =  \frac{3\pi}{2 l^2_{\text{Pl}}}
\hat{L}\,, 
\end{equation}
so that
\begin{equation}
\hat{M} \phi_n = e_n \phi_n
\end{equation}
with $e_n =\frac{3}{2}\pi l^{-2}_{\text{Pl}} \lambda_n$. One can now find the
Green function $G(z,z')$ corresponding to the operator $\hat{M}$, which
satisfies the  relation
\begin{equation}
\hat{M}_z G(z,z') = \delta(z-z')\,.
\end{equation}  
The Green function can be expressed in terms of the eigenfunctions and the
corresponding eigenvalues using the formula:
\begin{equation}
G(z,z') = \sum_{n=1}^{\infty} \frac{\phi_n(z)\phi_n(z')w(z')}{e_n}\,. 
\end{equation}  
The contribution from the weight $w(z)$ is due to the fact that the
completeness relation is
\begin{equation}
 \sum_{n=1}^{\infty}\phi_n(z)\phi_n(z')w(z')=\delta(z-z')
\end{equation}
in the case of Sturm--Liouville operators.

\begin{figure}[h]
\begin{center}
\includegraphics[width=10cm]{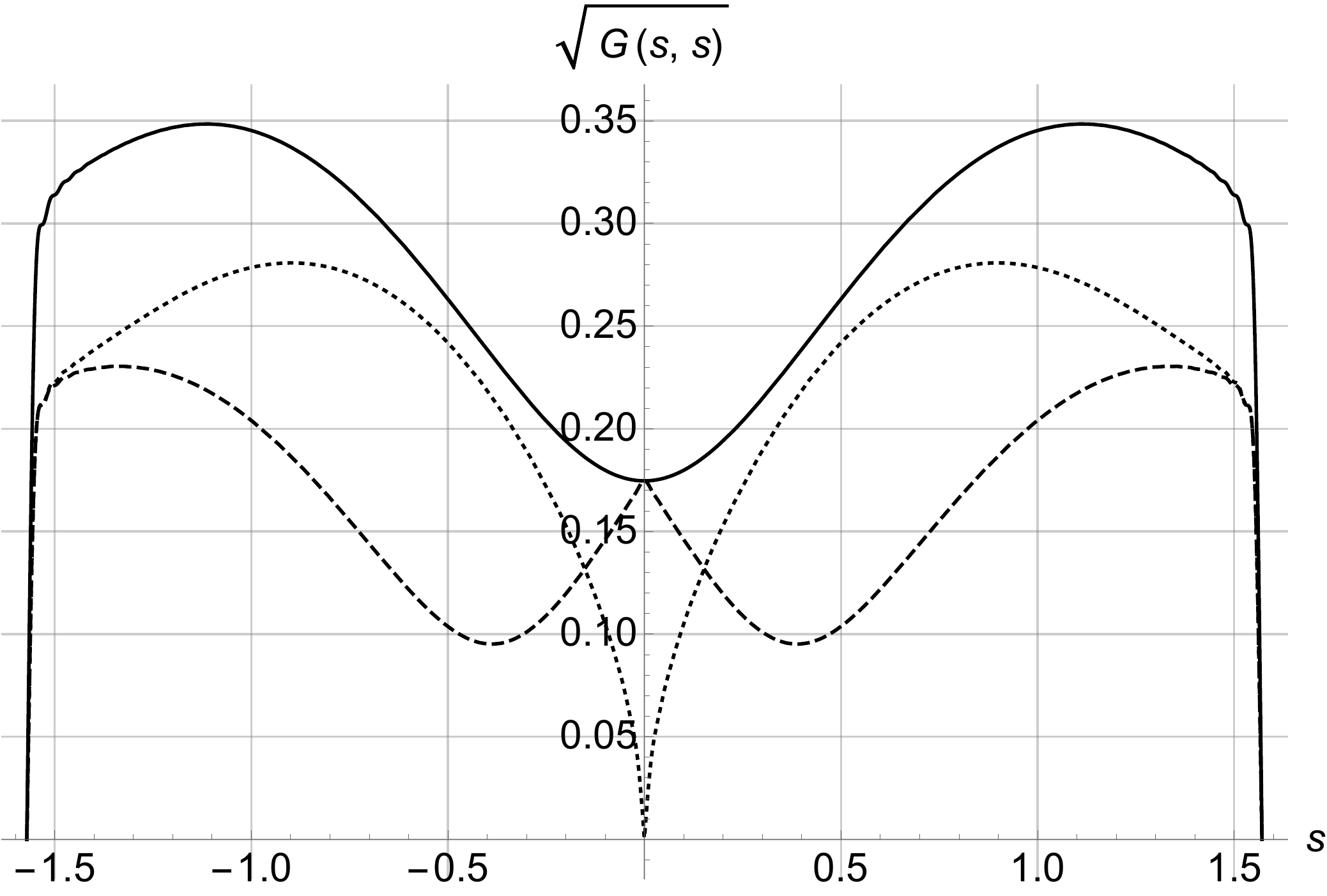}
\caption{Diagonal elements of the Green function for fluctuations of the scale
  factor: $\sqrt{G^{\text{odd}}(\sigma,\sigma)}$ (dotted line),
  $\sqrt{G^{\text{even}}(\sigma,\sigma)}$ (dashed line) and
  $\sqrt{G(\sigma,\sigma)}$ (solid line). Contribution from the negative
  eigenvalue $\phi^e_0$ was not taken into account in $G^{\text{even}}$. }
\label{GreenDeSitter}
\end{center}
\end{figure}

The Green function can be decomposed in odd and even parts according to
parity of the eigenvalues:
\begin{eqnarray} \label{GSph} 
G(z,z') &=&G^{\text{odd}}(z,z')+G^{\text{even}}(z,z'), \\
G^{\text{odd}}(z,z') &=&  \frac{2}{3\pi} l^2_{\text{Pl}}  \sum_{m=1}^{\infty}
\frac{\phi^o_m(z)\phi^o_m(z')w(z')}{-3+2m(2m+1)}\,,      \\
G^{\text{even}}(z,z') &=& \frac{2}{3\pi} l^2_{\text{Pl}}  \sum_{m=1}^{\infty}
\frac{\phi^e_m(z)\phi^e_m(z')w(z')}{-3+(2m+1)(2m+2)}\,.    \\
\end{eqnarray}
In Fig.~\ref{GreenDeSitter} we show diagonal elements of the Green function. 

While it is tempting to relate the Green function with the correlator $\langle
y(z) y(z')\rangle$, such identifications cannot be made directly due to the
nontrivial measure $w(z)$.  Therefore, while the behavior of
$\sqrt{G(\sigma,\sigma)}$ shares some qualitative similarity with the quantum
fluctuations of the 3-volume measured in CDT (in particular when
  transformed to volume fluctuations; see Sec.~\ref{s:Rel}), the two
quantities cannot be directly compared.

It is interesting to note that the amplitude of $\sqrt{G(\sigma,\sigma)}$ does
not depend on the value of the cosmological constant $\Lambda$, and is
dimensionally related to the Planck length only. In particular,
  fluctuations of the {\em scale factor} are independent of the radius $a_0$
  or the 4-volume $V_4$ of the spherical universe. In CDTs, by comparison, a
  non-trivial scaling behavior of {\em volume} fluctuations with respect to
  the number $N_4$ of 4-simplices has been found \cite{CDTDeSitter,CDTEffAc},
  which implies a non-trivial scaling behavior with respect to $V_4$.

\subsubsection{Relations between moments}
\label{s:Rel}

For a detailed comparison, we should transform the scale-factor fluctuations
derived here to volume fluctuations. Because the volume is non-linearly
related to the scale factor, relating their fluctuations or moments in an
exact form is possible only if we know the quantum state or the statistical
ensemble in which they are computed.  In a given state, we define moments of
an observable $O$ as $\Delta(O^n)=\langle(\Delta\hat{O})^n\rangle$, where
$\Delta \hat{O}=\hat{O}-\langle\hat{O}\rangle$. For the observable $V=V_0a^3$,
we can use polynomial expansions to obtain
\begin{eqnarray}
 \frac{\Delta(V^2)}{V_0^2} &=& \langle(\Delta \hat{V})^2\rangle = \langle
 (\hat{a}^3-\langle\hat{a}^3\rangle)^2\rangle\\
&=& \langle\hat{a}^6\rangle- \langle\hat{a}^3\rangle^2= \langle(\Delta
\hat{a}+\langle\hat{a}\rangle)^6\rangle-
\langle(\Delta\hat{a}+\langle\hat{a})^3\rangle^2\\
&=& \Delta(a^6)-\Delta(a^3)^2+6\langle\hat{a}\rangle
 \left(\Delta(a^5)- \Delta(a^3)\Delta(a^2)\right)+ 3\langle\hat{a}\rangle^2
   \left(4\Delta(a^4)-3\Delta(a^2)^2\right)\nonumber\\
&&+ 18\langle\hat{a}\rangle^3
   \Delta(a^3)+ 
   9\langle\hat{a}\rangle^4 \Delta(a^2)\,. \label{Va}
\end{eqnarray}

This equation is true in any state, but it requires higher-order moments of
$a$ in order to compute fluctuations of $V$.  Since we used a quadratic action
for perturbations, it is sufficient to assume a Gaussian state of
fluctuations. For a Gaussian wave function, repeated integrations by parts
show that
\begin{equation}
 \Delta(a^{2n})=(2n-1)\rho^2\Delta(a^{2n-2})= (2n-1)!! \rho^{2n-2}
 \Delta(a^2)= (2n-1)!! \Delta(a^2)^n
\end{equation}
with the variance $\rho^2=(\Delta a)^2=\Delta(a^2)$. Therefore,
\begin{equation} \label{VaGaussian}
 \frac{\Delta(V^2)}{V_0^2}= 3\left(3\langle\hat{a}\rangle^4+
   12\langle\hat{a}\rangle^2\Delta(a^2)+ 5\Delta(a^2)^2\right) \Delta(a^2)\,.
\end{equation}

  Applying (\ref{VaGaussian}) to the fluctuations
  $\Delta(a^2)=G(\sigma,\sigma)$ given in (\ref{GSph}), which as noted are
  independent of $\Lambda$ or $a_0$, the leading contribution to volume
  fluctuations is
\begin{equation}
 \frac{\Delta(V^2)}{V_0^2}\approx 9\langle\hat{a}\rangle^4\Delta(a^2)\propto
 a_0^4
\end{equation}
if we identify $\langle\hat{a}\rangle$ with the background solution $\bar{a}$
in (\ref{bara}). Written as a function of $\sigma=\tau/a_0\propto
\tau/V_4^{1/4}$ in (\ref{stau}), $\Delta(V^2)(s)/V_4$ is therefore a universal
function independent of the 4-volume $V_4$ or $a_0$. This result confirms the
scaling behavior found in CDTs \cite{CDTDeSitter,CDTEffAc}, where the role of
$V_4$ is played by the total number $N_4$ of 4-simplices.

\begin{figure}[htbp]
\label{flucs}
	\centering
	\includegraphics[scale=0.615]{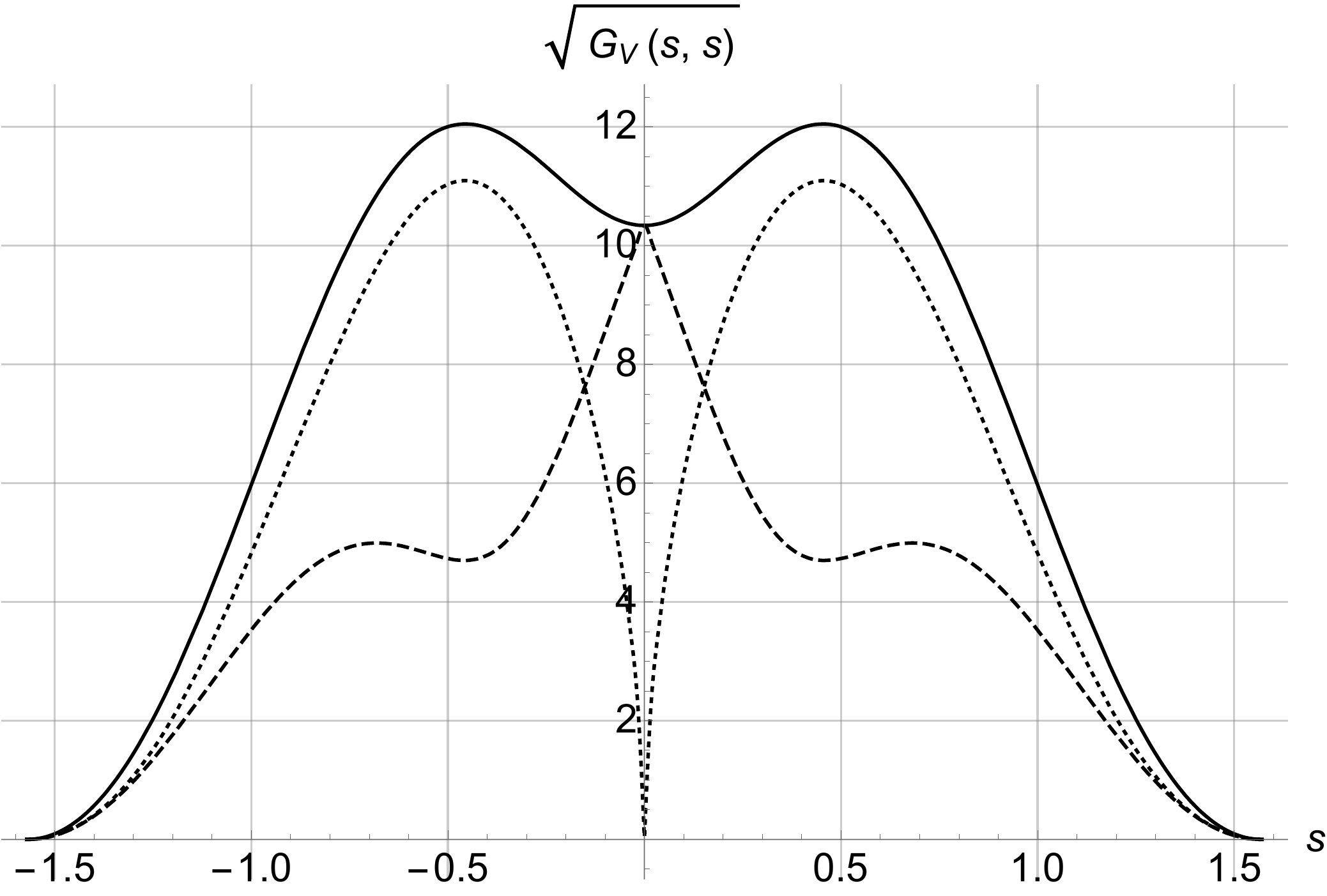}
	\caption{Diagonal elements of the Green function representing volume
          fluctuations $\Delta V$, using the transformation (\ref{VaGaussian})
          with $V_0=2\pi^2$ and $a_0=1$. (For this value of $a_0$,
          $\sqrt{G_V(\sigma,\sigma)}=\sqrt{\Delta(V^2)/a_0^4}$.) Here, the
          dotted line is $\sqrt{G^{\text{odd}}_V(\sigma,\sigma)}$, the dashed
          line is $\sqrt{G^{\text{even}}_V(\sigma,\sigma)}$ and the solid line
          is $\sqrt{G_V(\sigma,\sigma)}=
          \sqrt{G^{\text{even}}_V(\sigma,\sigma)+G^{\text{odd}}_V(\sigma,\sigma)
          }$.
\label{Transform}}   
\end{figure}

In Fig.~\ref{Transform}, we apply the transformation (\ref{VaGaussian}) to the
$a$-fluctuations shown in Fig.~\ref{GreenDeSitter}.  For comparison, CDT
results are usually shown for $\sqrt{\langle(\Delta N_3)^2\rangle/N_4}$ using
the correlation function of the number $N_3$ of 3-simplices as well as the
number $N_4$ of 4-simplices (which is fixed). These variables are related to
the 3-volume $V$ and the 4-volume $V_4$, respectively, by $N_3=V/C_4$ and
$N_4=V_4/C_4$ with $C_4=\sqrt{5}/96$ \cite{CDTDeSitter}. Moreover,
$V_4=(8/3)\pi^2 a_0^4$. Therefore,
\begin{equation} \label{NV}
 \sqrt{\frac{\langle(\Delta   N_3)^2\rangle}{N_4}}= \sqrt{\frac{3}{8\pi^2C_4}}
 \sqrt{\frac{\Delta(V^2)}{a_0^4}}\approx 0.1
 \sqrt{\frac{\Delta(V^2)}{a_0^4}}
\end{equation}
with an order-of-magnitude agreement with the maximum in
Fig.~\ref{Transform}. 

However, compared with CDT results, our fluctuations do not show
characteristic ``shoulders'' where the increase slows down briefly between an
endpoint of evolution and the midpoint. (This behavior is closer to the shape
shown by the even contribution to the Green function, rather than the full
function.) There is therefore an indication that path integral calculations in
a minisuperspace model do not capture the correct state selected by CDT
simulations. We will now analyze this issue using canonical effective methods,
which give us more control on the evolving state. 

\subsubsection{The spherical minisuperspace model: Canonical effective methods} 
\label{s:SphMin}

We will now take the minisuperspace perspective and see what it has to say
about fluctuations in the spherical model.  For later purposes, it is useful
to introduce a general parameterization \cite{ROPP} of canonical variables
which goes beyond the examples --- $a$ and $V$ --- employed so far:
\begin{equation} \label{QP}
 Q=\frac{3(\ell_0a)^{2(1-x)}}{8\pi G(1-x)} \quad\mbox{and}\quad
 P=-\ell_0^{2x+1}a^{2x}\dot{a}
\end{equation}
where $\ell_0=V_0^{1/3}$ with $V_0$ (for now) fixed to $2\pi^2$.  In these
canonical variables, the Hamiltonian constraint for Lorentzian signature
reads
\begin{eqnarray}
 \frac{8\pi G}{3}C&=&-\left(\frac{8\pi
       G}{3}(1-x)Q\right)^{(1-4x)/2(1-x)}P^2- k\ell_0^2 \left(\frac{8\pi
       G}{3}(1-x)Q\right)^{1/2(1-x)} \nonumber\\
&&+ \frac{1}{3}\Lambda \left(\frac{8\pi
       G}{3}(1-x)Q\right)^{3/2(1-x)}\,, \label{C}
\end{eqnarray}
including a curvature term with $k=0$ or $k=\pm 1$.  We will not always impose
the constraint strictly, but rather use $C=-c$ constant, as suitable for
path-integral theories such as CDTs in which the lapse function is not varied,
as explained in Sec.~\ref{s:Fix}.

As developed in \cite{EffAc,Karpacz} for quantum cosmology, canonical
effective methods are based on an extended phase space on which, in addition
to $Q$ and $P$, we consider moments
\begin{equation}
 \Delta(Q^aP^b):=
 \langle(\hat{Q}-\langle\hat{Q}\rangle)^a
(\hat{P}-\langle\hat{P}\rangle)^b\rangle_{\rm symm}
\end{equation}
taking the product in totally symmetric ordering. These moments, together with
the expectation values of $\hat{Q}$ and $\hat{P}$, are equipped with a Poisson
bracket that is derived from the commutator,
\begin{equation}
 \{\langle\hat{A}\rangle,\langle\hat{B}\rangle\} =
 \frac{\langle[\hat{A},\hat{B}]\rangle}{i\hbar} \,,
\end{equation}
and extended to all moments using linearity and the Leibniz rule. In
particular, for second-order moments we have
\begin{eqnarray}
 \{\Delta(Q^2),\Delta(QP)\} &=& 2\Delta(Q^2)\\
 \{\Delta(Q^2),\Delta(P^2)\} &=& 4\Delta(QP)\\
 \{\Delta(QP),\Delta(P^2)\} &=& 2\Delta(P^2)\,,
\end{eqnarray}
while $\{\langle\hat{Q}\rangle,\langle\hat{P}\rangle\}=1$ and both expectation
values have zero Poisson brackets with all moments. For brackets of higher
moments, see \cite{EffAc,HigherMoments}.

Following \cite{GaussianDyn,QHDTunneling}, it is useful to choose
Casimir--Darboux coordinates on the phase space of second-order moments, given
by the canonical pair
\begin{equation} \label{sps}
 s=\sqrt{\Delta(Q^2)} \quad,\quad p_s = \frac{\Delta(QP)}{\sqrt{\Delta(Q^2)}}
\end{equation}
and the Casimir variable
\begin{equation} \label{U}
 U = \Delta(Q^2)\Delta(P^2)-\Delta(QP)^2\geq \frac{\hbar^2}{4}
\end{equation}
which is bounded from below by the uncertainty relation.  These variables have
brackets $\{s,p_s\}=1$ and $\{s,U\}=0=\{p_s,U\}$.

We compute an effective constraint as a function of these canonical
variables. To this end, the effective constraint is defined as the expectation
value of an operator $\hat{C}$ quantizing $C$, assuming totally symmetric
ordering of all products in the constraint. (If a different ordering is
desired, it can always be written as a combination of totally symmetric terms,
possibly with an explicit dependence on $\hbar$ from applying commutators.)
Expectation values of products of basic operators can then be written as
functions of $Q=\langle\hat{Q}\rangle$, $P=\langle\hat{P}\rangle$, $s$, $p_s$,
and $U$ as follows: We first write the basic operators $\hat{Q}$
  and $\hat{P}$ in the form $\hat{A}=A+ \Delta \hat{A}$ and then perform a
  formal Taylor expansion around $A=\langle\hat{A}\rangle$ for small
  $\Delta\hat{A}=\hat{A}-A$. Applied to the constraint $C(Q,P)$, the expansion
  yields
\begin{eqnarray}
 \langle C(\hat{Q},\hat{P})\rangle &=& \langle
 C(Q+\Delta\hat{Q},P+\Delta\hat{P})\rangle\\
&=& C(Q,P)+\sum_{a+b=2}^{\infty} \frac{1}{a!b!} \frac{\partial^{a+b}
  C(Q,P)}{\partial^aQ\partial^bP} \Delta(Q^aP^b)\,.
\end{eqnarray}
For the purpose of an approximation to first order in $\hbar$, we then
truncate all moment terms to $a+b=2$, and finally insert the inverse
\begin{equation}
 \Delta(Q^2)=s^2\quad,\quad \Delta(QP)=sp_s\quad,\quad \Delta(P^2)=p_s^2+
 \frac{U}{s^2}
\end{equation}
of (\ref{sps}), (\ref{U}). In its general form, the semiclassical constraint is
\begin{eqnarray}
 \frac{8\pi G}{3}C_{\rm s} &=& - \left(\frac{8\pi
       G}{3}(1-x)Q\right)^{(1-4x)/2(1-x)}
   \left(\left(1-
\frac{(1-4x)(1+2x)}{8(1-x)^2}\frac{s^2}{Q^2}\right)P^2\right.\nonumber\\
 &&\qquad\qquad\qquad\qquad\qquad\qquad\qquad+\left. \frac{1-4x}{1-x}
\frac{s}{Q} Pp_s+ p_s^2 +\frac{U}{s^2}\right)\nonumber\\
&& -k\ell_0^2
   \left(\frac{8\pi 
       G}{3}(1-x)Q\right)^{1/2(1-x)}\left(1+
\frac{2x-1}{8(1-x)^2}     \frac{s^2}{Q^2}\right)\nonumber\\
&&+ \frac{1}{3}\Lambda \left(\frac{8\pi
       G}{3}(1-x)Q\right)^{3/2(1-x)}  \left(1+\frac{3(1+2x)}{8(1-x)^2}
     \frac{s^2}{Q^2}\right)\,.
\end{eqnarray}

It is possible to extend this canonical effective formulation to higher orders
of moments \cite{Bosonize}. Results obtained up to fourth order
  \cite{EffPotRealize} suggest the generic behavior
  $\Delta(Q^{2n})\sim s^{2n}$ and $\Delta(Q^{2n+1})=0$ for a suitable class of
  states. (There are additional independent degrees of freedom in higher
  moments, which we do not take into account here.) Using these values, an
  ``all-orders'' effective constraint is obtained by replacing any
  potential-like term $W(Q)$ in the classical constraint by
\begin{eqnarray} \label{W}
 \langle W(\hat{Q})\rangle&=& W(Q)+\sum_{n=2}^{\infty} \frac{1}{n!}
 \frac{\partial^nW(Q)}{\partial^nQ} \Delta(Q^n)= W(Q)+\sum_{m=1}^{\infty}
 \frac{1}{(2m)!}  \frac{\partial^{2m}W(Q)}{\partial^{2m}Q} s^{2m}\nonumber\\
&=&  \frac{1}{2}\left(W(Q+s)+W(Q-s)\right)\,.
\end{eqnarray}
These all-orders effective potentials go beyond a finite-order
truncation. They have successfully been tested in various tunneling situations
\cite{EffPotRealize,Ionization}.

Not all the terms in constraints (\ref{C}) of interest here are
potential-like, but expansions similar to (\ref{W}) can be applied to any
function of $Q$ and $P$. The resulting all-orders constraint for (\ref{C}) is
given by
\begin{eqnarray} 
 \frac{8\pi G}{3}C_{\rm all}&=&-\left(\left(\frac{8\pi
       G}{3}(1-x)(Q+s)\right)^{(1-4x)/2(1-x)}+\left(\frac{8\pi
       G}{3}(1-x)(Q-s)\right)^{(1-4x)/2(1-x)}\right)\frac{P^2}{2}\nonumber\\
&&- \left(\frac{8\pi
       G}{3}(1-x)Q\right)^{(1-4x)/2(1-x)} \left(p_s^2+\frac{U}{s^2}
+ \frac{1-4x}{1-x}  \frac{sPp_s}{Q}\right)\nonumber\\
&&- 
\frac{k\ell_0^2}{2}\left(\left(\frac{8\pi
       G}{3}(1-x)(Q+s)\right)^{1/2(1-x)}+\left(\frac{8\pi
       G}{3}(1-x)(Q-s)\right)^{1/2(1-x)}\right) \nonumber\\
&&+ \frac{1}{6}\Lambda \left(\left(\frac{8\pi
       G}{3}(1-x)(Q+s)\right)^{3/2(1-x)}+\left(\frac{8\pi
       G}{3}(1-x)(Q-s)\right)^{3/2(1-x)}\right)\,. \label{Cao}
\end{eqnarray}

\subsubsection{Minisuperspace fluctuations}
\label{s:MiniFluct}

We now consider the minisuperspace model of the spherical universe with
positive spatial curvature, $k=1$, in which case $Q\propto V$ is convenient
($x=-1/2$). The Wick rotated classical constraint is
\begin{equation} \label{CSph}
C=6 \pi G V P_V^2+\frac{\Lambda V}{8\pi G}-\frac{3\ell_0^2V^{1/3}}{8\pi G}
\end{equation}
where $V=\ell_0^3 a^3$ with $\ell_0=\sqrt[3]{2\pi^2}$, and $P_V$ is the
conjugate momentum.

To the first order the semiclassical  Hamiltonian  is
\begin{eqnarray}
C_{\rm s} &=&6 \pi G V P_V^2+\frac{\Lambda V}{8\pi
  G}-\frac{3\ell_0^2V^{1/3}}{8\pi G}\nonumber\\
&&+6 \pi G V \Delta(P_{V}^2)+12 \pi G \Delta(V
P_{V})P_V+\frac{\ell_0^2 V^{-5/3}\Delta(V^2)}{24\pi G} \,.
\end{eqnarray}
In terms of canonical variables (\ref{sps}),
\begin{eqnarray}
C_{\rm s} &=&6 \pi G V P_V^2+\frac{\Lambda V}{8\pi
  G}-\frac{3\ell_0^2V^{1/3}}{8\pi G}\nonumber\\
&&+6 \pi G V \left(p_s^2+\frac{U}{s^2}\right)+16 \pi G s p_s
P_V+\frac{\ell_0^2V^{-5/3}s^2}{24\pi G}\,. 
\end{eqnarray}

 At this order, $\Delta(P_V^2)\Delta(V^2)-\Delta(V P_V)^2=U$ is a
  constant. For a Gaussian, for instance, we have the minimum value
  $U=\hbar^2/4$. All four variables, $V$, $P_V$, $s$ and $p_s$, are
  dynamical. (Since we consider a gauge-fixed treatment, we do not impose
  effective constraints \cite{EffCons,EffConsRel,TwoTimes} which would
  otherwise determine a relationship of $s$ and $p_s$ with $V$ and $P_V$.)
  Unique evolution requires initial values for all four variables,
  and therefore partial knowledge about the state through $s$ and $p_s$. The
  state derived by CDT simulations is a statistical ensemble, which we can
  represent by a thermal state with inverse temperature $\beta$ given by the
  time period, $\hbar\beta=\pi a_0$. (As shown by (\ref{Sphere}), the Euclidean
  coordinate $\tau$ is related to the 4-sphere angle $\eta$ by $\tau=a_0\eta$,
  and $\eta$ takes values in the range from $-\pi/2$ to $\pi/2$.) 

  In order to compute fluctuations in such a state, we perturb the Euclidean
  action
\begin{equation}
 S=\frac{3}{8\pi G}\int \left(a\dot{a}^2-\frac{1}{3}\Lambda a^3\right){\rm
   d}\tau
\end{equation}
by a homogeneous mode, $a=\bar{a}+\tilde{v}$. (We ignore the curvature term in
the action because it is not significant near the maximum of $a(\tau)$.) The
quadratic perturbation of the action is
\begin{equation}
 S_{\delta}=\frac{3}{8\pi G} \int \left(\dot{\tilde{v}}^2-
   \frac{\ddot{\bar{a}}}{\bar{a}} \tilde{v}^2-\Lambda
   \tilde{v}^2\right)\bar{a}{\rm d}\tau 
\end{equation}
and implies the Hamiltonian
\begin{equation}
 H_{\delta}= \frac{3}{8\pi G} \int\left(\left(\frac{4\pi G}{3}\right)^2
   \frac{\tilde{p}^2}{\bar{a}}+ \ddot{\bar{a}}\tilde{v}^2+
   \Lambda\bar{a}\tilde{v}^2\right){\rm d}\tau
\end{equation}
 with the momentum 
\begin{equation}
 \tilde{p}=\frac{3\bar{a}}{4\pi G} \dot{\tilde{v}}
\end{equation}
conjugate to $\tilde{v}$.

After a canonical transformation
\begin{equation} \label{vpcan}
 v=\sqrt{\frac{3\bar{a}}{4\pi G}}\tilde{v} \quad,\quad p=\sqrt{\frac{4\pi
   G}{3\bar{a}}}\tilde{p}\,,
\end{equation}
we obtain the Hamiltonian
\begin{equation}
  H_{\delta}= \frac{1}{2} \int\left( p^2+ \frac{\ddot{\bar{a}}}{\bar{a}}v^2+
   \Lambda v^2\right){\rm d}\tau\,.
\end{equation}
Given the background solution $\bar{a}(\tau)=a_0\cos(\tau/a_0)$ (or the
Raychaudhuri equation), we have $\ddot{\bar{a}}/\bar{a}=-1/a_0^2=-\Lambda/3$,
and therefore
\begin{equation}\label{Hamv}
  H_{\delta}= \frac{1}{2} \int\left( p^2+\frac{2}{3}
   \Lambda v^2\right){\rm d}\tau\,.
\end{equation}
is equivalent to a harmonic oscillator with frequency 
\begin{equation}
 \omega= \sqrt{2\Lambda/3}=\frac{\sqrt{2}}{a_0}\,.
\end{equation}

As shown in \cite{EffPotRealize}, it is possible to compute statistical
quantities by using a semiclassical version of (\ref{Hamv}) in which $(v,p)$
is accompanied by fluctuation variables $(s_v,p_{s_v})$. In particular, the
partition function
\begin{eqnarray}
  \mathcal{Z}(\beta,\omega,\lambda)&=&
  \int_{0}^{\infty}\int_{-\infty}^{\infty}\int_{U_{\rm min}}^{\infty}\, 
  {\rm d}s_v \,{\rm d}p_{s_v} \, {\rm d}U
 \exp{\left(-\beta \left(\frac{1}{2}p_{s_v}^2+\lambda
        \frac{U}{2 
          s_v^2}+\frac{1}{8}\omega^2 s_v^2\right)\right)} \nonumber \\
&=& 4\pi \frac{2+\beta \omega 
  \sqrt{U_{\rm min} \lambda }}{\lambda\omega^3\beta^3}
\exp{\left(-\frac{1}{2}\beta \omega
    \sqrt{U_{\rm min}\lambda }\right)} 
\end{eqnarray}
can be calculated exactly, where $\beta=1/k_{\rm B}T$, $U_{\rm
  min}=\hbar^2/4$, and $\lambda$ is a multiplier that allows us to compute the
expected uncertainty product
\begin{eqnarray}
\langle U \rangle&=&\frac{8}{\beta^2
  \omega}\frac{1}{\mathcal{Z}}\left.\frac{\partial^2 \mathcal{Z}}{\partial
  \omega \partial \lambda}\right|_{\lambda=1}\nonumber\\
&=&U_{\rm min}+\frac{24}{\beta^2
  \omega^2}+\frac{4 U_{\rm min}}{2+\sqrt{U_{\rm min}}\beta \omega}\,.
\end{eqnarray}
Moreover,
\begin{equation}
\langle
s_v^2\rangle=\left.\frac{\partial\mathcal{Z}}{\partial\omega}\right|_{\lambda=1}
=   
\frac{12}{\omega^2 
  \beta}+\frac{U_{\rm min}\beta}{1+\frac{1}{2}\sqrt{U_{\rm min}}\omega \beta}\,.
\end{equation}

We evaluate these expressions by identifying $\hbar\beta$ with the time period
of the Euclidean model, $\hbar\beta=\pi a_0$. With $U_{\rm min}=\hbar^2/4$, we
obtain
\begin{equation} \label{UIn}
 \langle U\rangle=
 \hbar^2\left(\frac{1}{4}+\frac{12}{\pi^2}+\frac{\sqrt{2}}{\pi+2\sqrt{2}}\right)
\end{equation}
and
\begin{equation}
 \langle s_v^2\rangle= \left(\frac{6}{\pi}+
 \frac{\pi}{4+\pi \sqrt{2}}\right) \hbar a_0\approx 2\hbar a_0\,.
\end{equation}
Inverting the canonical transformation (\ref{vpcan}), the average
$v$-fluctuations $\langle s_v^2\rangle$ imply fluctuations
\begin{equation}
 \langle\Delta(a^2)\rangle= \frac{4\pi G}{3\bar{a}} \langle s_v^2\rangle\approx
 \frac{8\pi}{3} \ell_{\rm P}^2
\end{equation}
at maximum $\bar{a}(\tau)=a_0$. These fluctuations are independent of $a_0$,
in agreement with the scaling behavior observed in our path-integral
derivation. The transformation (\ref{VaGaussian}) then implies volume
fluctuations
\begin{equation}\label{VIn}
 \frac{\Delta(V^2)}{V_0^2}= 24\pi \ell_{\rm P}^2 a_0^4\,.
\end{equation}

\begin{figure}[!tbp]  
  \begin{center}
    \includegraphics[scale=0.7]{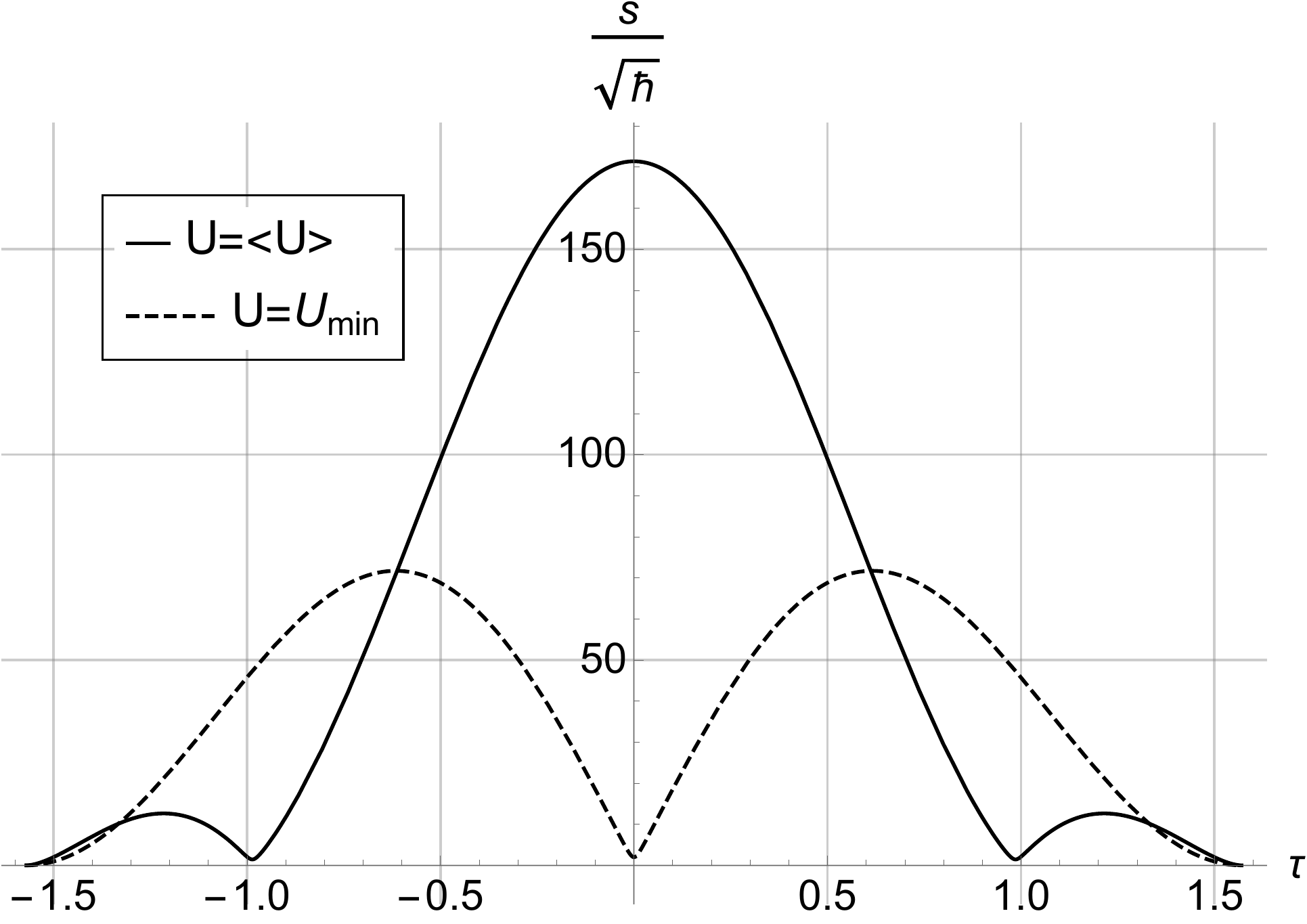}
    \caption{Volume fluctuations in a thermal state (solid), compared with a
      state with minimal uncertainty (dashed). For the solid curve we
      determine $s_0$ from \eqref{VIn} and use the ensemble average for the
      uncertainty. In the dashed curve we use $s_0=2 a_0^2$ and $U=U_{\rm
        min}=\hbar^2/4$. We also take $\Lambda=3$ and therefore
      $a_0=\sqrt{3/\Lambda}=1$.  \label{f:Flucs}}
  \end{center}
\end{figure}

The relationship (\ref{NV}) implies that the ratio
$\sqrt{\langle\Delta(N_3^2)\rangle/N_4}\approx 17$ (in units in which
$\ell_{\rm P}=1$) is almost one order of magnitude larger than seen in CDTs,
but it is encouraging to see that our new state may lead to the development of
shoulders in a plot of fluctuations: Since these fluctuations were derived at
maximum $a(\tau)$, we can use them as initial fluctuations for the volume
fluctuation $s$, in addition to $P_V=0$ and $p_s=0$.  Figure~\ref{f:Flucs}
shows good agreement with the results of our path-integral calculations in
Sec.~\ref{s:Green}, in particular after applying the transformation
(\ref{VaGaussian}); see Fig.~\ref{Transform}.  After setting symmetric initial
conditions at $\tau=0$, the evolution automatically closes the universe in the
sense that volume fluctuations approach zero at two opposite points on the
time axis. In the effective treatment, this behavior is quite non-trivial
because volume fluctuations approaching zero imply diverging momentum
fluctuations. It would therefore have been difficult to select a well-defined
initial state at one of the two endpoints of evolution.  CDTs
\cite{CDTDeSitter,CDTEffAc} show fluctuations with a local maximum at the
midpoint of evolution, where we set initial conditions, as well as
``shoulders'' between the local maximum and the two zeros. Although this
behavior is not exactly reproduced in quantitative details by our results,
Fig.~\ref{f:Flucs} indicates that there is at least qualitative agreement
provided we use the average fluctuations and uncertainty in a thermal state,
rather than the minimal values possible in a pure state.

Detailed future studies may reveal additional properties of relevant
states. In particular, there is a conceptual difference between the thermal
state we have been able to derive, which is thermal at the local maximum of
the volume, and a CDT ensemble, whose entire history is in equilibrium with
respect to local moves.

\subsection{Toroidal model}
\label{s:Toroidal}

In models with toroidal topology, a new subtlety arises because CDTs have
revealed an additional non-classical term in an effective action that would be
masked by the curvature term in a spherical model. The toroidal model also has
a more complicated relationship between the scale of fluctuations and the time
range. In this subsection, we display path-integral and minisuperspace
derivations of correlation functions for actions that include the new term
found in CDTs, while the next section will explore possible quantum origins of
this term.

\subsubsection{Correlation function}

We begin with a simple model, defined by isotropy, zero spatial curvature and
just a cosmological constant. In terms of the scale factor $a$, the classical
action is given by
\begin{equation}
 S[a] = \frac{3V_0}{8\pi G} \int {\rm d}\tau
 Na^3\left(\left(\frac{\dot{a}}{Na}\right)^2+ \frac{1}{3}\Lambda\right)
\end{equation}
where $\dot{a}$ is the derivative of $a$ by a time coordinate defined
implicitly through the lapse function $N$. (We use here the opposite sign of
the cosmological constant compared with (\ref{ClassAc}), following the
convention in CDT papers on the toroidal model. In this context,
  $\Lambda$ is interpreted as a Lagrange multiplier enforcing a fixed
  4-volume. It can therefore be derived from a simulation and turns out to
  have opposite signs in the spherical and toroidal models, respectively. The
  convention used here is such that $\Lambda>0$ is always positive. In the
  toroidal case, the absence of a curvature term then implies that the
  Euclidean model is equivalent to a Lorentzian model with the opposite sign
  of $\Lambda$. In particular, background solutions will be given by
  hyperbolic rather than trigonometric functions, as we will see.) Moreover,
$V_0=\int{\rm d}^3x$ is the volume of a finite region chosen to define the
averaging of the full action to a homogeneous version. Unlike in the spherical
model, there is no preferred value for this parameter, which does not affect
classical homogeneous solutions but plays an important role upon quantization;
see \cite{MiniSup,Infrared}.

Using the variable $w:=\sqrt{V}$, with the volume $V=V_0a^3$ of the averaging
region, we have a quadratic action
\begin{equation} \label{Sw}
 S[w]=\frac{3}{8\pi G} \int{\rm d}\tau \left(\frac{4}{9N} \left(\frac{{\rm
         d}w}{{\rm d}\tau}\right)^2+ \frac{1}{3}N\Lambda w^2\right)
\end{equation}
with a standard kinetic term. The corresponding Friedmann equation, $\delta
S/\delta N=0$, has solutions $\bar{w}(\sigma)=w_0e^{\sigma}$ with
$\sigma=\frac{1}{2}\sqrt{3\Lambda}\tau$, using from now on $N=1$ and proper
time $\tau$.

Perturbing in the form 
\begin{equation} \label{wpert}
 w(\sigma)=w_0(e^{\sigma}+y(\sigma))\,,
\end{equation}
the action for $y$ is, up to boundary terms, given by
\begin{equation} \label{Sy}
 S[y] = \frac{\sqrt{\Lambda/3}w_0^2}{4\pi G} \int{\rm d}\sigma\:
 y\left(-\frac{{\rm 
       d}^2y}{{\rm d}\sigma^2}+y\right)\,.
\end{equation}
Setting zero boundary values at $\sigma=0$ and $\sigma=\sigma_1$ (to be fixed
later), eigenfunctions of the linear operator $\hat{L}=-{\rm d}^2/{\rm
  d}\sigma^2+1$ with eigenvalues $\lambda_n=1+n^2\pi^2/\sigma_1^2$ are given
by $y_n(\sigma)=\sqrt{2/\sigma_1}\sin(n\pi \sigma/\sigma_1)$. We can therefore
compute $y$-fluctuations from the Green's function as in Sec.~\ref{s:Green}:
\begin{eqnarray}
 &&G_y(\sigma,\sigma)=\frac{8\pi G\hbar}{\sqrt{\Lambda/3}w_0^2\sigma_1} 
\sum_{n=1}^{\infty} \frac{\sin^2(n\pi \sigma/\sigma_1)}{1+n^2\pi^2/\sigma_1^2}\\
&=& =
- \frac{\pi G\hbar}{\sqrt{\Lambda/3}w_0^2}
\left(\frac{2}{\sigma_1}-2\coth(\sigma_1)\right. \\ 
&&\qquad\qquad+
e^{-2i\pi \sigma/\sigma_1}
\left(\frac{{}_2F_1(1,1-i\sigma_1/\pi,2-i\sigma_1/\pi,e^{-2\pi i 
    \sigma/\sigma_1})}{\sigma_1+i\pi}\right.\nonumber\\ 
&&\qquad\qquad\qquad\qquad\left.+
  \frac{{}_2F_1(1,1+i\sigma_1/\pi,2+i\sigma_1/\pi,e^{-2\pi i
      \sigma/\sigma_1})}{\sigma_1-i\pi}\right)\nonumber\\ 
&&\qquad\qquad+ e^{2i\pi \sigma/\sigma_1}\left(
    \frac{{}_2F_1(1,1-i\sigma_1/\pi,2-i\sigma_1/\pi,e^{2\pi i
        \sigma/\sigma_1})}{\sigma_1+i\pi}\right.\nonumber\\ 
&&\qquad\qquad\qquad\qquad+ \left.\left.
    \frac{{}_2F_1(1,1+i\sigma_1/\pi,2+i\sigma_1/\pi,e^{2\pi i
      \sigma/\sigma_1})}{\sigma_1-i\pi}\right)\right)\,. \nonumber
\end{eqnarray}

\begin{figure}
\begin{center}
\includegraphics[width=10cm]{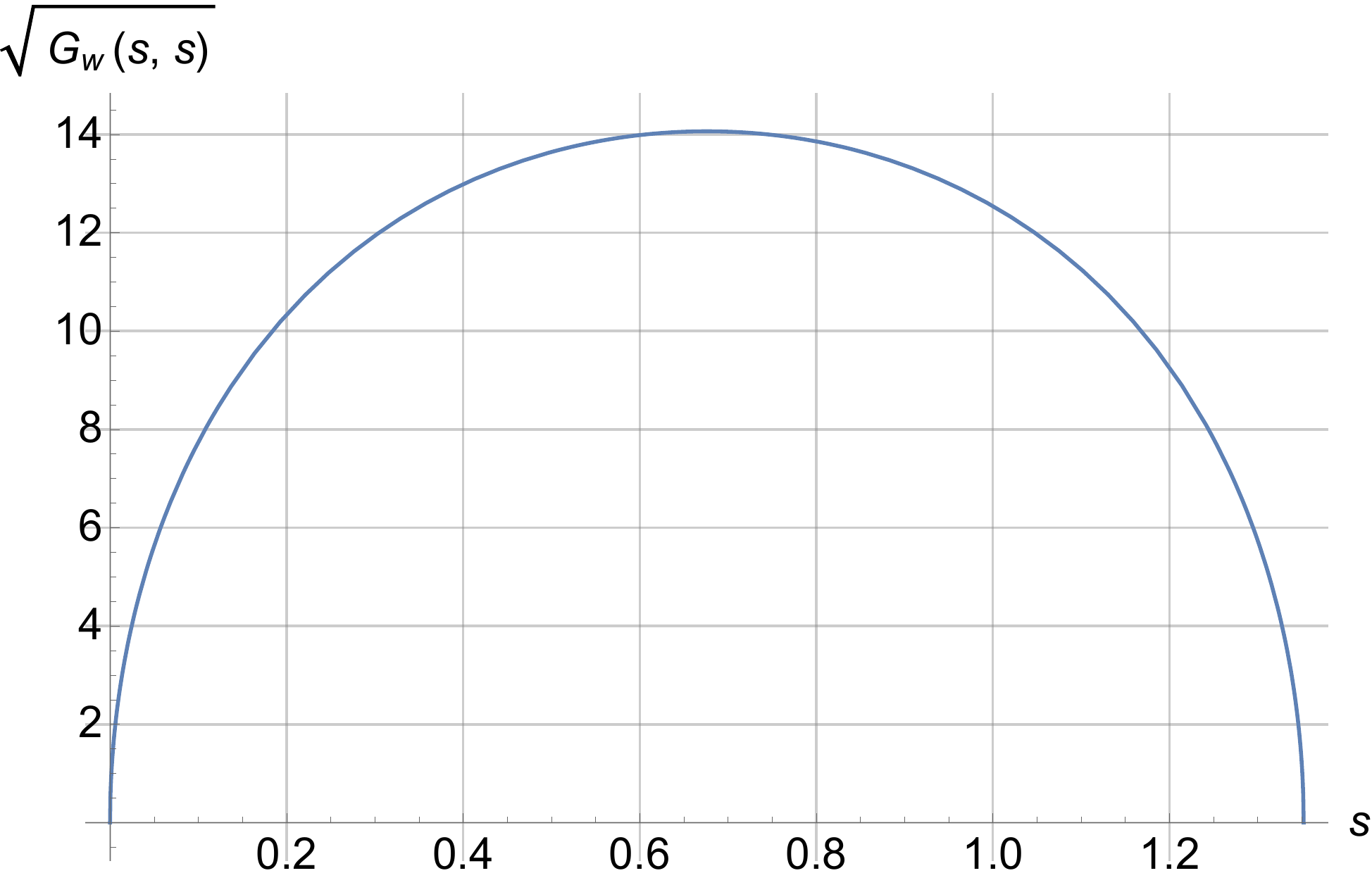}
\caption{Fluctuations of $w(\sigma)$: $\sqrt{G_w(\sigma,\sigma)}$ for an action
  (\ref{Sw}). \label{f:Gw0}}
\end{center}
\end{figure}

\begin{figure}[htbp]
\begin{center}
\includegraphics[width=12cm]{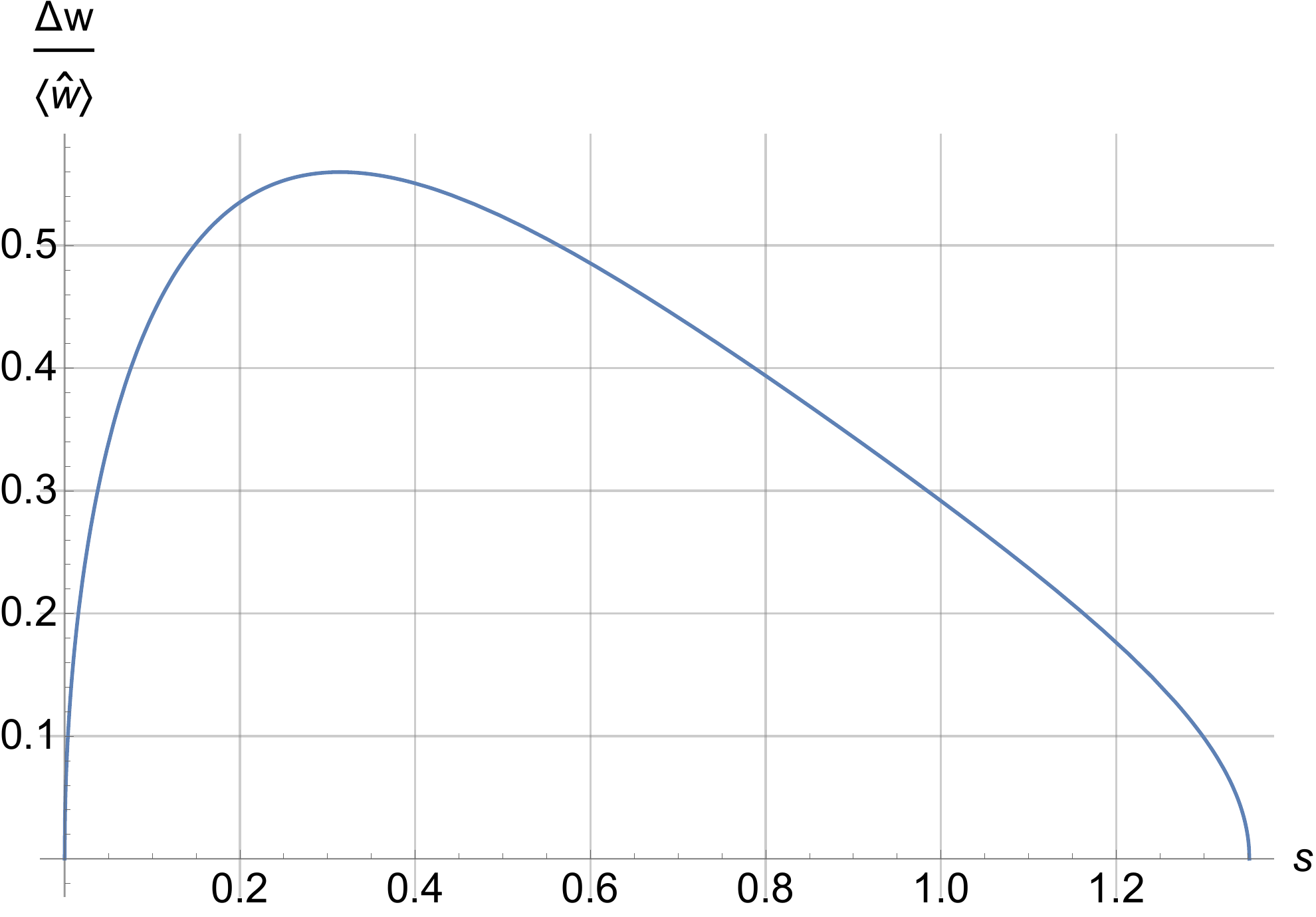}
\caption{Relative fluctuations of $w(\sigma)$ for an action
  (\ref{Sw}).  \label{f:relw0}}
\end{center}
\end{figure}

Fluctuations of $w$ are related to $y$-fluctuations by
\begin{equation} \label{Gwy}
 G_w(\sigma,\sigma)=w_0^2 G_y(\sigma,\sigma)\,
\end{equation}
shown in Fig.~\ref{f:Gw0}.  The result, unlike $w$, is independent of
$w_0$. Therefore, the ratio $(\Delta w)/w$ (shown in Fig.~\ref{f:relw0} for
one example) depends on $w_0$ and the averaging volume $V_0$.

In \cite{CDTTopology,CDTTorus}, an isotropic model with $k=0$ has been studied
using CDTs. As one of the results, the effective minisuperspace action has an
additional term proportional to $V^{\gamma}$ where $\gamma$ is close to
$-3/2$.  With such a term in the action,
\begin{equation} \label{Smu}
 S[w]= \frac{3}{8\pi G}\int{\rm d}\tau \left(\frac{4}{9N} \left(\frac{{\rm
         d}w}{{\rm 
         d}\tau}\right)^2+\frac{1}{3}N\left(\mu w^{2\gamma}+ \Lambda
     w^2\right)\right) \,,
\end{equation}
it is still possible to solve the corresponding Friedmann equation
analytically, by
\begin{equation} \label{wbar}
 \bar{w}(\sigma)= (\mu/\Lambda)^{1/2(1-\gamma)}
 \sinh(\sigma+\sigma_0)^{1/(1-\gamma)}\,,
\end{equation}
now with $\sigma=\frac{3}{2}(1-\gamma)\sqrt{\Lambda/3}\tau$.  Notice that the
action is no longer homogeneous in $w$, and therefore $\mu$ should scale with
the averaging volume $V_0$ used to define homogeneous variables such that the
volume is $V=V_0a^3$. Since $\mu w^{2\gamma}$ should scale like $w^2=V=V_0a^3$
if we change $V_0$, $\mu$ should be proportional to $V_0^{1-\gamma}$. The
solution (\ref{wbar}) is then proportional to $\sqrt{V_0}$, which is the
correct scaling behavior of $w$.

\begin{figure}
\begin{center}
\includegraphics[width=10cm]{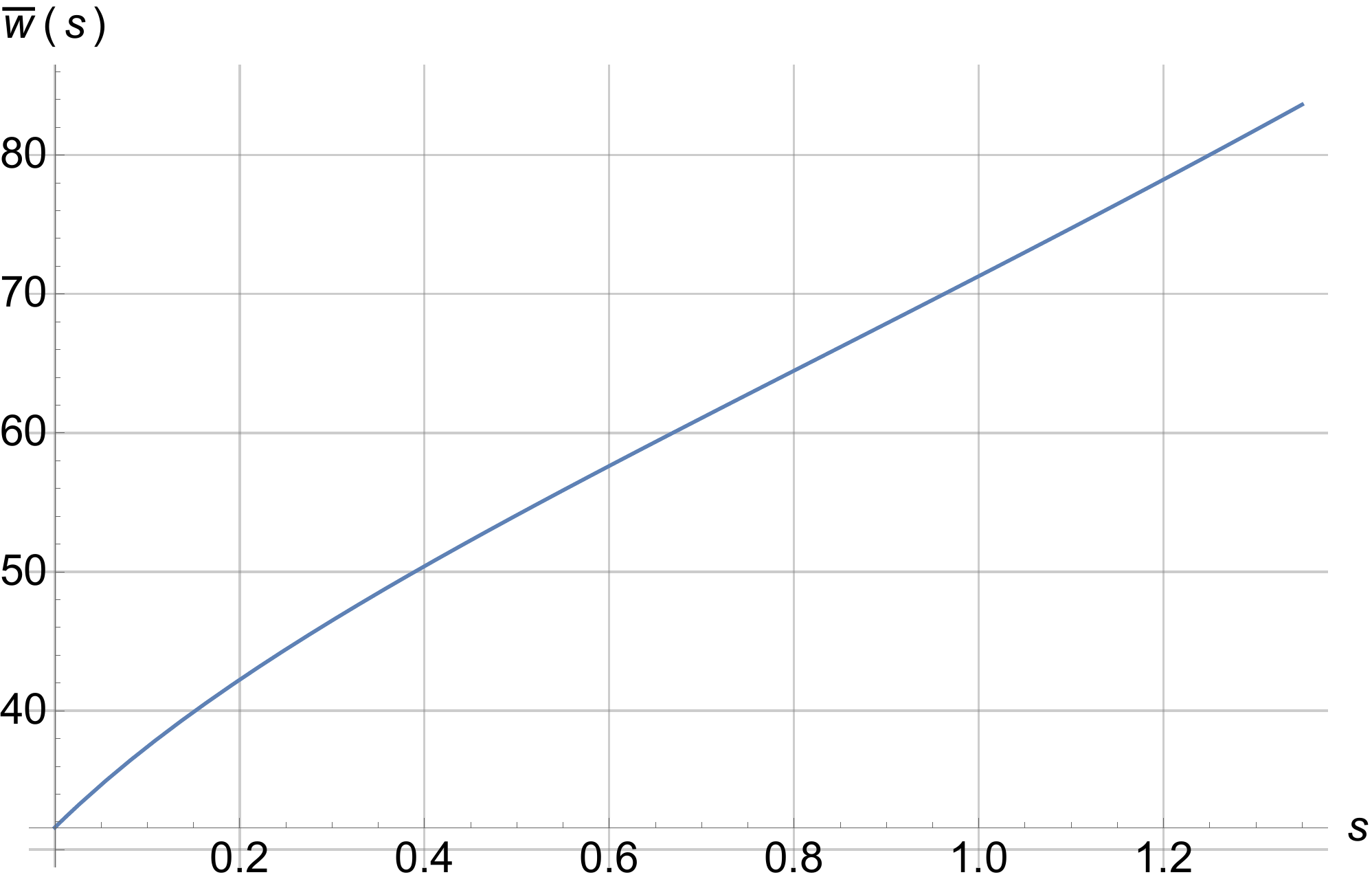}
\caption{The function $\bar{w}(\sigma)$ for boundary values as used in toroidal
  models of CDTs.  \label{f:w}} 
\end{center}
\end{figure}

The integration constant $\sigma_0$ is determined by the value of $\bar{w}$ at
the first boundary point, $\sigma=0$:
\begin{equation}
 \sinh(\sigma_0)=\sqrt{\frac{\Lambda}{\mu}}\bar{w}(0)^{1-\gamma}\,.
\end{equation}
The second boundary point, $\sigma=\sigma_1$, is then fixed so as to obtain a
desired value of $\bar{w}(\sigma_1)$:
\begin{equation}
  \sinh(\sigma_1+\sigma_0) = \sqrt{\frac{\Lambda}{\mu}}
  \bar{w}(\sigma_1)^{1-\gamma} 
\end{equation}
or
\begin{eqnarray}
 \sinh(\sigma_1) &=& \sinh(\sigma_1+\sigma_0)\cosh(\sigma_0)-
 \cosh(\sigma_1+\sigma_0)\sinh(\sigma_0)\nonumber\\
&=& \frac{\Lambda}{\mu} \bar{w}(0)^{1-\gamma} \bar{w}(\sigma_1)^{1-\gamma}
\left(\sqrt{1+\frac{\mu}{\Lambda} \bar{w}(0)^{2(\gamma-1)}}-
  \sqrt{1+\frac{\mu}{\Lambda} \bar{w}(\sigma_1)^{2(\gamma-1)}}\right)\,.
\end{eqnarray}
For a comparison with results from CDTs \cite{CDTTopology,CDTTorus}, we will
use $\bar{w}(0)=\sqrt{10^3}=31.6$ and $\bar{w}(\sigma_1)=\sqrt{7000}=83.7$,
and thus $\sigma_0=0.195$ and $\sigma_1=1.352$. Moreover, $\mu=2.86\cdot 10^5$
and $\Lambda=3.5\cdot 10^{-4}$, such that $\sqrt{\mu/\Lambda}=2.86\cdot 10^4$.
For these values, $\bar{w}(s)$ is shown in Fig.~\ref{f:w}.

We perturb (\ref{wbar}) by
\begin{equation} \label{wpertmu}
 w(\sigma)= (\mu/\Lambda)^{1/2(1-\gamma)}
 \left(\sinh(\sigma+\sigma_0)^{1/(1-\gamma)}+y(\sigma)\right)
\end{equation}
such that $y$ is independent of $V_0$. (The $V_0$-scaling in (\ref{wpert}) is
now provided by $\mu^{1/2(1+\gamma)}$.)  Up to boundary terms, the perturbed
action (with $N=1$) expanded to second order is
\begin{equation} \label{symu}
 S[y] = \frac{\sqrt{\Lambda/3}}{4\pi G(1-\gamma)}
 \left(\frac{\mu}{\Lambda}\right)^{1/(1-\gamma)}\!\!\! \int{\rm d}\sigma\:
 y\!\left(-(1-\gamma)^2\frac{{\rm d}^2y}{{\rm d}\sigma^2}+
   1+\gamma(\gamma-1)
     \sinh(\sigma+\sigma_0)^{2(2\gamma-1)/(1-\gamma)}\right)\,.
\end{equation}

\begin{figure}
\begin{center}
\includegraphics[width=10cm]{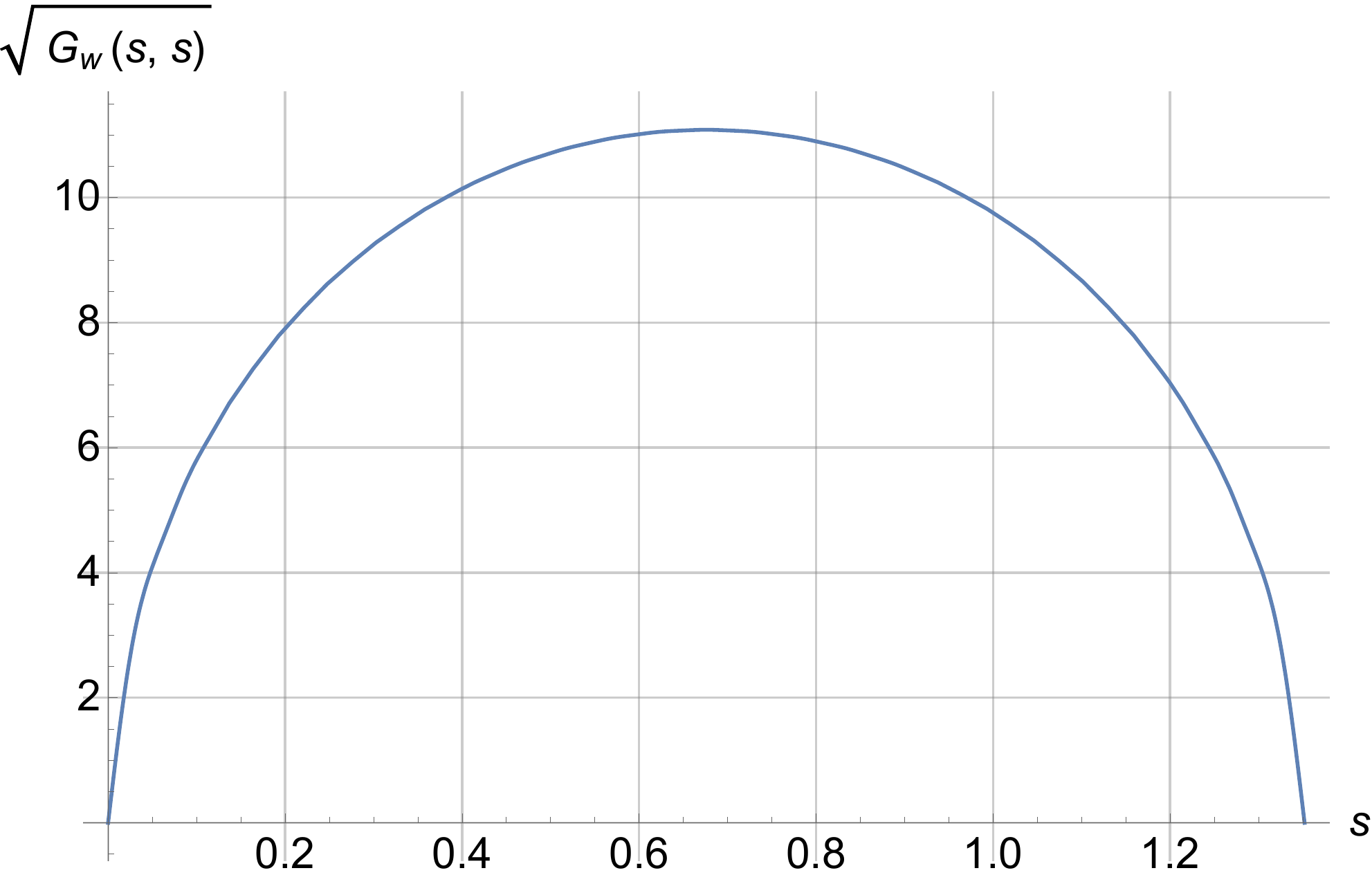}
\caption{Fluctuations of $w(\sigma)$: $\sqrt{G_w(\sigma,\sigma)}$ for
  $\gamma=-3/2$, based on a numerical integration of (\ref{lambdatilde}) and
  summing the first 25 terms in the corresponding series
  (\ref{Gw}).  \label{f:Gw}}
\end{center}
\end{figure}

\begin{figure}
\begin{center}
\includegraphics[width=12cm]{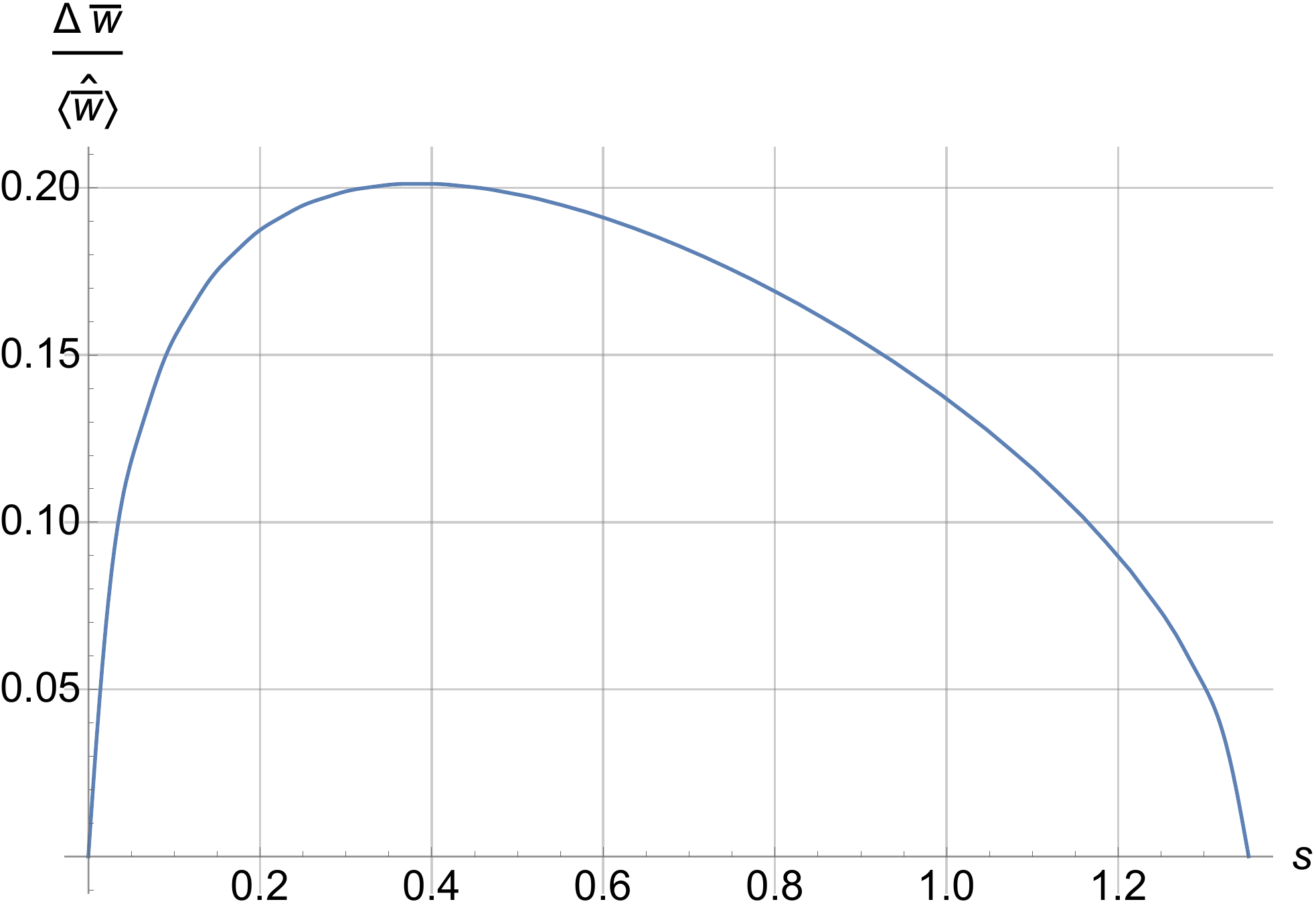}
\caption{Relative fluctuations of $w(\sigma)$ for $\gamma=-3/2$, based on a
  numerical integration of (\ref{lambdatilde}) and summing the first 25 terms
  in the corresponding series (\ref{Gw}). \label{f:relw}}
\end{center}
\end{figure}

The new linear operator leads to an eigenvalue problem with a potential given
by a power of the sinh function, which in general is difficult to
solve. However, for the values of $\gamma$ indicated by CDTs, the
$\sigma$-dependent potential in (\ref{symu}) can be treated as a
perturbation. For instance, for $\gamma\approx -1$, the potential is
proportional to $\bar{w}^{2(2\gamma-1)}=V^{2\gamma-1}=V^{-3}$, while for
$\gamma\approx -3/2$ it is proportional to $V^{-4}$ in terms of the volume
$V$, which is large in the parameter range considered in toroidal models of
CDTs.  Using quantum-mechanical perturbation theory, the eigenvalues are then
approximately given by
\begin{equation} \label{lambdatilde}
 \tilde{\lambda}_n = 1+(\gamma-1)^2\frac{n^2\pi^2}{\sigma_1^2}+
 \frac{2}{\sigma_1}\frac{\gamma}{\gamma-1} \int_0^{\sigma_1}{\rm 
   d}\sigma  \sin^2(n\pi \sigma/\sigma_1)
 \sinh(\sigma+\sigma_0)^{2(2\gamma-1)/(1-\gamma)}\,. 
\end{equation}
In this case, we have to multiply $y$-fluctuations with
$(\mu/\Lambda)^{1/(1-\gamma)}$ in order to obtain $w$-fluctuations:
\begin{equation} \label{Gw}
 G_w(\sigma,\sigma) = \frac{8\pi G(1-\gamma)\hbar}{\sqrt{\Lambda/3}\sigma_1}
 \sum_{n=1}^{\infty} 
 \frac{\sin^2(n\pi \sigma/\sigma_1)}{\tilde{\lambda}_n}\,,
\end{equation}
see Fig.~\ref{f:Gw}. 
These fluctuations are independent of $\mu$, but they are sensitive to the
$\mu w^{2\gamma}$-term in the action through the value of $\gamma$. Formally,
the previous result (\ref{Gwy}) is obtained for $\gamma=0$, in which case
$\mu$ just adds a constant to the action (\ref{Smu}).  Relative fluctuations,
shown in Fig.~\ref{f:relw}, depend on $\mu$ through (\ref{wpertmu}).

We note that the curvature term of the spherical model cannot be treated as a
perturbation around the cosmological-constant model. In terms of $w$, the
action is then
\begin{equation}
 S=\frac{3\pi}{4G} \int{\rm d}\tau \left(\frac{4}{9}\dot{w}^2-\frac{1}{3}\Lambda
   w^2+\ell_0^2w^{2/3}\right)
\end{equation}
with $\Lambda>0$. Therefore, the cosmological-constant model in this case does
not lead to a discrete spectrum of $\lambda_n$. Moreover, the curvature term,
which corresponds to $\mu=1$ and $\gamma=1/3$ when written as $\mu
w^{2\gamma}$, implies a perturbative correction in the analog of (\ref{symu})
which is proportional to $\bar{w}^{-1/3}$. Although the exponent is still
negative, small $w$ are important in the spherical model near the end points
of its evolution. For such values, the curvature term is not just a
perturbation.

\subsubsection{Volume fluctuations}

We have computed fluctuations of $w=\sqrt{V}$. In order to obtain volume
fluctuations as in Sec.~\ref{s:Rel}, we expand
\begin{eqnarray}
 \Delta(V^2) &=& \langle(\Delta \hat{V})^2\rangle = \langle
 (\hat{w}^2-\langle\hat{w}^2\rangle)^2\rangle\\
&=& \langle\hat{w}^4\rangle- \langle\hat{w}^2\rangle^2= \langle(\Delta
\hat{w}+\langle\hat{w}\rangle)^4\rangle-
\langle(\Delta\hat{w}+\langle\hat{w})^2\rangle^2\\
&=& \Delta(w^4)- \Delta(w^2)^2 + 4\langle\hat{w}\rangle\Delta(w^3)+
4\langle\hat{w}\rangle^2\Delta(w^2)\,. \label{Vw}
\end{eqnarray}
For a Gaussian state,
\begin{equation}
 \Delta(w^{2n})=(2n-1)!! \Delta(w^2)^n\,,
\end{equation}
as before, and (\ref{Vw}) can be simplified to
\begin{equation} \label{Vw2}
 \Delta(V^2) = 2\left(2\langle\hat{w}\rangle^2+\Delta(w^2)\right)
 \Delta(w^2)\,.
\end{equation}
For relative fluctuations, we obtain
\begin{eqnarray}
 \frac{\Delta V}{\langle\hat{V}\rangle} &=&
 \frac{\sqrt{\Delta(V^2)}}{\langle\hat{w}^2\rangle}=
   \frac{\sqrt{\Delta(V^2)}}{\Delta(w^2)+\langle\hat{w}\rangle^2}\nonumber\\
&=&
2\frac{\sqrt{\langle\hat{w}\rangle^2+
\frac{1}{2}\Delta(w^2)}}{\langle\hat{w}\rangle^2+\Delta(w^2)}
\Delta w \approx 2\frac{\Delta w}{\langle\hat{w}\rangle}\,.
\end{eqnarray}

\begin{figure}[htbp]
\begin{center}
\includegraphics[width=10cm]{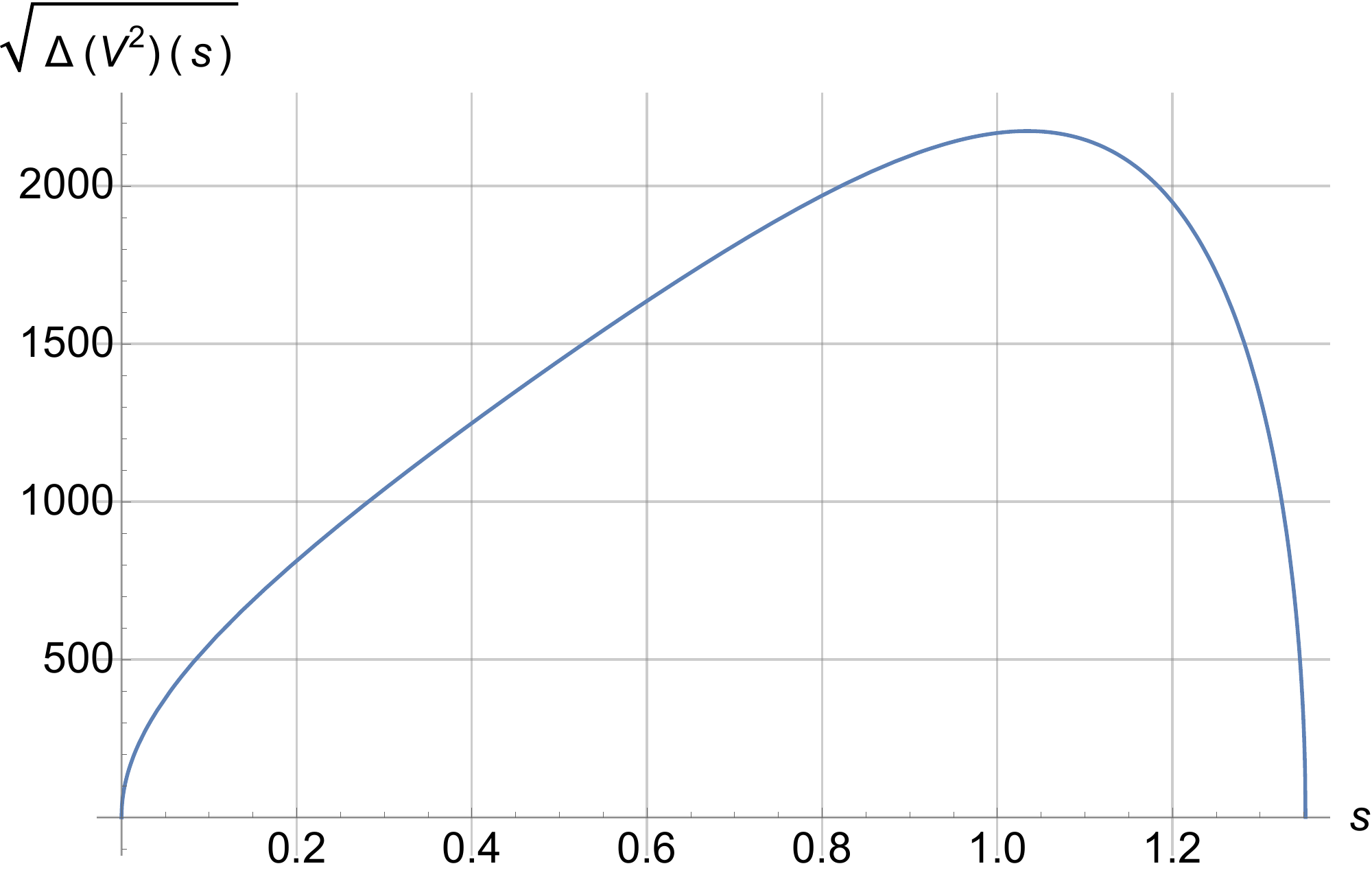}
\caption{Fluctuations of $V(\sigma)$: $\sqrt{\Delta(V^2)(\sigma)}$ for
  $\mu=0$, where the initial volume is $V(0)=10^3$. \label{f:GV0}}
\end{center}
\end{figure}

\begin{figure}
\begin{center}
\includegraphics[width=10cm]{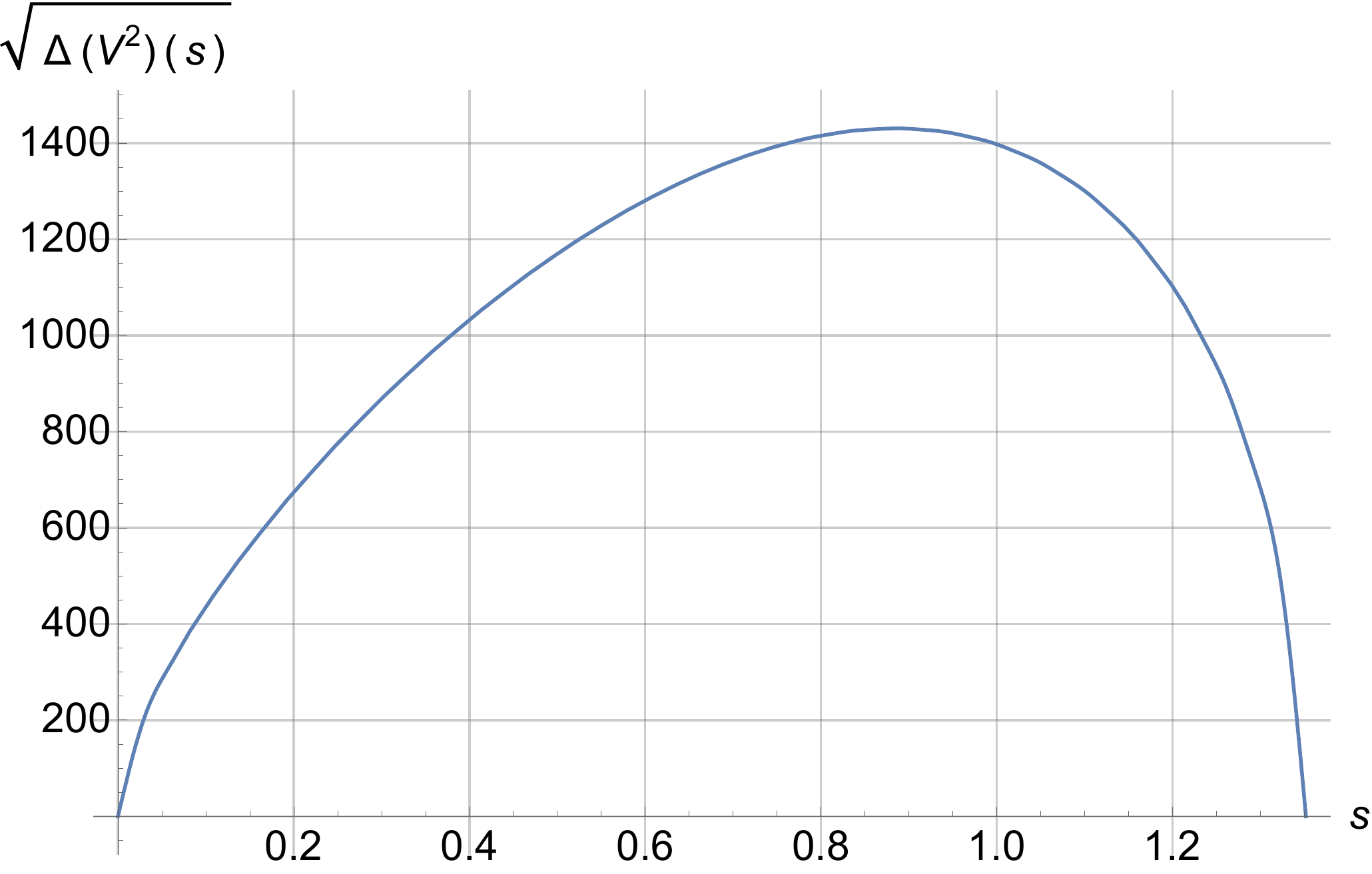}
\caption{Fluctuations of $V(\sigma)$: $\sqrt{\Delta(V^2)(\sigma)}$ for $\gamma
  =-3/2$, based on a numerical integration of (\ref{lambdatilde}) and summing
  the first 25 terms in the corresponding series (\ref{Gw}). \label{f:GV}}
\end{center}
\end{figure}

The volume fluctuations obtained from $w$-fluctuations are shown in
Figs.~\ref{f:GV0} and \ref{f:GV}, demonstrating qualitative agreement with CDT
results \cite{CDTTopology,CDTTorus}. Moreover, we can confirm the
  same scaling behavior of fluctuations as seen in the spherical model,
  Sec.~\ref{s:Rel}, except that there is a different universal function that
  describes the shape of the fluctuation curve. To this end, we first compute
  the $\Lambda$-dependence of the 4-volume
\begin{eqnarray}
 V_4&=& V_0\int\bar{w}(\tau)^2{\rm d}\tau= 
\frac{2V_0}{(1-\gamma)\sqrt{3\Lambda}}
   \int_0^{\sigma_1}  \bar{w}(\sigma)^2{\rm d}\sigma\nonumber\\
 &=&  \frac{2V_0}{(1-\gamma)\sqrt{3\Lambda}}
   \left(\frac{\mu}{\Lambda}\right)^{1/(1-\gamma)} \int_0^{\sigma_1}
   \sinh(\sigma+\sigma_0)^{2/(1-\gamma)}{\rm d}\sigma \propto
   \Lambda^{(3-\gamma)/(2(1-\gamma))}\,. 
\end{eqnarray}
Using (\ref{Vw2}), volume fluctuations
\begin{eqnarray}
 G_V(\sigma,\sigma)&\approx& 4\bar{w}(\sigma)^2 G_w(\sigma,\sigma)\nonumber\\
&=& \frac{32\sqrt{3}\pi
   G(1-\gamma)\hbar}{\sqrt{\Lambda}\sigma_1}
 \left(\frac{\mu}{\Lambda}\right)^{1/(1-\gamma)}
 \sinh(\sigma+\sigma_0)^{1/(1-\gamma)}
 \sum_{n=1}^{\infty} \frac{\sin^2(n\pi
   \sigma/\sigma_1)}{\tilde{\lambda}_n}\nonumber\\ 
& \propto& \Lambda^{(3-\gamma)/(2(1-\gamma))}\propto V_4 \label{GVTor}
\end{eqnarray}
have the same $\Lambda$-dependence, such that $\Delta(V^2)/V_4$ as a function
of
\begin{equation} \label{stauTor}
 \sigma\propto \sqrt{\Lambda}\tau\propto V_4^{(\gamma-1)/(3-\gamma)}\tau
\end{equation}
is independent of $V_4$. Provided the rescaling of the time coordinate is
adjusted from $\tau/V_4^{1/4}$ in the spherical model to
$V_4^{(\gamma-1)/(3-\gamma)}\tau$, a universal scaling behavior is
obtained. This result assumes a fixed $\gamma$, while some CDT fits suggest
that $\gamma$ may be running \cite{CDTTopology,CDTTorus} in which case there
could be violations of the universal behavior. Comparing the scaling behavior
(\ref{GVTor}), (\ref{stauTor}) derived here with CDT simulations may therefore
shed further light on the role of $\gamma$.

\subsubsection{Fluctuations}
\label{s:TorFluc}

We now apply the minisuperspace methods used in Sec.~\ref{s:SphMin} to the
toroidal model. First ignoring the term $\mu V^{\gamma}$, the
constraint
\begin{equation} \label{CTor}
C=-6 \pi G V P_V^2+\frac{\Lambda V}{8\pi G}
\end{equation}
with $\Lambda>0$ generates equations of motion
\begin{eqnarray}
 \dot{V} &=& -12\pi GVP_V \label{VTor} \\
 \dot{P}_V &=& 6\pi GP_V^2-\frac{\Lambda}{8\pi G}= -\frac{C}{V}\,. \label{PVTor}
\end{eqnarray}
(Again, we have a different sign convention compared with (\ref{CSph}),
following \cite{CDTTopology,CDTTorus}.)

For $C=0$, $4\pi GP_V= \pm \sqrt{\Lambda/3}$ is constant, such that
$V(\tau)\propto \exp(\mp \sqrt{3\Lambda}\:\tau)$. 
For $C\not=0$, we combine the two first-order equations of motion to obtain
\begin{equation}
 \ddot{V}=-12\pi G \frac{{\rm d}(VP_V)}{{\rm d}\tau}= 12\pi G(12\pi
 GVP_V^2+C)=   3(\Lambda V-4\pi G C)\,.
\end{equation}
This inhomogeneous equation is solved by
\begin{equation}
 V(\tau)= 4\pi G\frac{C}{\Lambda}+ A\exp\left(\sqrt{3\Lambda}\:\tau\right)+
 B\exp\left(-\sqrt{3\Lambda}\:\tau\right)\,.
\end{equation}
Therefore,
\begin{equation}
 P_V(\tau)=-\frac{\dot{V}}{12\pi GV}= -\frac{1}{4\pi
   G}\sqrt{\frac{\Lambda}{3}} 
 \frac{A\exp\left(\sqrt{3\Lambda}\:\tau\right)- B\exp\left(-
     \sqrt{3\Lambda}\:\tau\right)}
{A\exp\left(\sqrt{3\Lambda}\:\tau\right)+ 
 B\exp\left(-\sqrt{3\Lambda}\:\tau\right)+4\pi GC/\Lambda}\,.
\end{equation}
The constraint equation then implies $AB=4\pi^2G^2C^2/\Lambda^2$, solutions of
which can be parameterized by a single constant $D$ such that
\begin{equation}
 A=2\pi G\frac{CD}{\Lambda}\quad,\quad B=2\pi G\frac{C}{\Lambda D}\,.
\end{equation}
In order for $V(\tau)$ to have positive late-time regimes, we need $A>0$ and
$B>0$. Therefore, ${\rm sgn}D={\rm sgn}C$. With these conditions, we can write
the solutions as
\begin{eqnarray}
 \frac{V(\tau)}{8\pi G} &=& \left\{\begin{array}{cl} C\Lambda^{-1}\cosh^2\left(
       \frac{1}{2}\sqrt{3\Lambda}\:\tau+\log|D|\right) & \mbox{if }C>0\\ 
     |C|\Lambda^{-1}\sinh^2\left(\frac{1}{2}\sqrt{3\Lambda}\:\tau+\log|D|\right)
     & \mbox{if }C<0\end{array}\right.\\ 
 4\pi GP_V(\tau)&=& \left\{\begin{array}{cl} -\sqrt{\Lambda/3}
     \tanh\left(\frac{1}{2}\sqrt{3\Lambda}\:\tau+\log|D|\right) & \mbox{if
     }C>0\\ 
 -\sqrt{\Lambda/3}
     \coth\left(\frac{1}{2}\sqrt{3\Lambda}\:\tau+\log|D|\right) & \mbox{if
     }C<0\end{array}\right. 
\end{eqnarray}
Notice that $P_V(\tau)$ depends on $C$ only through ${\rm sgn}D={\rm
  sgn}C$. Moreover, the two independent solutions in the limiting case of
$C=0$ are obtained for $D=0$ and $D\to\infty$, respectively.

We introduce fluctuation terms to second order, using the quantum constraint
\begin{equation} \label{CQTor}
C_{\rm s}=-6 \pi G V P_V^2+\frac{\Lambda V}{8\pi G} -6 \pi G V
\left(p_s^2+\frac{U}{s^2}\right)-12 
\pi G P_Vs p_s  \,.
\end{equation}
The dependence on $s$ is not quadratic owing to the term $U/s^2$, where $U\geq
\hbar^2/4$. However, we can formally interpret this term as a centrifugal
potential of a system with two auxiliary fluctuation variables, $X$ and $Y$,
such that (\ref{CQTor}) is the spherically symmetric reduction with
\begin{equation} \label{sxy}
s^2=X^2+Y^2\,.
\end{equation}
The extended constraint,
\begin{equation} \label{CQTorXY}
C_{\rm s}=-6 \pi G V P_V^2+\frac{\Lambda V}{8\pi G} -6 \pi G V
\left(p_X^2+p_Y^2\right)-12 \pi G 
P_V\left(X p_X+Y p_Y \right) \,,
\end{equation}
is quadratic in the new variables and such that $X$ and $Y$ decouple.
Focusing on $X$ and its momentum $p_X$, we have equations of motion
\begin{eqnarray}
\dot{p}_X&=&12 \pi G P_V p_X \label{dotpX} \\
\dot{X}&=&-12 \pi G (V p_X- P_V X)
\end{eqnarray}
linear in these variables, but coupled to the expectation values $V$ and
$P_V$. 

If we first assume small quantum back-reaction, we can solve for $X$ and $p_X$
by using the classical equations of motion and solutions for $V$ and $P_V$. In
particular, replacing $12\pi GP_V$ in (\ref{dotpX}) by $-\dot{V}/V$, using
(\ref{VTor}), this equation turns into
\begin{equation}
 \dot{p}_X=-\frac{\dot{V}}{V}p_X
\end{equation}
and is solved by
\begin{equation}
p_X=\frac{B_X}{V}
\end{equation}
with constant $B_X$. The second equation can then be written as
\begin{equation}
\dot{X}=-32 \pi^2 G^2 B_X +\frac{\dot{V}}{V}X
\end{equation}
with solution
\begin{equation}
X(\tau)=A_X V(\tau)-12 \pi G B_X V(\tau) \int^{\tau}_{0}\frac{1}{V(t)}{\rm d}t
\end{equation}
with another constant $A_X$. Again using a classical equation, (\ref{PVTor}),
this solution can be simplified to
\begin{equation} \label{xsol}
X(\tau)=V(\tau)\left(A_X +12 \pi G \frac{B_X}{C}  P_V(\tau)\right)
\end{equation}
using (\ref{PVTor}), assuming $C\not=0$. For $C=0$, we have 
\begin{equation} \label{xC}
 X(\tau)=A_XV\mp 4\pi G\sqrt{3/\Lambda}B_X\,.
\end{equation}

In terms of auxiliary variables, we have related the semiclassical constant of
motion $U$ to angular momentum in the $XY$-plane. The additional condition
\begin{equation}
U=(X p_Y - Y p_X)^2\geq \frac{\hbar^2}{4}
\end{equation}
should therefore be imposed, which evaluates to
\begin{equation}
\left(A_X B_Y-A_Y B_X\right)^2\geq \frac{\hbar^2}{4}
\end{equation}
Moreover, the phase of the auxiliary variables, related to
\begin{equation}
\cot{(\phi(\tau))}=\frac{X}{Y}=\frac{A_XC+12 \pi GB_X
  P_V(\tau)}{A_YC+12 \pi G  B_Y P_V(\tau)} \,,
\end{equation}
is spurious. At any time $\tau$ at which $P_V(\tau)\not=0$, we can eliminate
the spurious phase by fixing the ratio of $B_X/B_Y$. For simplicity, we choose
$B_X=B_Y=B$ and therefore
\begin{equation} \label{UB}
U=B^2\left(A_X-A_Y\right)^2\geq \frac{\hbar^2}{4} \,.
\end{equation}
(This ratio can formally be identified with the spurious phase at a
singularity: The background solutions contain a $\tau_{\infty}$
where $P_V(\tau_{\infty})\to\infty$, and
\begin{equation}
\cot{(\phi(\tau_{\infty}))}=\frac{B_X}{B_Y}\,.
\end{equation}
However, around a singularity it may not be safe to assume weak quantum
back-reaction.)

For a comparison with CDT results or our preceding path-integral calculations,
we are interested in solutions that have zero fluctuations at two different
times, the endpoints of a CDT universe. Using (\ref{sxy}), $s=0$ if and only
if $X=0$ and $Y=0$. Equation~(\ref{xsol}) and its analog for $Y(\tau)$ then
imply that at least at one $\tau$, we have
\begin{equation}
 P_V(\tau)=-\frac{A_XC}{12\pi G B}=-\frac{A_YC}{12\pi G B}\,.
\end{equation}
Since the resulting $A_X=A_Y$ is in violation of (\ref{UB}), it is
impossible to have zero fluctuations even at a single time, let
alone two times for the endpoints of a CDT universe. Similarly, if $C=0$ the
solution (\ref{xC}) implies that $A_X=A_Y$ following the same
arguments. Quantum back-reaction therefore seems relevant. An analysis then
requires numerical solutions, in which we will consider also a possible term
$\mu V^{\gamma}$.

Including the term $\mu V^{\gamma}$, the constraint
\begin{equation}
  C=-6\pi GVP_V^2+\frac{\Lambda V}{8\pi G}+\frac{\mu V^{\gamma}}{8\pi G}
\end{equation}
generates the equations of motion
\begin{eqnarray}
 \dot{V} &=& -12\pi GVP_V \\
 \dot{P}_V &=& 6\pi GP_V^2-\frac{\Lambda}{8\pi G}-\frac{\gamma\mu
 V^{\gamma-1}}{8\pi G}= -\frac{C}{V}+ 
 \frac{(1-\gamma)\mu}{8\pi G} V^{\gamma-1}\,. \label{PVTorgamma}
\end{eqnarray}
In terms of $w=\sqrt{V}$, the constraint equation can be written as
\begin{equation}
 \frac{1}{6\pi G} \left(\frac{{\rm d}w}{{\rm d}\tau}\right)^2= -
 C+\frac{\Lambda w^2}{8\pi G}+ \frac{\mu w^{2\gamma}}{8\pi G}\,.
\end{equation}
As in the example of the Appendix, solutions can be written in terms of
Weierstrass' function, but we will not use them explicitly.

Including fluctuations to second order, the constraint is
\begin{align}
\label{eq:constraint_gamma}
C_{\gamma} = C_{\rm s} + \frac{\mu V^{\gamma}}{8\pi G} + \frac{\mu \gamma
\, (\gamma  -1)}{16\pi G} V^{\gamma-2}  \, s^2 
\end{align}
where $C_{\rm s}$ is given in (\ref{CQTor}). Alternatively, we may use
(\ref{CQTorXY}) with $s^2=X^2+Y^2$. In the latter case, the equations of
motion are
\begin{eqnarray}
 \dot{p}_X&=& 12\pi GP_Vp_X- \frac{\mu\gamma(\gamma-1)}{8\pi G}V^{\gamma-2}X\\
 \dot{X}&=& -12\pi G(Vp_X+P_VX) \label{Xdot}
\end{eqnarray}
and, using the background equations (not ignoring quantum back-reaction at
this point) imply the second-order equation
\begin{equation} \label{ddotX}
 \ddot{X}= (12\pi G)^2 \left(Vp_X+\frac{1}{2}P_VX\right)P_V+ \frac{3}{2}
\left(\Lambda+\mu\gamma^2 V^{\gamma-1}\right)X\,.
\end{equation}

As in the spherical model, we need a suitable state to produce
  appropriate $X$-solutions. For the sign of $\Lambda$ realized in the
  toroidal model, there is no stable thermal state according to the methods
  used in the spherical model. Instead, we use the existence of a local
  maximum of fluctuations in order to restrict possible initial values of $X$
  and its momentum: For fluctuations as shown by CDTs as well as our
  path-integral calculations, using Dirichlet boundary conditions at two
  times, we need a local maximum of $X(\tau)$ and therefore a range with
  $\ddot{X}<0$. (Without loss of generality, we assume $X>0$ at this
  point. The sign of $X$ is irrelevant for fluctuations $s$, and our argument
  here works equally for negative $X$. In this case we would need a local
  minimum of $X$ and therefore $\ddot{X}>0$ for a local maximum of $s$.) For
  the relevant background solutions, $P_V\propto
  -\dot{V}/V<0$. Equation~(\ref{ddotX}) then implies that $p_X$ has to be
  positive and sufficiently large for $\ddot{X}$ to be negative, such that
  $\dot{X}\approx 0$ is possible near a local maximum. Using (\ref{Xdot}), we
  need $p_X\approx -XP_V/V$ which is possible for positive $p_X$ thanks to
  $P_V<0$.  In this range, therefore,
\begin{equation} \label{Xddot}
 \ddot{X}\approx -\frac{1}{2}(12\pi G)^2 VP_Vp_X+ \frac{3}{2}
\left(\Lambda+\mu\gamma^2 V^{\gamma-1}\right)X>0\,.
\end{equation}

We can find generic constraints on our initial conditions in order to get a
local maximum. Supposing we are at the local maximum, we need $p_X=-X
P_V/V$. Using (\ref{Xddot}), this equation implies
\begin{equation}\label{condish}
(12 \pi G)^2P_V^2=(12 \pi G)^2\left(\frac{Vp_X}{X}\right)^2> 3
\left(\Lambda+\mu \gamma^{2}V^{\gamma-1}\right)\,. 
\end{equation}
Therefore, the classical constraint
\begin{equation}
 C=-6\pi GV P_V^2+\frac{\Lambda V}{8\pi G}+ \frac{\mu V^{\gamma}}{8\pi G}<
 \frac{\mu(1-\gamma^2) V^{\gamma}}{8\pi G}
\end{equation}
cannot be zero unless $\gamma=\pm 1$. Because $X$ does not have a simple
algebraic relationship with $V$, the semiclassical constraint $C_{\rm s}$
cannot be satisfied either. The existence of a local maximum of fluctuations
in the toroidal model is therefore an indication that the constraint, unlike
in the spherical model, is not imposed by CDT simulations.  

\begin{figure}
\begin{center}
\includegraphics[scale=0.5]{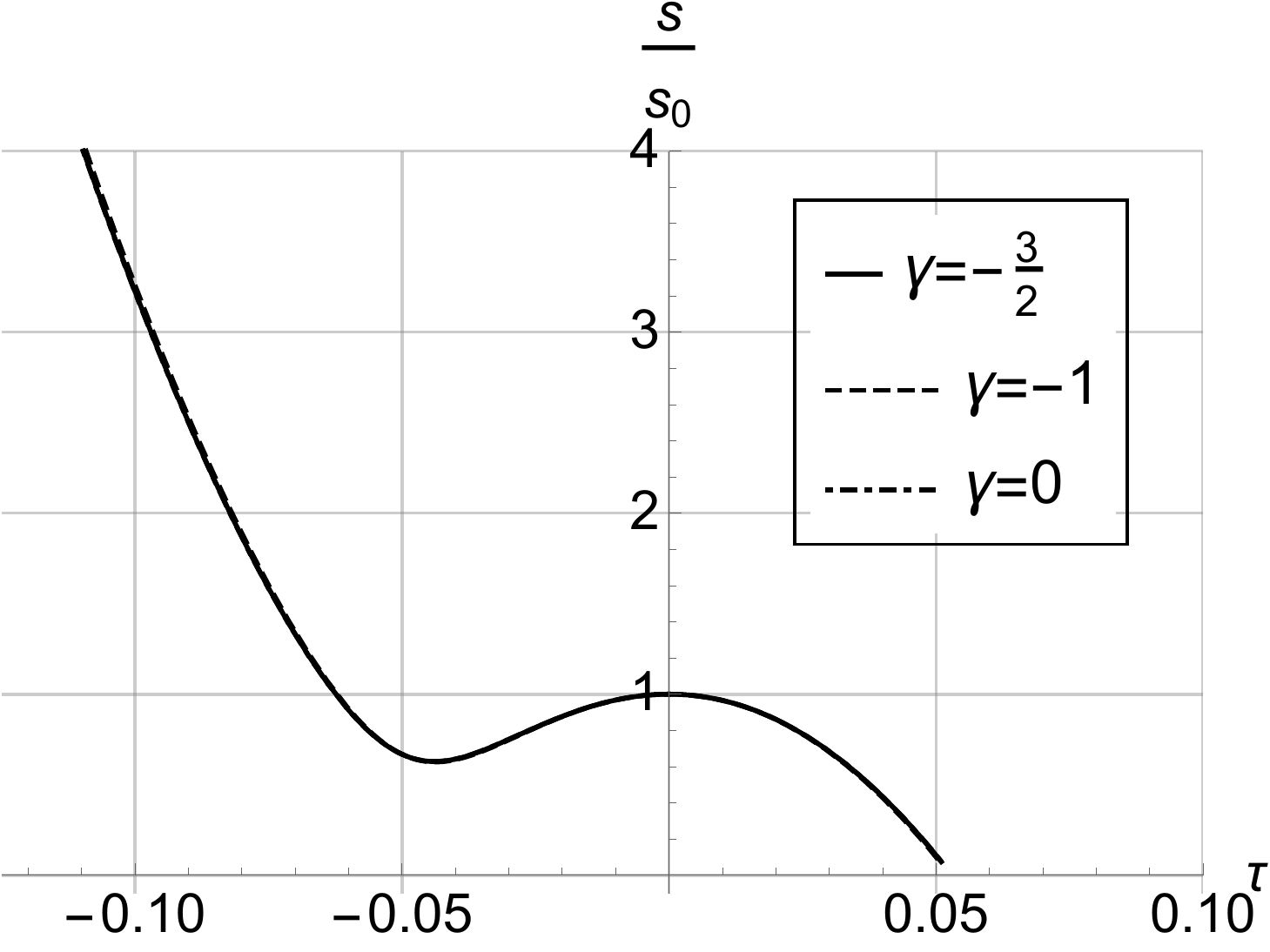} \includegraphics[scale=0.5]{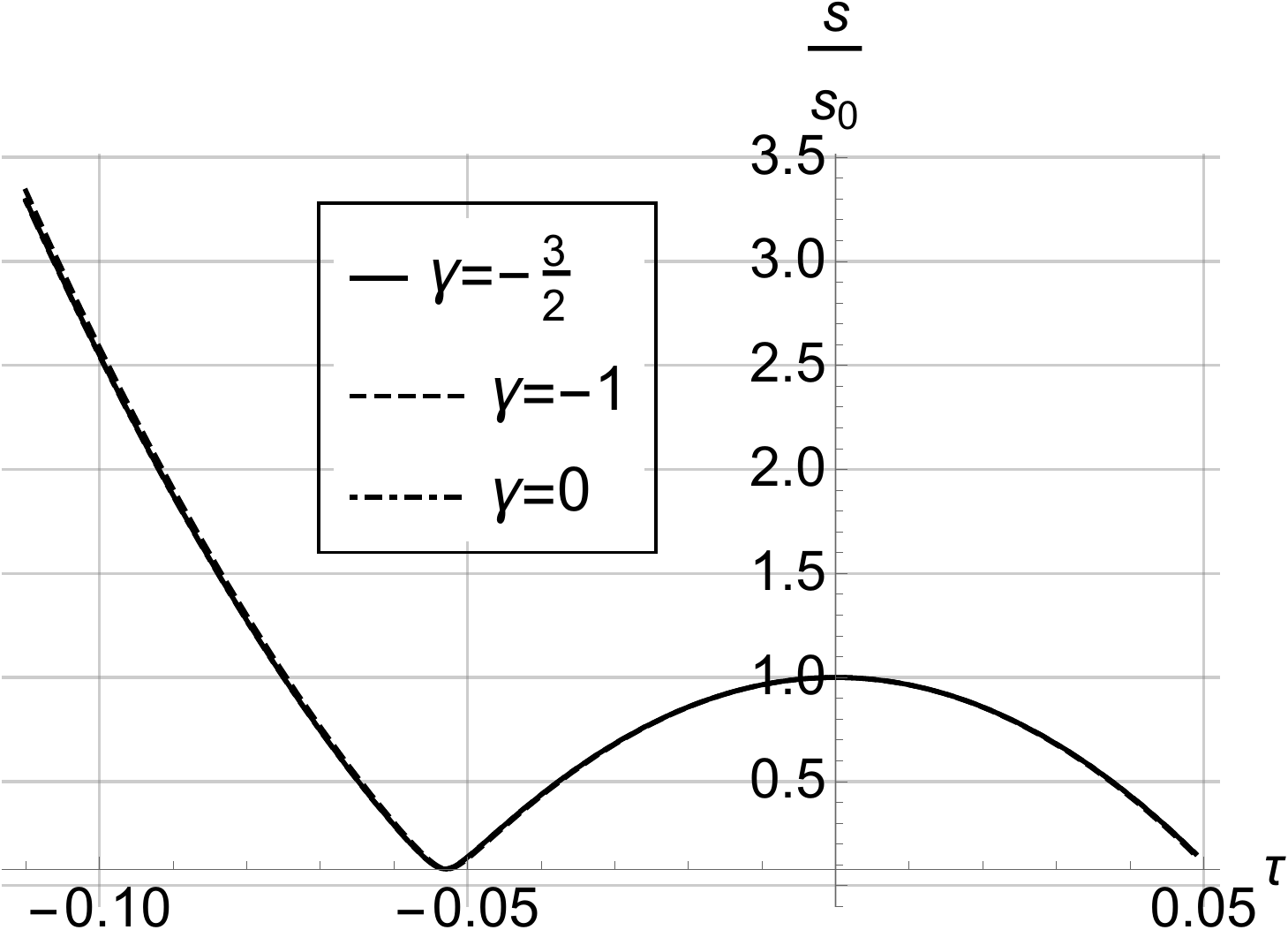}
\caption{Volume fluctuations as a function of time in a situation where the
  inequality (\ref{condish}) is satisfied.  The parameters $\Lambda=3.5\cdot
  10^{-4}$ and $\mu=2.86\cdot 10^5$ are as they appear in the CDT toroidal
  model. Moreover, according to Figs.~\ref{f:w} and \ref{f:Gw}, the volume
  close to the local maximum of fluctuations is approximated as
  $V(0.7)=\bar{w}(0.7)^2\approx 60^2$, which we use in (\ref{condish}) to
  obtain an estimate of $P_{V}$ at our initial time here, $\tau=0$. For
  initial fluctuations, we chose $s_0=100 \sqrt{\hbar}$ (left) and $s_0=300
  \sqrt{\hbar}$ (right), respectively, and then computed $p_{s,0}=-s_0
  P_{V}(0)/V(0)$. For larger fluctuations, it is possible to have vanishing
  fluctuations at two different times. The value of $\gamma$, however, has
  only little influence on local extrema or zeros, see also
  Fig.~\ref{f:Flucs_Sing}, making an origin of the $\mu V^{\gamma}$-term in
  the boundary conditions unlikely. \label{f:Norm} }
\end{center}
\end{figure}

\begin{figure}
\begin{center}
\includegraphics[scale=0.8]{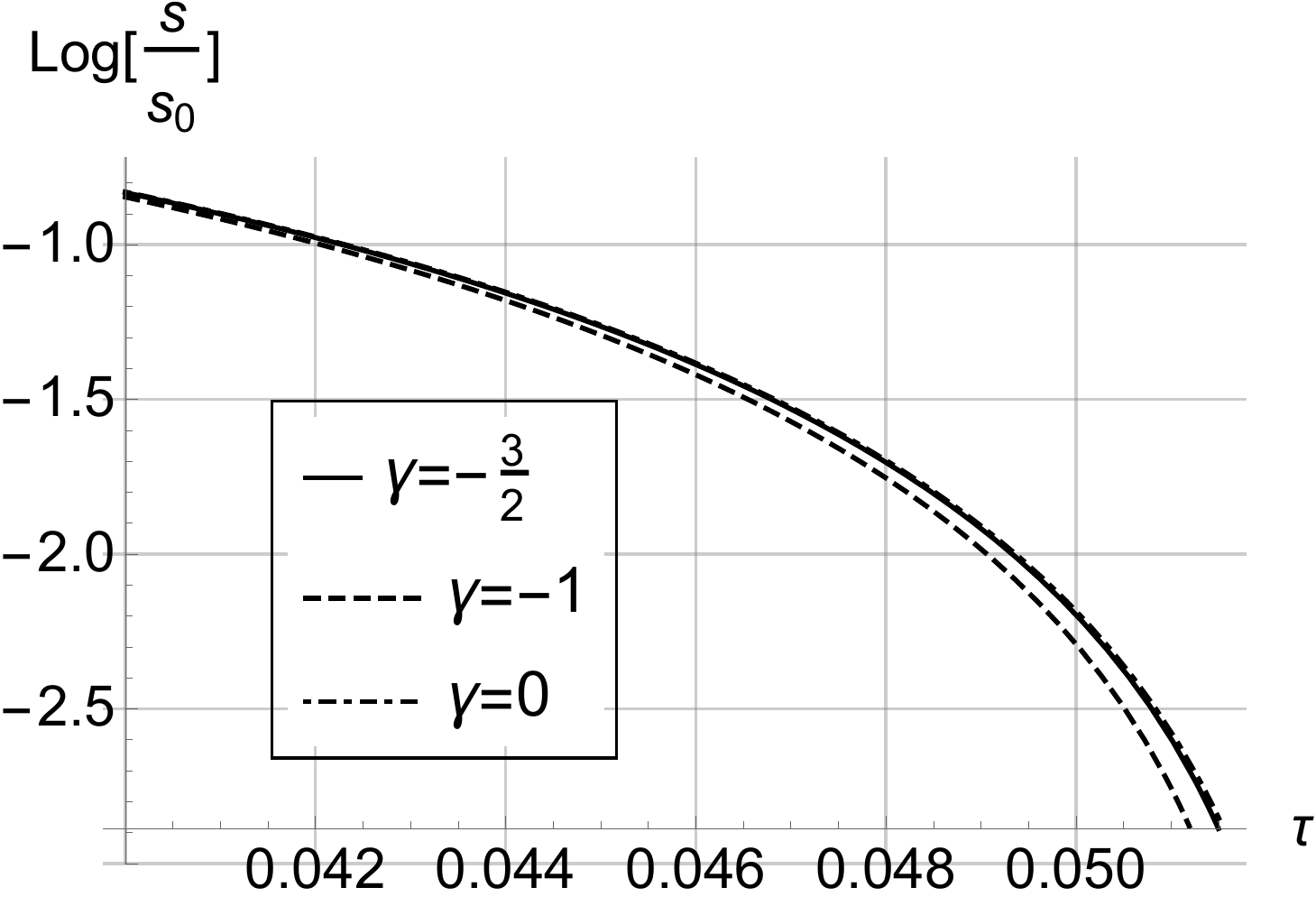}
\caption{Logarithmic plot of fluctuations as in Fig.~\ref{f:Norm}, zoomed in
  on the zero at 
  positive times.
  \label{f:Flucs_Sing} }
\end{center}
\end{figure}

In Fig.~\ref{f:Norm} we show an example of a solution with a local maximum of
$X$. For suitable fluctuations, it is possible to have two zeros and be in
agreement with Dirichlet boundary conditions as imposed in CDT simulations of
the toroidal model. The large maximum value found in our example,
$s_0=300\sqrt{\hbar}$, is in good agreement with fluctuations derived from CDT
simulations \cite{CDTTopology,CDTTorus}. The influence of $\gamma$ on the
existence of two zeros is, however, minor; see
Fig.~\ref{f:Flucs_Sing}. Imposing Dirichlet boundary conditions therefore
cannot explain the origin of a term $\mu V^{\gamma}$ in an effective action.

\section{Quantum origin}

The new term $\mu V^{\gamma}$ with $\gamma\sim -3/2$ in an effective action
derived from CDTs has no classical analog, and so
far it has been difficult to find possible interpretations in terms of known
quantum effects.  We will now address possible origins of this new term. We
will discuss mainly two general possibilities, quantum back-reaction of
fluctuations on expectation values, or factor-ordering choices. It turns out
that each could explain the new term, but not in an easy manner. The term
therefore seems to contain much information about specific effects, which
could be exploited in further studies.

We note that not setting the Hamiltonian constraint equal to zero leads to an
additional term, but, as derived in Sec.~\ref{s:Fix}, it is dust-like in the
Friedmann equation (\ref{EffectiveFriedmann}) and does not have the
$a^{-15/2}$-behavior shown by the CDT term.

\subsection{Non-Gaussianity}

Before we propose our two main proposals to explain the origin of the new term
in the action, we briefly mention the option that such a term could indicate
the influence of a non-Gaussian state put into the straightjacket of a
Gaussian distribution by using quadratic expansions to extract the effective
action.

We start with the cosmological term proportional to $Q^{3/2(1-x)}$ in a
Hamiltonian, using the generic canonical variables (\ref{QP}). It contributes
a semiclassical correction through its second derivative, which is also
proportional to $s^2$, defined in (\ref{sps}). If there is a non-zero
third-order moment, there would be an additional contribution proportional to
the third derivative of $Q^{3/2(1-x)}$. Assuming that a third-order moment of
$Q$ is proportional to $s^3$ (as suggested by \cite{Bosonize,EffPotRealize}),
this new term could be of the form $V^{-3/2}$ provided the third derivative of
$Q^{3/2(1-x)}$ has the same form as the second derivative of $V^{-3/2}\propto
Q^{-9/4(1-x)}$, or
\begin{equation}
 \frac{2x+1}{2(1-x)} = -\frac{9}{4(1-x)}
\end{equation}
which implies $x=-11/4$. The basic configuration variable would then be
$Q\propto a^{15/2}\propto V^{5/2}$, a rather unusual choice.

\subsection{Fluctuation couplings}
\label{s:OrFluc}

We have already seen in Sec.~\ref{s:TorFluc} that quantum back-reaction is
likely relevant in the toroidal model. An effective potential or force as
extracted from CDTs could be a consequence of fluctuations back-reacting on
the volume expectation value. Such terms depend sensitively on quantization
choices, such as what one considers the basic canonical variables
corresponding to some fixed $x$ in $Q$ and $P$, defined in (\ref{QP}). Here,
we illustrate fluctuation couplings based on a few sample solutions.

In (at least) two cases, the system of $Q$, $s$ and their canonical momenta
can be decoupled and solved completely.  For $x=1/4$, we have the canonical
pair
\begin{equation}
 Q=\frac{\sqrt{V_0}a^{3/2}}{2\pi G}\propto w \quad,\quad P=-\sqrt{V_0a}\:\dot{a}
\end{equation}
and the constraint
\begin{equation}
 \frac{8\pi G}{3}C=-P^2-(2\pi GQ)^{2/3}k\ell_0^2+ \frac{1}{3}\Lambda(2\pi GQ)^2\,.
\end{equation}
For $k=0$, the classical constraint is quadratic and therefore the all-orders
constraint agrees with the second-order quantum constraint. For Euclidean
signature and the opposite sign of the cosmological-constant term (in
accordance with \cite{CDTTopology,CDTTorus}), we have
\begin{equation}
  \frac{8\pi G}{3}C=P^2- \frac{1}{3}\Lambda(2\pi GQ)^2\,.
\end{equation}
The semiclassical (or all-orders) constraint is then
\begin{equation}
 \frac{8\pi G}{3}C_{\rm s}=P^2+p_s^2+\frac{U}{s^2}-
 \frac{4\pi^2G^2}{3}\Lambda(Q^2+s^2)\,.
\end{equation}
The background equations of motion
\begin{equation}
 \dot{Q}= \frac{3}{4\pi G}P \quad,\quad \dot{P}=\pi G\Lambda Q
\end{equation}
can be solved in a standard way. They imply that the contribution
\begin{equation}
 \frac{3}{8\pi G}P^2- \frac{\pi G}{2}\Lambda Q^2=c_1
\end{equation}
in $C$ is conserved separately of $C=c$ itself. Moreover, the equations
\begin{eqnarray}
 \dot{s} &=& \frac{3}{4\pi G} p_s\label{sdot}\\
 \dot{p}_s &=&\frac{3}{4\pi G} \frac{U}{s^2} +\pi G\Lambda s  \label{psdot}
\end{eqnarray}
imply that
\begin{eqnarray}
 (sp_s)^{\bullet} &=& \frac{3}{4\pi G} \left(p_s^2+\frac{U}{s^2}\right)+ \pi
 G\Lambda s^2= -\frac{3}{4\pi G}(P^2+2C_s)+ \pi G\Lambda (Q^2+2 s^2) \nonumber\\
 &=& 2(C_s-c_1)+2\pi
 G\Lambda s^2
\end{eqnarray}
and therefore
\begin{equation}
 (sp_s)^{\bullet\bullet}= 4\pi G\Lambda p_s\dot{p}_s= 3\Lambda sp_s
\end{equation}
can be solved for $sp_s$. We obtain
\begin{equation}
sp_s = A\sinh\left(\sqrt{3\Lambda} (\tau-\tau_0)\right)\,,
\end{equation}
and 
\begin{equation}
 p_s=\frac{A\sinh\left(\sqrt{3\Lambda} (\tau-\tau_0)\right)}{s}\,,
\end{equation}
inserted in (\ref{sdot}), implies
\begin{equation}
 s^2 = \frac{A}{2\pi G}\sqrt{3/\Lambda}
 \left(\cosh\left(\sqrt{3\Lambda}(\tau-\tau_0)\right)+B\right)\,.
\end{equation}
Combining our solutions for $sp_s$ and $p_s$, we obtain
\begin{equation}
 p_s=\sqrt{2\pi G A}(\Lambda/3)^{1/4}
 \frac{\sinh\left(\sqrt{3\Lambda}(\tau-\tau_0)\right)}
{\sqrt{\cosh\left(\sqrt{3\Lambda}(\tau-\tau_0)\right)+B}} 
\end{equation}
where
\begin{equation}
 B=\sqrt{1-\frac{U}{A^2}}
\end{equation}
in order to fulfill (\ref{psdot}). Owing to the decoupling, these solutions do
not have back-reaction of moments on expectation values, and are therefore not
consistent with an effective force other than that implied by the
cosmological constant.

The situation is only slightly different for $x=-1/2$. In this case, 
\begin{equation}
 Q=\frac{V_0a^3}{4\pi G}\quad\mbox{and}\quad P=-\frac{\dot{a}}{a}
\end{equation}
and
\begin{equation}
 \frac{8\pi G}{3}C=-4\pi G QP^2-k\ell_0^2(4\pi G)^{1/3} Q^{1/3}+\frac{4\pi
 G}{3}\Lambda Q\,.
\end{equation}
The all-orders quantum constraint (\ref{Cao}) is
\begin{eqnarray}
 C_{\rm all} &=& -\frac{3}{2}
 \left(QP^2+2sPp_s+Qp_s^2\right)-\frac{3}{2}\frac{Q}{s^2}\\
&&- \frac{3}{16\pi G}
 k\ell_0^2(4\pi G)^{1/3} \left((Q+s)^{1/3}+(Q-s)^{1/3}\right) +\frac{1}{2} \Lambda
 Q\,.
\end{eqnarray}
By the canonical trasformation to
\begin{eqnarray}
 X=\sqrt{2(Q+s)} \quad&,&\quad P_X=\sqrt{Q+s}\;(P_Q+p_s)\,,\\
 Y=\sqrt{2(Q-s)}\quad&,&\quad P_Y=\sqrt{Q-s}\;(P_Q-p_s)\,,
\end{eqnarray}
the kinetic term is brought to standard form:
\begin{eqnarray}
 C_{XY} &=& -\frac{3}{2}(P_X^2+P_Y^2)- 6U\frac{X^2+Y^2}{(X^2-Y^2)^2}\\
 && -\frac{3}{16\sqrt[3]{2}\pi G}k\ell_0^2(4\pi G)^{1/3} (X^{2/3}+Y^{2/3})+
 \frac{1}{4}\Lambda(X^2+Y^2)\,.
\end{eqnarray}
The flat Euclidean version
\begin{equation}
 C=\frac{3}{2}(P_X^2+P_Y^2)+ 6\frac{X^2+Y^2}{(X^2-Y^2)^2}-\frac{1}{4}
\Lambda (X^2+Y^2)
\end{equation}
generates equations of motion
\begin{eqnarray}
 \dot{X} &=& 3 P_X\\
\dot{Y} &=& 3 GP_Y\\
\dot{P}_X&=& 12U X\frac{X^2+3Y^2}{(X^2-Y^2)^3}+\frac{1}{2}\Lambda X\\
\dot{P}_Y&=& -12UY\frac{3X^2+Y^2}{(X^2-Y^2)^3}+ \frac{1}{2}\Lambda Y\,.
\end{eqnarray}
For $\frac{1}{2}(X^2+Y^2)=2Q=V/12\pi G$, we obtain the decoupled equation
\begin{equation}
 \frac{1}{2}(X^2+Y^2)^{\bullet\bullet} = 3\left(2C_s+\Lambda(X^2+Y^2)\right) \,.
\end{equation}
The volume 
\begin{equation}
 \frac{1}{2}(X^2+Y^2)= \frac{C_s}{\Lambda} +A\sinh(\sqrt{6\Lambda}\tau)+
 B\cosh(\sqrt{6\Lambda}\tau)
\end{equation}
therefore does not couple to
fluctuations. However, this system is not fully decoupled because the
fluctuation, $\frac{1}{2}(X^2-Y^2)=2s$, obeys the equation
\begin{eqnarray}
 \frac{1}{2}(X^2-Y^2)^{\bullet\bullet} &=& 9
 \left(8U\frac{(X^2+Y^2)^2}{(X^2-Y^2)^3}- \frac{4U}{X^2-Y^2}+
   \frac{1}{6}\Lambda(X^2-Y^2)\right)\\ 
&&+ 9(P_X^2-P_Y^2) \nonumber
\end{eqnarray}
coupled to the volume and the momenta.

As a more complicated example, we consider the case of $x=1/2$, or
$Q=3\ell_0a/4\pi G$ proportional to the scale factor. The semiclassical
constraint
\begin{equation}
 C_{\rm s}=\frac{9}{32\pi^2 G^2}
 \left(\frac{1}{Q}\left(1+\frac{s^2}{Q^2}\right)P^2- 
   2\frac{s}{Q^2}Pp_s+ \frac{1}{Q}p_s^2+\frac{U}{Qs^2}\right)-
 \frac{8\pi^2G^2}{27}\Lambda (Q^3+3Qs^2)
\end{equation}
now generates the equations of motion
\begin{eqnarray}
 \dot{Q} &=& \frac{9}{16\pi^2 G^2} \left(\frac{1}{Q}
   \left(1+\frac{s^2}{Q^2}\right)P-\frac{s}{Q^2}p_s\right)\\
\dot{s} &=& \frac{9}{16\pi^2 G^2} \left(-\frac{s}{Q^2}P+\frac{1}{Q}p_s\right)\\
\dot{P} &=& \frac{9}{32\pi^2 G^2}
\left(\frac{1}{Q^2}\left(1+3\frac{s^2}{Q^2}\right)P^2- 4\frac{s}{Q^3}Pp_s+
  \frac{1}{Q^2}p_s^2+ \frac{U}{Q^2s^2}\right)\nonumber\\
&&+ \frac{8\pi^2G^2}{9}\Lambda (Q^2+s^2)\\
\dot{p}_s &=& -\frac{9}{16\pi^2 G^2}
\left(\frac{s}{Q^3}P^2-\frac{1}{Q^2}Pp_s-\frac{U}{Qs^3}\right)+
\frac{16\pi^2G^2}{9}\Lambda Qs\,.
\end{eqnarray}
A combination of these equations  leads to the rather messy equation
\begin{eqnarray}
 \ddot{Q} &=& \left(\frac{81}{256\pi^4G^4}\right)^2 \left(-\frac{1}{2Q^3}
   \left(1+6\frac{s^2}{Q^2}+ 3\frac{s^4}{Q^4}\right)P^2+ \frac{3s}{Q^4}
   \left(1+\frac{s^2}{Q^2}\right)Pp_s\right.\nonumber\\
&&\left.-
   \frac{1}{2Q^3}\left(1+3\frac{s^2}{Q^2}\right)p_s^2-
   \frac{1}{2Q^3s^2}\left(1-\frac{s^2}{Q^2}\right)U\right)
+\frac{1}{2}\Lambda \frac{Q^4+s^4}{Q^3}\,.
\end{eqnarray}
For $Q^n$, it implies
\begin{eqnarray}
 (Q^n)^{\bullet\bullet} &=& nQ^{n-4} (Q^3\ddot{Q}+ (n-1)Q^2\dot{Q}^2)\nonumber\\
&=& nQ^{n-4}\left(\frac{9}{16\pi^2G^2}\right)^2 \left(\frac{1}{2}
  \left(2n-3+2(2n-5) 
    \frac{s^2}{Q^2}+(2n-5) \frac{s^4}{Q^4}\right)P^2\right.\nonumber\\
&&- \frac{s}{Q}(2n-5)
  \left(1+\frac{s^2}{Q^2}\right)Pp_s- \frac{1}{2}
  \left(1-(2n-5)\frac{s^2}{Q^2}\right)p_s^2\nonumber\\
&&\left. -\frac{1}{2s^2}
  \left(1-\frac{s^2}{Q^2}\right) U\right) +\frac{1}{2}n \Lambda
(Q^4+s^4) Q^{n-4}\,.
\end{eqnarray}
In particular, for $n=2$, the resulting equation for the square of the scale
factor can be simplified because it has the same combination of $Pp_s$ and
$p_s^2$ as the constraint:
\begin{eqnarray} \label{Q2}
 (Q^2)^{\bullet\bullet}
&=& 2\frac{9}{16\pi^2G^2} \left(\frac{P^2}{Q^2}+\frac{U}{Q^4}\right)
-\frac{9}{8\pi^2 G^2} \left(1+\frac{s^2}{Q^2}\right)\frac{c}{Q}\nonumber\\
&&+\frac{2}{3} \Lambda (Q^2-2s^2)\,.
\end{eqnarray}

\begin{figure}[!tbp]
    \includegraphics[scale=0.4]{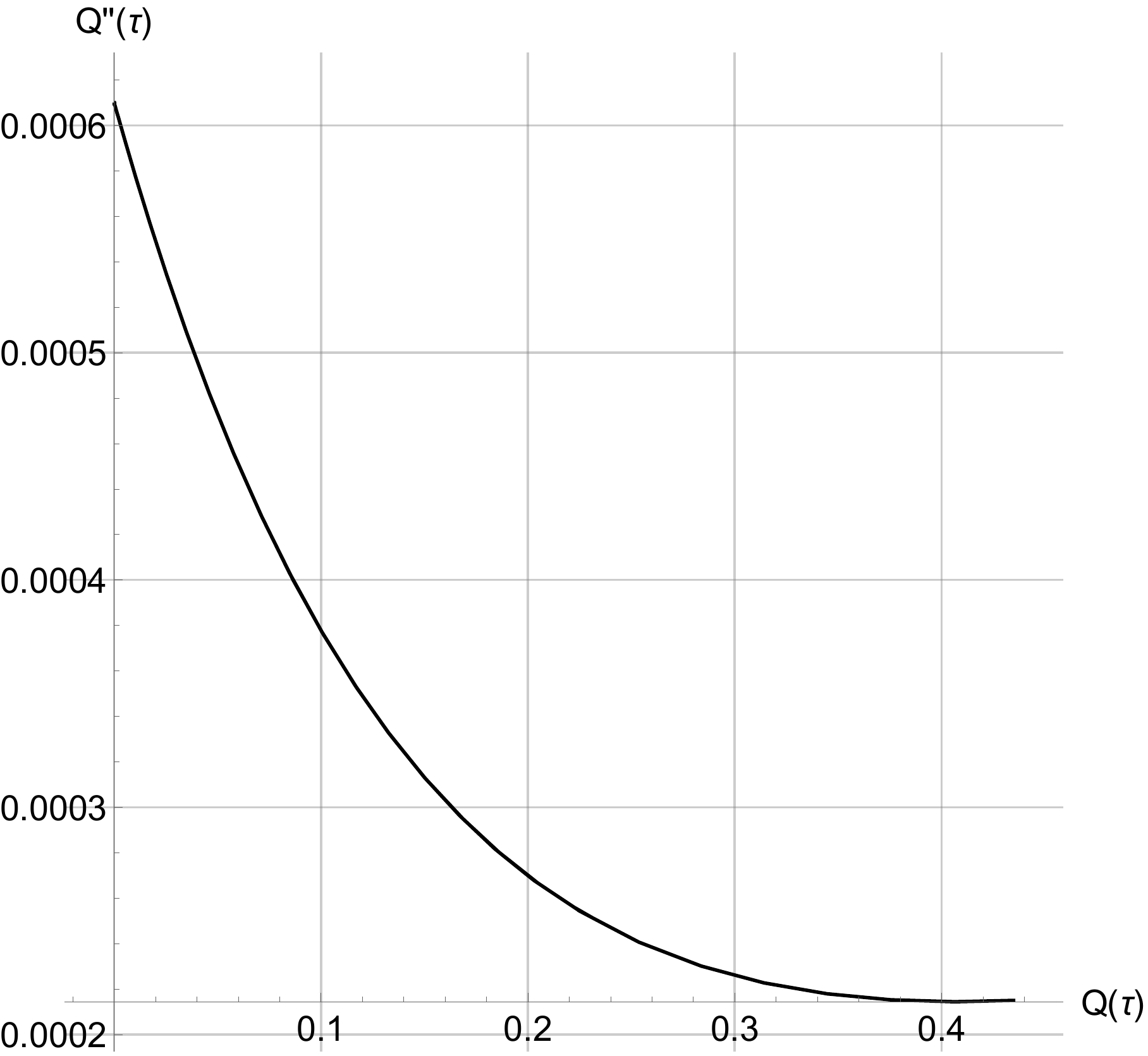}
  \hfill
    \includegraphics[scale=0.4]{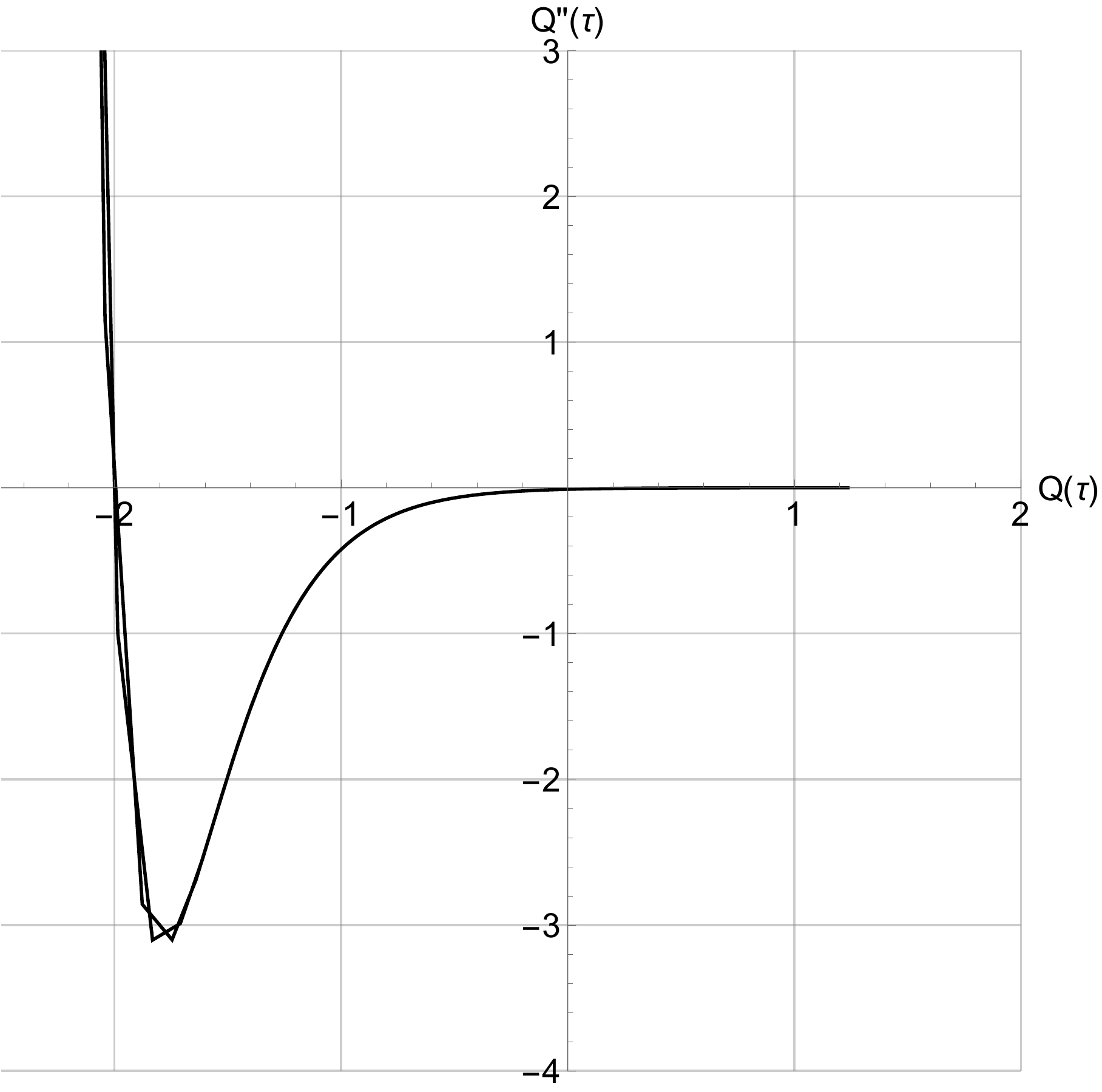}
    \caption{The effective force (\ref{EffForce}) in terms of $Q$ for
      $U=\hbar^2/4$ with $C_{\rm s}=0$ (left) and $C_{\rm s}=0.05$
      (right). The cosmological constant $\Lambda=3.5 \cdot 10^{-4}$ is chosen
      as in 
      the toroidal model of CDTs.
\label{fig:f1}}
\end{figure}

\begin{figure}[htbp]
\begin{center}
\includegraphics[width=15cm]{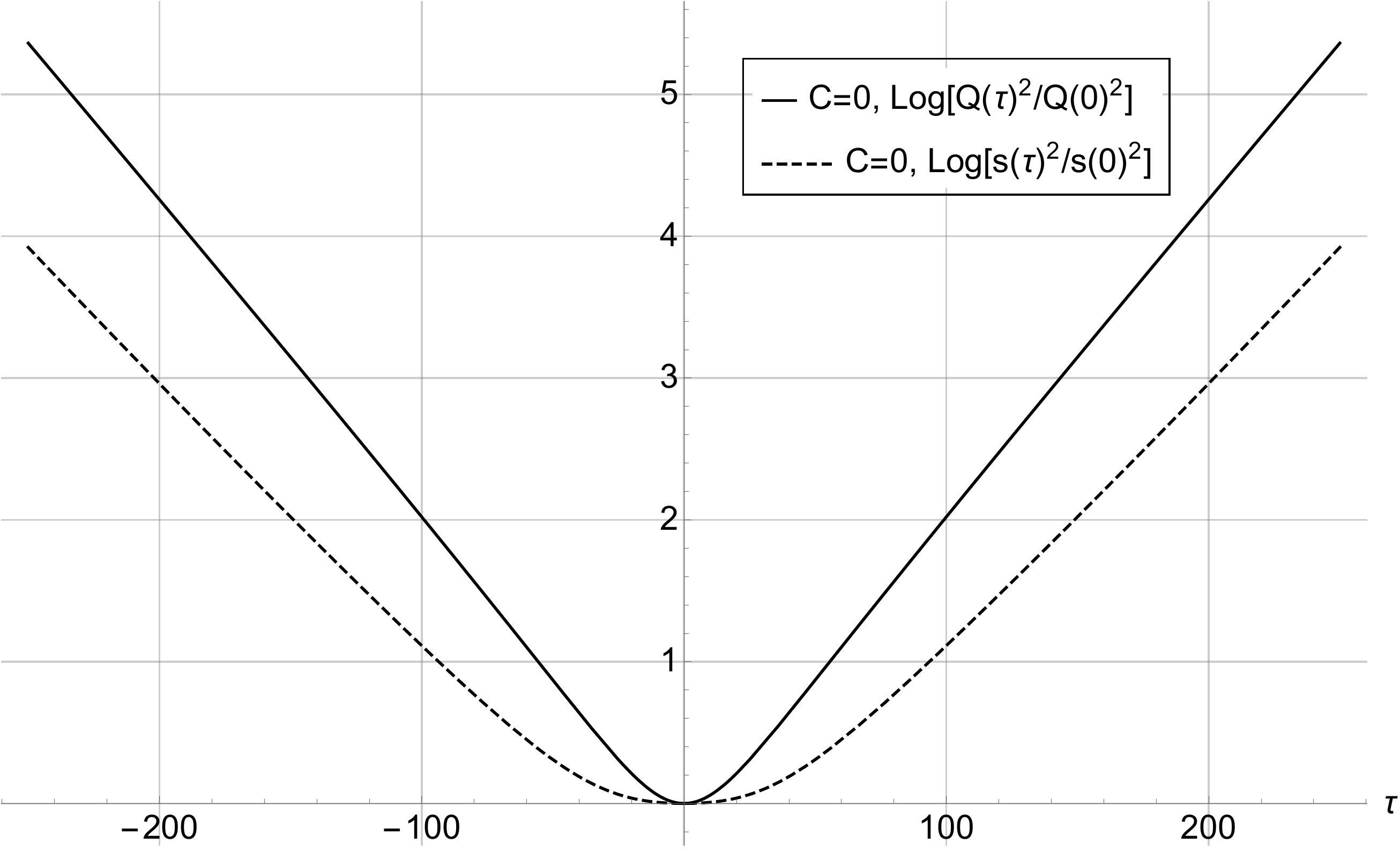}

\includegraphics[width=15cm]{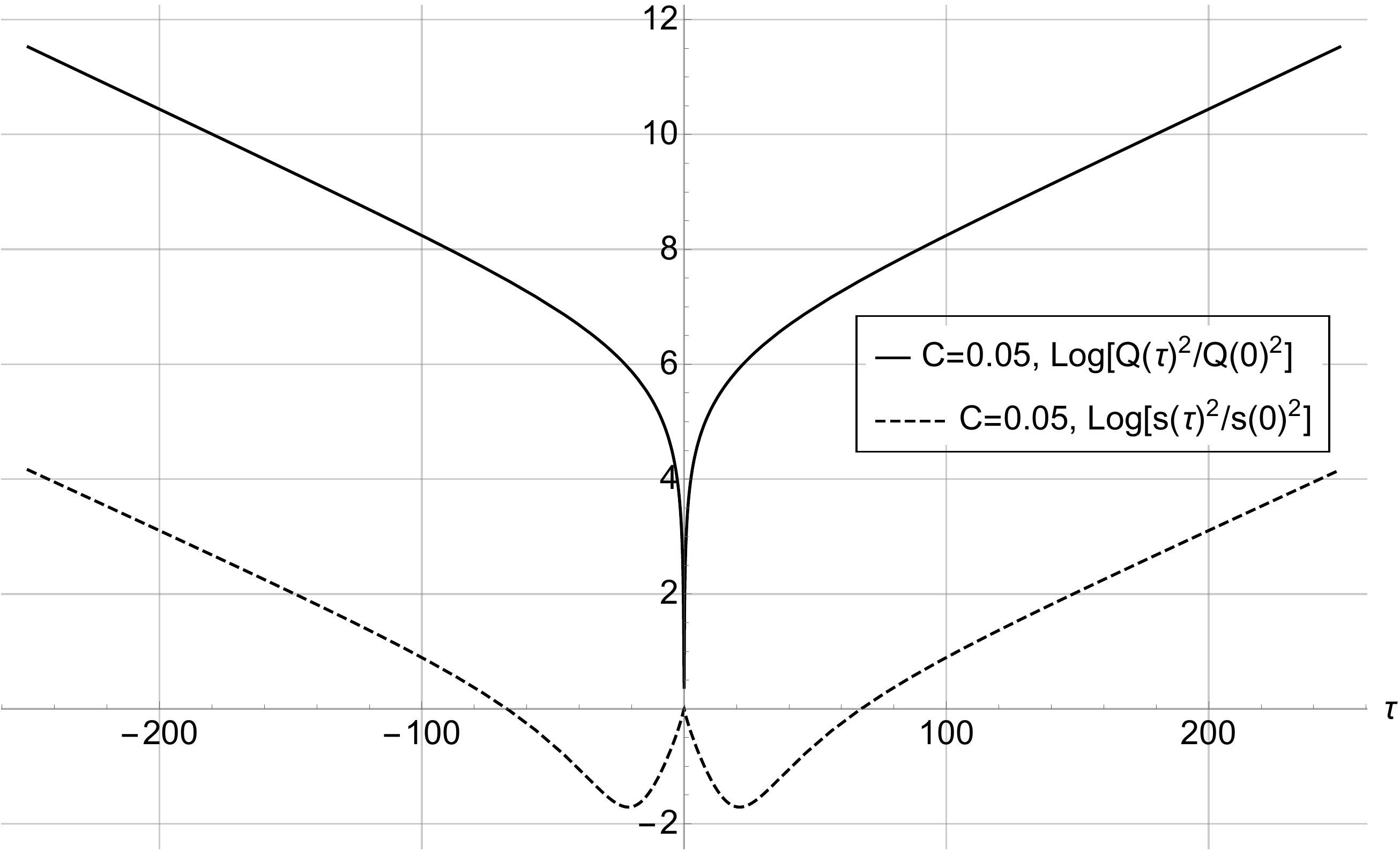}
\caption{Logarithmic plots of $Q(\tau)$ (solid) and $s(\tau)$ (dashed) for
  $C_{\rm s}=0$ (top) and $C_{\rm s}=0.05$ (bottom). For small $Q$, $\ddot{Q}$
  is negative for $C_{\rm s}=0.05$ but not for $C_{\rm s}=0$. Not imposing the
  constraint could therefore be related to a new potential such as
  $\mu V^{\gamma}$.  \label{f:Q-s}}
\end{center}
\end{figure}

Here, we can identify several effective force terms. Most of them are
positive, but for small $Q$ with $c>0$ and $s^2c>2QU$ the dominant effective
force 
\begin{equation} \label{EffForce}
F_C= -\frac{9}{8\pi^2 G^2} \frac{s^2C_{\rm s}}{Q^3}
\end{equation}
is negative. Qualitatively, this expectation agrees with the effective
potential obtained from CDTs. Numerically, the effective force ${\rm d}^2
Q/{\rm d} \tau^2$ can be deduced from $Q(\tau)$. It is plotted parameterically
in Fig.~\ref{fig:f1} for both $C_{\rm s}=0$ and $C_{\rm
  s}=c\not=0$. Intriguingly, $\ddot{Q}$ is negative for small $Q$, as would be
expected for a force from an inverse-power potential such as the effective
force $-U'(V)=-\Lambda-\mu\gamma V^{\gamma-1}$ deduced in \cite{CDTTorus}.  In
Fig.~\ref{f:Q-s}, we show the evolution of $Q$ and its fluctuations $s$ with
both options: strictly imposing the constraint $C_{\rm s}=0$ and with $C_{\rm
  s}=c\not=0$, respectively. This plot further demonstrates that an effective
force from quantum back-reaction can lead to negative $\ddot{Q}$ at small $Q$,
but only if the constraint is not imposed. The phase during which $\ddot{Q}<0$
is accompanied by decreasing fluctuations $s$. This behavior may be difficult
to resolve in CDT simulations where discretization effects are larger at small
$Q$. Such decreasing fluctuations in a toroidal model have not been seen in
\cite{CDTTopology,CDTTorus}.  

A negative term only exists if we do not strictly impose the constraint,
$C_{\rm s}=c\not=0$. The term is then proportional to $\hbar$, which is the
typical behavior of the variance $s^2$. Moreover, if $s$ does not depend on
$V_0$, as indicated by our preceding path-integral results (\ref{Gw}), the new
term in an effective potential for (\ref{Q2}) scales like
\begin{equation} \label{EQ}
 s^2\frac{c}{Q^2}\propto \frac{\hbar V_0}{V_0^{2/3}}=\hbar\ell_0\,.
\end{equation}
There is an interesting contrast with the scaling in our second potential
origin of the $\mu V^{\gamma}$-term, discussed in the next subsection.

\subsection{Ordering terms}
\label{s:OrOr}

Factor-ordering choices are another potential source of non-classical terms in
an effective action.  For easier reference to the volume, we now write
canonical variables as
\begin{equation}
 Z=V^p\quad,\quad P_Z=\frac{1}{p} V^{1-p} P_V
\end{equation}
where $V=V_0a^3$ and $P_V=-(4\pi G)^{-1}\dot{a}/a$.  For a symmetric
Hamiltonian quadratic in momenta, the main factor ordering ambiguity arises
from the fact that there are two standard symmetric orderings, $Z^qP_Z^2Z^q$
and $\frac{1}{2}(Z^{2q}P_Z^2+P_Z^2Z^{2q})$, that can be used to quantize
$Z^{2q}P_Z^2$ for some $q$ determined by the classical constraint. Any
linear combination of these two orderings can be obtained by adding a multiple
of 
\begin{equation} \label{Ord1}
 2Z^qP_Z^2Z^q- (Z^{2q}P_Z^2+P_Z^2Z^{2q}) = 2q^2\hbar^2Z^{2q-2}
\end{equation}
to a given ordering. A new ordering term $2q^2\hbar^2Z^{2q-2}$ is then
implied,  which can be compared with an effective potential such as $\mu
V^{\gamma}$. 

Other, less standard ordering choices, such as using
\begin{equation}
 P_ZZ^{2\epsilon}P_Z-Z^{\epsilon}P_Z^2Z^{\epsilon}=-\epsilon(\epsilon-1)\hbar^2
 Z^{2\epsilon-2}
\end{equation}
or
\begin{equation} \label{Zq}
 Z^{q+\epsilon}P_Z^2Z^{q-\epsilon}+Z^{q-\epsilon}P_Z^2Z^{q+\epsilon}- 
(Z^{2q}P_Z^2+P_Z^2Z^{2q}) = 2(q^2-\epsilon^2)\hbar^2Z^{2q-2}
\end{equation}
lead to the same power law as in (\ref{Ord1}). For a constraint of the form
(\ref{C}) with $x=-1/2$, we need
\begin{equation}
 Z^{2q}P_Z^2 = \frac{1}{p^2} V^{2pq-2p+2}P_V^2 \sim V P_V^2\,,
\end{equation}
which implies $2p(q-1)=-1$. However, the ordering term is then
\begin{equation}
 Z^{2q-2}=V^{2p(q-1)}=V^{-1}
\end{equation}
rather than $V^{\gamma}$ with $\gamma$ close to $-3/2$ as suggested by
\cite{CDTTorus}.

If the constraint is not quadratic in $P_Z$, owing to higher-order corrections
in the momentum which are possibly indicated by \cite{CDTEffAc}, there is more
freedom in ordering terms. For instance, we have
\begin{equation} \label{Order4}
 2Z^qP_Z^4Z^q-(Z^{2q}P_Z^4+P_Z^4Z^{2q})=12 q^2\hbar^2Z^{q-1}P_Z^2Z^{q-1}+
 2q^2(q^2-1)\hbar^2Z^{2q-4}\,.
\end{equation}
The first term shows quantum corrections to the kinetic term. The last term
has a different power from the quadratic result in (\ref{Zq}) and can lead to
\begin{equation}
 Z^{2q-4}=V^{2p(q-2)}=V^{-3/2}
\end{equation}
if $p=1/4$ and $q=-1$. In this estimate, we still use $2p(q-1)=-1$ assuming
that the kinetic term is quantum corrected as in
$Z^q(P_Z^2+\ell^2P_Z^4+\cdots)Z^q$. This value of $p$ implies
$x=1-\frac{3}{2}p=5/8$ in the previous parameterization. However, the
coefficient $q^2-1$ in (\ref{Order4}) is then equal to zero.

A final modification gives the desired term: We have
\begin{eqnarray}
 &&Z^{q+\epsilon}P_Z^4Z^{q-\epsilon}+Z^{q-\epsilon}P_Z^4Z^{q+\epsilon}-
(Z^{2q}P_Z^4+P_Z^4Z^{2q})\nonumber\\
&=&12 \hbar^2  (q^2-\epsilon^2)Z^{q-1}P_Z^2Z^{q-1}- 
 2\hbar^2(q^2-\epsilon^2)(1-q^2+\epsilon^2+12\epsilon)Z^{2q-4}\,.
\end{eqnarray}
The power of $Z$ in the last term has not changed, but we have a new
coefficient. Still using $q=-1$, the final ordering term is
\begin{equation}
 - 2\hbar^2(1-\epsilon^2)(\epsilon+12)\epsilon Z^{-6}\,.
\end{equation}
It is non-zero for generic $\epsilon$, and can be rather large for large
$\epsilon$. In order to obtain a value around $\mu=2.86\cdot 10^5$ (with
$\hbar=1$), we need $\epsilon\approx 17$. This value may seem unnaturally
large, but it is encouraging that it is close to an integer.

A possible Hamiltonian operator could have the kinetic term
\begin{eqnarray}
 H_{\rm kin}&=& -\frac{3}{2} Z^{-1}(P_Z^2+\ell^2P_Z^4)Z^{-1}\nonumber\\
&&-\frac{3}{2}
   \ell^2\left(Z^{-1+\epsilon}P_Z^4Z^{-1-\epsilon}+
     Z^{-1-\epsilon}P_Z^4Z^{-1+\epsilon}
     -Z^{-2}P_Z^2-P_Z^2Z^{-2}\right)\nonumber\\
&=& -\frac{3}{2} Z^{-1}(P_Z^2+\ell^2P_Z^4)Z^{-1}
  -18\ell^2\hbar^2(1-\epsilon^2) Z^{-2}P_Z^2Z^{-2}\nonumber\\
&&-
  3\ell^2\hbar^2(1-\epsilon^2)(\epsilon+12)\epsilon Z^{-6}\,.
\end{eqnarray}
Using $p=1/4$, we have 
\begin{eqnarray}
 H_{\rm kin}
&=& -\frac{3}{2}
\left(V^{-1/4}\left(4(V^{3/4}P_V+P_VV^{3/4})^2+
16\ell^2(V^{3/4}P_V+P_VV^{3/4})^4\right)V^{-1/4}\right)\nonumber\\
&&  -72\ell^2\hbar^2(1-\epsilon^2)
V^{-1/2}(V^{3/4}P_V+P_VV^{3/4})^2V^{-1/2}\nonumber\\
&&-  3\ell^2\hbar^2(1-\epsilon^2)(\epsilon+12)\epsilon V^{-3/2} \label{HOrd}
\end{eqnarray}
in terms of $V$. Here, the symmetric ordering of $P_Z^2$ is equal to
\begin{equation} \label{ordering}
 (V^{3/4}P_V+P_VV^{3/4})^2 =  4V^{3/4}P_V^2V^{3/4}- \frac{15}{16}\hbar^2
 V^{-1/2}\,.
\end{equation}

The factor-ordering term to be compared with the new contribution to the
action is the last contribution in (\ref{HOrd}), proportional to $\hbar^2$ and
scaling like $V_0^{-3/2}=\ell_0^{-9/2}$. This behavior is rather different
from the potential term (\ref{EQ}) seen from quantum back-reaction. Varying
$\hbar$ and $V_0$ in causal dynamical triangulations can therefore distinguish
between these two options.

\section{Conclusions}

We have derived several minisuperspace results for fluctuations
  in models studied previously in CDTs. We have found qualitative agreement
  and potential explanations of subtle features such as the issue of fixing
  time and imposing the constraints, the scaling behavior of fluctuations with
  respect to $\Lambda$, or the possible origin of new non-classical terms in
  effective actions. However, in all these issues there is room for further
  explorations.

In the spherical model, Sec.~\ref{s:GaugeFixedSol}, we have been able to
identify a crucial difference between background solutions of the constraint
(the Friedmann equation) compared with solutions in which the constraint is
assumed to be non-zero (but then remains constant). This difference may
explain why the background solution of the volume extracted from CDTs agrees
with solutions of the Friedmann equation, even though fixing the time gauge
would seem to relax the Friedmann equation and only impose the less
restrictive Raychaudhuri equation. In the toroidal model, however, the
difference between imposing the constraint and not doing so is much less
pronounced. We have found several indications that it may not be imposed, in
contrast to the spherical model.

In fact, not imposing the constraint may be one reason why CDTs in toroidal
models have indicated the presence of an unexpected non-classical term, $\mu
V^{\gamma}$ with $\gamma$ close to $-3/2$, in an effective
action. Unfortunately, the detailed derivation through quantum back-reaction
in minisuperspace models, shown in Sec.~\ref{s:OrFluc}, indicates that such a
term, though possible, does not seem natural. Another potential origin,
through factor-ordering choices shown in Sec.~\ref{s:OrOr}, appears perhaps
more natural but also requires some work to obtain the required power-law
behavior. In conclusion, it seems difficult to explain the term in a unique
fashion. Our derivations indicate how this issue could be explored further:
Quantum corrections that could account for terms seen in CDTs have different
dependencies, (\ref{EQ}) compared with (\ref{ordering}), on the averaging
volume $V_0$ or $\hbar$. These constants are usually fixed in CDT simulations,
but, as we suggest, running several simulations with different choices for
these values can shed additional light on possible quantum corrections.

Another suggestion based on the scaling behavior, this time of volume
fluctuations with respect to the cosmological constant, follows from our
derivations of minisuperspace fluctuations using two different methods:
path-integral calculations and moment dynamics. In the spherical model, we
have been able to rederive the universal behavior found in CDTs. In the
toroidal model, we have found a new universal behavior that suggests plotting
relative volume fluctuations as a function of time multiplied by a specific
power of the cosmological constant, (\ref{stauTor}). The exponent depends on
$\gamma$, which we assumed to be constant in our calculation. More detailed
investigations of the scaling behavior in CDTs could therefore show additional
features such as a potential running of $\gamma$.

Our results have therefore suggested several ``CDT experiments'' which,
motivated by detailed analytical calculations, have the potential of further
illuminating some of the main open questions in this framework. Open questions
also remain on the minisuperspace side, for instance related to finer details
in the plots of CDT fluctuations that we have not been able to reproduce in
the spherical model, or to the imposition of Dirichlet boundary conditions
which appears somewhat unnatural in the toroidal minisuperspace model. With
further studies on both sides of the correspondence between CDTs and
minisuperspace models analyzed here, it may become possible to use CDTs to
test the minisuperspace approximation, or to extend calculations to
midisuperspace models \cite{CDTMidi}.  

\section*{Acknowledgements}

We thank Suddhasattwa Brahma for comments. This work was supported in part by
NSF grant PHY-1607414. JM is supported by the Sonata Bis Grant
DEC-2017/26/E/ST2/00763 of the National Science Centre Poland and the
Mobilno\'s\'c Plus Grant 1641/MON/V/2017/0 of the Polish Ministry of Science
and Higher Education.

\begin{appendix}

\section{Solving the Raychaudhuri equation}

The aim of this appendix is to derive the general solution to the instanton
equation
\begin{equation}
\left( \frac{{\rm d}a}{{\rm
      d}\tau}\right)^2=1-\frac{\Lambda}{3}a^2-\frac{c}{a}, \label{insteq1} 
\end{equation}
for $a \geq 0$. For this purpose, let us introduce a new variable 
\begin{equation}
u := \frac{a_0}{a}\,,  \label{defa}
\end{equation}
together with the rescaled imaginary time $s:=\tau/a_0$, which transforms
Eq.~(\ref{insteq1}) to
\begin{equation}
\left( \frac{{\rm d}u}{{\rm d}s}\right)^2=u^4-u^2-\tilde{c}
u^5\,, \label{insteq2} 
\end{equation}
where $\tilde{c} := c/a_0$. The next step is to introduce a new time variable
$w$, defined such that
\begin{equation}
{\rm d}s = \frac{{\rm d}w}{u},
\end{equation}
which is well-defined since we consider $u$ being positive definite, and when
applied to (\ref{insteq2}) gives
\begin{equation}
\left( \frac{{\rm d}u}{{\rm d}w}\right)^2=u^2-1-\tilde{c} u^3\,, \label{insteq3}
\end{equation}
which resembles the case of an oscillator with cubic anharmonicity. A further
change of variable,
\begin{equation}
u = -\frac{4}{\tilde{c}}v+\frac{1}{3\tilde{c}}\,, \label{uvchange}
\end{equation}
transforms Eq.~(\ref{insteq3}) into the Weierstrass equation
\cite{Weierstrass}:
\begin{equation}
\left(\frac{{\rm d}v}{{\rm d}w}\right)^2=4v^3-g_2v-g_3=4(v-e_1)(v-e_2)(v-e_3)\,, 
\label{insteq4} 
\end{equation}
where 
\begin{equation}
g_2 = \frac{1}{12} \ \ \text{and} \ \ g_3=
\frac{\tilde{c}^2}{16}-\frac{1}{216}\,.  
\end{equation}
The constants $e_1$, $e_2$ and $e_3$ are roots of the polynomial equation
$4e_i^3-g_2e_i-g_3=0$, which for further convenience are ordered such that
$e_1>e_2>e_3$. Equation~(\ref{insteq4}) has solutions in the form of
Weierstrass elliptic $\wp$ function
\begin{equation}
v(w)= \wp(w-w_0; g_2, g_3)\,,  \label{WeierstrassSol1}
\end{equation}
where $w_0$ is a constant of integration. The Weierstrass $\wp$ function is a
doubly periodic function with the two half-periods $\omega_1$ and $\omega_2$:
\begin{equation}
\omega_1 = \int_{e_1}^{+\infty} \frac{{\rm d}v}{\sqrt{4v^3-g_2v-g_3}} \ \
\text{and} \ \ 
\omega_1 = i \int_{-\infty}^{e_3} \frac{{\rm d}v}{\sqrt{4v^3-g_2v-g_3}}\,. 
\end{equation}
Due to the third-order form of the polynomial in Eq.~(\ref{insteq4}), there
are in general two branches of solutions, for positive and negative values of
$v$.  In the considered case for $\tilde{c} \in \left[0, \frac{2}{3\sqrt{3}}
\right]$ there are two types of solutions: (i) unbounded solutions in the range
$v \in [e_1, +\infty)$ and (ii) oscillatory solutions in the range $v \in [e_3,
e_2]$. The solution corresponding to the branch (i) is the one given by
Eq.~(\ref{WeierstrassSol1}), while the solution corresponding to the second
branch is obtained by taking $v(w+\omega_3)$, where
$\omega_3:=\omega_1+\omega_2$:
\begin{equation}
v(w+\omega_3)= e_2+\frac{2e_2^2+e_3e_1}{v(w)-e_2}=
e_2+\frac{2e_2^2+e_3e_1}{\wp(w-w_0; g_2, g_3)-e_2}\,,  
\label{WeierstrassSol2}
\end{equation}   
where the addition theorem for the Weierstrass elliptic function has been used
\cite{Weierstrass}. The second solution is the one we are interested in since
it corresponds to the solution in the positive domain of the scale
factor. Namely, applying Eq.~(\ref{WeierstrassSol2}) to Eq.~(\ref{uvchange})
and then to Eq.~(\ref{defa}) we find the solution:
\begin{equation} 
a(w)= \frac{3 \tilde{c}a_0}{1-12v(w+\omega_3)}=
\frac{3 \tilde{c}a_0}{1-12 e_2-12(2e_2^2+e_3e_1)/(\wp(w-w_0; g_2,
    g_3)-e_2)}\,. \label{WeierstrassSol3} 
\end{equation}
In the special case when $\tilde{c}=0$, the solution to Eq.~(\ref{insteq3}) can
easily be found to be $u(w) = \cosh(w-w_0)$, which leads to
\begin{equation}
a(w) = \frac{a_0}{\cosh(w-w_0)}. \label{solawc0}
\end{equation}
The new time variable $w$ used above can be related with the $s =\tau/a_0$
time variable through the integral
\begin{equation}
 s = \int_0^w\frac{{\rm d}w'}{u(w')}.  
\end{equation} 
In the case of $\tilde{c}=0$ the integration can be performed in a
straightforward manner, giving (for $w_0=0$):
\begin{equation}
 s = 2 \arctan \left( \tanh (w/2)\right),
\end{equation}
which can be rewritten into the form $ \cosh (w) = 1/\cos(s)$, with the use of
which Eq.~(\ref{solawc0}) can be expressed as
\begin{equation}
a(s) = a_0\cos(s),
\end{equation}
which is correctly the Wick rotated version of Eq.~(\ref{dSSolution}). 
\end{appendix}


\providecommand{\href}[2]{#2}\begingroup\raggedright\endgroup

\end{document}